\documentclass[aps,prx,reprint,superscriptaddress,longbibliography, floatfix]{revtex4-2}

\usepackage[utf8]{inputenc}
\usepackage{hyperref}
\usepackage{graphicx}
\usepackage{adjustbox}	
\usepackage{bm}
\usepackage{comment}
\usepackage{array}

\usepackage{amsmath,amssymb}
\newcommand\norm[1]{\left\lVert#1\right\rVert} 
\begin{document}

\title{Efficient Computation of Stellarator Coils with an Augmented Lagrangian Optimization Method}

\author{Pedro F. Gil}
\affiliation{Max-Planck-Institut für Plasmaphysik, 85748 Garching, Germany}

\author{Weiping Li}
\affiliation{BASIS Independent Silicon Valley, California, CA 95126, USA}
\affiliation{Courant Institute of Mathematical Sciences, New York University, New York, NY 10012, USA}

\author{Julianne Stratton}
\affiliation{Courant Institute of Mathematical Sciences, New York University, New York, NY 10012, USA}

\author{Alan A. Kaptanoglu}
\affiliation{Courant Institute of Mathematical Sciences, New York University, New York, NY 10012, USA}

\author{Eve V. Stenson}
\affiliation{Max-Planck-Institut für Plasmaphysik, 85748 Garching, Germany}


\begin{abstract}
Finding feasible coils for stellarator fusion devices is a critical challenge of realizing this concept for future power plants. 
Current design efforts struggle to navigate the highly nonconvex optimization landscape, spend considerable resources scanning the parameter space, and may produce suboptimal coils. In this work, we present an augmented Lagrangian approach to tackle the ill-posed problem of coil optimization. 
We illustrate its effectiveness and versatility by generating coils for five stellarators with very different symmetries and magnetic field shaping. In all cases, we find Pareto-optimal coil solutions that in various ways outperform published coil sets.
\end{abstract}

\maketitle


\section{Introduction}
The stellarator, a toroidal magnetic confinement device, stands as a promising candidate for achieving controlled nuclear fusion. Unlike its axisymmetric counterpart, the tokamak, the stellarator relies on complex, three-dimensional magnetic fields to confine the plasma and therefore does not require the use of induced currents that can lead to plasma instabilities~\cite{Schuller_1995, Jahns_1978, Helander_2012}. This also allows for steady-state operation, which provides a clear advantage over other technologies to become an energy source. The confining magnetic fields are generated by a system of external coils, the design of which is a critical determinant of the stellarator's performance and viability.
 
Early stellarators were built from relatively straightforward coils (circular and helical shapes); unfortunately, the plasma confinement turned out to be much worse than anticipated.  In recent decades, there has been a resurgence in stellarator research -- first in the public and now also in the private sector -- that has been driven by advances in theory and optimization, especially with respect to designing the magnetic field itself; this has come at the cost of considerably more complicated coils. 
Mathematically, these coils are arbitrary 3D curves in space, and current approaches to calculating them can easily produce infeasible shapes. 

Traditionally, stellarator design is subdivided into two stages: stage I, where the plasma boundary shape is optimized, and stage II, where the coil shapes and currents are optimized. This procedure has been used in essentially all previous fully developed stellarator designs, including for designing currently operational experiments such as Wendelstein 7-X and HSX~\cite{stageI_w7x, KLINGER2013461, anderson_almagri_anderson_matthews_talmadge_shohet_1995}. This two-stage process allows for stage-1 optimization to explore suitable physics properties for fusion reactors by solving the magnetohydrodynamic equations of force balance in fixed-boundary~\cite{vmec_paper,NUHRENBERG1988113,LandremanPaul,desc_code}. A separate stage-II optimization  focuses on finding a corresponding coil set. However, the optimization of stellarator coils is a magnetostatic ill-posed inverse problem; a multitude of distinct coil configurations can, in principle, generate similar magnetic fields, yet most of these solutions are impractical from an engineering standpoint~\cite{stellarator_intro}. 

Ultimately, coil complexity and requirements for larger build tolerances resulted in delays in the construction and assembly of Wendelstein 7-X and the cancellation of NCSX as it went 70 million dollars above its baseline cost of 102 million dollars ~\cite{w7x_scientific_basis,ncsx}. It has been determined that coil complexity and tolerances accounted for 66\% of the cost growth of NCSX. As for Wendelstein 7-X, it took 10 years to build the experiment, suffering from several delays due to lack of engineering capacity, strict margins and tight coil tolerances \cite{KLINGER2013461}. Overall it took $10^6$ hours to complete the 50 modular coils \cite{Bosch_2017}. 

Historically, the stellarator community has explored various approaches to tackle the challenge of finding simple yet accurate coils. These have ranged from employing stochastic optimization algorithms to prevent getting stuck in local minima and to produce robust configurations, performing large-scale scans capable of exploring vast parameter spaces, investigating diverse coil topologies, such as modular, helical or dipole coils, and even optimizing the plasma and the coils simultaneously~\cite{Wechsung_2022, Giuliani_2024, Wu_2025,Jorge_2024,SUZUKI2021112843,Fu_2025,drevlak_onset,Pomphrey_2001,Strickler01032002,coilopt++}. Despite these varied efforts, finding robust and efficient coil optimization techniques remains a challenge. 

Currently, various stellarator equilibrium concepts are being investigated also in the industry~\cite{LION2025114868,Anderson_Canik_Hegna_Mowry_2025,Gates_2025, Volpe2023Renaissance, helical_fusion}. Most of these concepts take advantage of the various flavors of omnigeneity, which is a magnetic field property allowing for the confinement of collisionless trapped particles~\cite{Goodman_Camacho, Dudt_Goodman_Conlin_Panici_Kolemen_2024, piecewise_omnigenous, matt_omnigeneity}. This property affects in particular alpha particles that contain a considerable portion of the energy (3.5 MeV) produced from D-T fusion reactions, essential for plasma heating. Among these flavors being pursued are the quasi-axisymmetric (QA), quasi-helical symmetric (QH)  and quasi-isodynamic (QI) configurations~\cite{LandremanPaul, goodman_prx}. Even with this restriction, there is remarkable freedom in the plasma shape and plasma properties~\cite{Nies_Paul_Panici_Hudson_Bhattacharjee_2024, Landreman_2022}. This represents an additional challenge for the development of a robust coil optimization scheme able to reliably deliver stellarator coil sets independently of the complexity of the magnetic configuration. 

Furthermore, it is estimated that the original Stellaris configuration by the start-up Proxima Fusion will have anisotropic deformations in the order of 15 mm, arising from electromagnetic loads \cite{LION2025114868}. This can potentially result in degraded field quality, asymmetries in the heat loads of island divertors, and so forth. The immense force requirements in this example are not unique to Stellaris, rather, they are emblematic of the inevitable challenges of reactor-scale coil design for stellarators.

Finding coil configurations that simultaneously achieve high magnetic field accuracy and adhere to these engineering requirements is a constrained and highly non-convex optimization problem. Nonconvex constraints are very challenging, so in practice this problem is converted into a multi-objective unconstrained optimization problem with weights attached to each objective. In stellarator optimization, these weights are manually tuned to find coil solutions that seem promising, but the values of these weights are hard to choose because there is no a priori knowledge of their optimal values. As many as 10-15 weights are typically used to balance the various engineering tradeoffs.

This manuscript describes the implementation of an augmented Lagrangian method for coil optimization, aimed at tackling this multifaceted challenge. 
We seek to identify best-to-date stellarator coil designs in terms of the plasma and engineering complexity metrics, while avoiding heuristic and manual tuning of the optimization weights. We demonstrate the value of this new method by finding a substantial set of coil solutions that, in various ways, outperform previously published coil sets. 
The entirety of this work is implemented and can be reproduced in the SIMSOPT optimization suite~\cite{landreman_simsopt_2021}. This publication expands on a preliminary account covered in \cite{auglag_prl}. The configuration files to generate the figures and results in the present work can be found at: \url{https://doi.org/10.5281/zenodo.18497939}

\section{Optimization Method}

\subsection{Prior developments in coil optimization}
Early stellarator coil optimization methods considered a single continuous toroidal surface surrounding the plasma, called a winding surface, where sheet currents could flow and a current potential was well-defined. A solution for the current potential would be found, and then coils would be ``cut'' from them. This approach has been used to find the Wendelstein 7-X coils and can be found in codes such as NESCOIL or REGCOIL~\cite{Merkel_1987,Landreman_2017}. More advanced winding surface methods are in development, especially for generating good initial coils and for performing single-stage optimizations~\cite{Fu_2025}. However, such an approach constrains the coils to sit on a single pre-defined toroidal surface and therefore limits the exploration of better coil sets. 

Recent efforts have introduced the concept of filamentary coils that can be described as a truncated Fourier series, allowing for the adjustment of the Fourier coefficients as the optimization is performed, and avoiding any surface restriction. This is the representation used in this paper within the SIMSOPT software framework, as described below in Cartesian coordinates for the i-th coil,
\begin{equation}
x^i(\theta)=\sum_{m=0}^{M}x_{c,m}^i\cos(m\theta)+\sum_{m=1}^{M}x_{s,m}^i\sin(m\theta),
\label{eq:coils_fourier}
\end{equation}
and similarly for $y^i$ and $z^i$. Here, $\theta$ corresponds to the coil arc length index, the order $M$ of the decomposition is to be defined by the user. The coefficients $[x_{c,m}, x_{s,m} ,y_{c,m}, y_{s,m}, z_{c,m}, z_{s,m}]$, along with the currents of the coils, constitute the $n =N(3(2M+1)+1)$ degrees of freedom for $N$ coils of the optimization problem. 
Traditional coil optimization fixes or experiments with a few values of $M$, but stage-1 optimization often uses a ``Fourier continuation'' method where a series of optimization sub-problems are solved with low $M$, and using those solutions to initialize sub-problems with higher and higher $M$ until convergence~\cite{landreman2021stellarator}.  We have performed a scan over $M$ in Appendix~\ref{sec:order_investigation} that indicates coil optimization routines could benefit from the same procedure. Setting $M\sim 15$ seems to yield the lowest field error and the final coil sets for $M > 15$ tend to worsen on average. This approximate trend seems to hold for traditional coil optimization algorithms as well as the augmented Lagrangian approach proposed here. 

Coil optimization can be formulated as an equality and inequality-constrained optimization problem:
\begin{equation}
    \begin{aligned}
& \underset{\bm x}{\text{minimize}} && f(\bm x) \\
& \text{subject to} && e_j(\bm x) = 0, \quad j \in \mathcal{E} \\ & && g_k(\bm x) \leq 0, \quad k \in \mathcal{I}
\end{aligned}
\label{eq: constrained_opt}
\end{equation}
Here $f: \mathbb{R}^n \to \mathbb{R}$ is the objective function to be minimized, $e_j:\mathbb{R}^n \to \mathbb{R}$ and $g_k:\mathbb{R}^n \to \mathbb{R}$ correspond to equality and inequality constraints, $\bm{x}\in\mathbb{R}^n$ are the degrees of freedom including the Fourier coefficients representing the coil filaments and currents, $\mathcal{E}$ and $\mathcal{I}$ are the set of indices for equality constraints and inequality constraints, respectively. Consider $N_c$ constraints in total. In previous work, the next step is to transform this very challenging constrained optimization into an unconstrained problem by constructing a weighted multi-objective function. This is the mechanism used in SIMSOPT, written in the form:
\begin{equation}
\begin{aligned}
& \underset{\bm x}{\text{minimize}} && f(\bm x) + \sum_{l=1}^{N_c} \omega_l c_l(\bm x),
\end{aligned}
\label{eq:unconstrained_opt}
\end{equation}
 where $c_l$ are the various inequality and equality constraints combined and $\omega_l \in\mathbb{R}$ for $l = 1,...,N_c$ are the weights. Inequality constraints are typically recast to equality constraints by $g_l\mapsto \max(g_l - g_\text{target}, 0)$, since this quantity vanishes if $g_l(\bm x)$ is below the target value. 
 As was pointed out in Conlin et al. for plasma optimization~\cite{Conlin_Kim_Dudt_Panici_Kolemen_2024}, the weights $\omega_l$ have to be set by the user, rarely represent an optimal tradeoff, and therefore have to be adjusted. It is also only possible to satisfy the constraints when $\omega_l \rightarrow \infty$. Moreover, in order to circumvent the struggle of weight adjustment, large-scale scans of targets/weights are used, which can be computationally expensive and suffer restriction to local regions of the parameter space.  
 
 \subsection{Augmented Lagrangian formalism}
 \label{sec:auglag_formalism}
 In order to tackle these issues, the following augmented Lagrangian formalism for coil optimization is proposed with the Lagrangian,
\begin{equation}
    \mathcal{L}_\text{A}(\bm x, \bm \lambda, \bm \mu) = f(\bm x) - \boldsymbol\lambda^\top \bm c(\bm x) + \frac{1}{2} \norm{\sqrt{\boldsymbol\mu}\circ\bm c(\bm x)}_2^2,
    \label{eq:aug_lag}
\end{equation}
where:
\begin{itemize}
    \item $\boldsymbol\lambda\in\mathbb{R}^{N_c}$ is the vector of Lagrange multipliers.
    \item $\boldsymbol\mu\in\mathbb{R}^{N_c}$ and $\mu_l > 0$ for all $l$, are the penalty parameters.
    \item $\bm c:\mathbb{R}^n\to\mathbb{R}^{N_c}$ is the vector of the original equality constraints.
\end{itemize}
The term $\boldsymbol\lambda^\top \bm c(\bm x)$ is the standard Lagrange multiplier term, and $\frac{1}{2} \norm{\sqrt{\boldsymbol\mu}\circ\bm c(\bm x)}^2$ is the penalty term defining the Hadamard product between $\boldsymbol\mu$ and $\bm c(\bm x)$. The latter penalizes constraint violations. Note that what previously were user-determined hyperparameters become now Lagrange multipliers that are automatically updated throughout the optimization, so time consuming manual updates are avoided. Moreover the penalty term prevents the necessity of infinite magnitudes for the $\omega_l$, avoiding the two main disadvantages of the previous formalisms. The algorithm employed to steer and update the Lagrange multipliers is inspired from~\cite{Conlin_Kim_Dudt_Panici_Kolemen_2024,ConnGouldToint2000,NocedalWright2006} and it is here presented and adapted for clarity:
\\

\textbf{Augmented Lagrangian Algorithm}
\par\noindent\rule{\columnwidth}{0.4pt}
\textbf{Initialize}: $p = 0$, $\bm x_0$, $\bm \mu_0> 0$, $\omega_\text{tol} > 0$, $\eta_\text{tol} > 0$, $\tau > 1$ \\
\indent $\bm \lambda_0 \leftarrow \text{array sampled from uniform distribution in [0,1]}$ \\
\indent $\bar{\mu}_0 \leftarrow \text{Average}(\bm \mu_0)$ \\
\indent $\omega_0 \leftarrow 1/\bar{\mu}_0$ \\
\indent $\eta_0 \leftarrow 1/\bar{\mu}_0^{0.1}$ \\
\indent \textbf{While}  $\norm{\nabla \mathcal{L}_\text{A}(\bm x_p, \bm \lambda_p, \bm \mu_p)}_2 > \omega_\text{tol}$ or $\norm{\bm c(\bm x_p)}_\infty > \eta_\text{tol}$ \\
\indent \indent $\bm x_{p+1} \leftarrow \text{argmin}_{\bm x} \mathcal{L}_\text{A}(\bm x, \bm \lambda_p, \bm\mu_p)$, solved to tol. $ \omega_p$ \\
\indent \indent \textbf{if} $\norm{\bm c(\bm x_{p+1}, \bm \lambda_p, \bm \mu_p)}_\infty < \eta_p$ then \\
\indent 
 \indent \indent $\bm \lambda_{p+1} \leftarrow \bm \lambda_p - \bm \mu_p \circ\bm c(\bm x_{p+1})$ \\
\indent \indent \indent $\bm \mu_{p+1} \leftarrow \bm \mu_p$ \\
\indent \indent \indent $\omega_{p+1} \leftarrow \text{max}( \omega_p / \bar{\mu}_{p+1}, \omega_\text{tol}) $ \\
\indent \indent \indent $\eta_{p+1} \leftarrow \text{max}( \eta_p / \bar{\mu}_{p+1}, \eta_\text{tol}) $ \\
\indent \indent \textbf{else:} \\
\indent \indent \indent $\bm \lambda_{p+1} \leftarrow \bm \lambda_p$ \\ 
\indent \indent \indent $\bm \mu_{p+1} \leftarrow \tau\bm \mu_p$ \\
\indent \indent \indent $\bar{\mu}_{p+1} \leftarrow \text{Average}(\bm \mu_p)$ \\ 
\indent \indent \indent $\omega_{p+1} \leftarrow \text{max}(1 / \bar{\mu}_{p+1}, \omega_\text{tol})$ \\
\indent \indent \indent $\eta_{p+1} \leftarrow \text{max}(1 / \bar{\mu}_{p+1}^{0.1}, \eta_\text{tol})$ \\
\indent \indent \textbf{end if} \\
\indent \indent $p \leftarrow p+1$ \\
\indent \textbf{end while}
\par\noindent\rule{\columnwidth}{0.4pt}

Here $w_{tol}$ is the tolerance of the Lagrangian gradient and $\eta_\text{tol}$ is the tolerance for the measure of constraint violation, and together they set the convergence condition. The argmin$(\mathcal{L}_\text{A})$ optimization is an internal minimization step where for fixed Lagrange multipliers the local optimizer L-BFGS-B~\cite{liu1989limited} is used, a quasi-Newton method suitable for problems with a high number of degrees of freedom. 
In the context of SIMSOPT the individual gradients of each constraint function $c_i(\bm x)$ are calculated using a combination of analytic derivatives and automatic differentiation, where the chain rule is applied using vector-Jacobian products allowing for efficient computations. 

\subsection{Treatment of squared flux}
As mentioned previously, the goal is to design feasible coils that are able to reproduce exactly the desired magnetic field for confining a fusion plasma. Therefore, the main objective of the optimization is to minimize the field error. This is achieved via the normalized squared flux, which is a measure of deviation between the coil magnetic field and the target magnetic surface,
\begin{equation}
    f_\text{SF} = \frac{1}{2}\frac{\int_{S} \vert \textbf{B} \cdot \textbf{n}\vert^2\text{d}S}{{\int_{S}\vert \textbf{B} \vert ^2}\text{d}S},
    \label{eq:squared_flux}
\end{equation}
where $S$ is the plasma target surface, $\bm{n}$ is its normal vector, and $\bm{B}$ is the magnetic field generated by the coils from the Biot-Savart law,
\begin{equation}
    \textbf{\textit{B}} = \frac{\mu_0}{4\pi}\sum_{i=1}^{N}I_i \int_{\Gamma_i} \frac{\text{d}\textbf{\textit{l}}_i\times \textbf{\textit{r}}}{r^3}.
    \label{eq:biot_savart}
\end{equation}
Here $\text{d}\textbf{\textit{l}}_i$ is the differential arc-length element of the filament, $\Gamma_i$ is the i-th coil curve and $\textbf{\textit{r}}$ is the distance between the corresponding arc-length element and the evaluation point. When $f_\text{SF} = 0$, the field produced by the stellarator coils is exactly tangential to the desired magnetic plasma surface. If this is achieved, then the plasma MHD equilibrium is ensured. However, computational methods cannot exactly produce $f_\text{SF} = 0$ for the entire plasma boundary. This necessity for zero errors seems to imply that $f_\text{SF}$ should serve as a candidate for the minimization function $f$ in Equation \ref{eq:aug_lag}.

Instead, in this work, the function $f$ is replaced with a dummy objective function that returns $0$, and the squared flux function is moved to the constraints with all the other objectives. This is motivated by the fact that the augmented Lagrangian methods are highly adapted to minimize constraint violation. We also found that for many devices $f_\text{SF} \lesssim 10^{-6}$ or so is sufficient for reproducing the plasma properties. In this case, we can convert the squared flux objective to $\max(f_\text{SF} - 10^{-6}, 0)^2$. Once this constraint is satisfied, the algorithm is free to use the extra ``slack'' to improve engineering constraints, rather than continuing to focus on the squared flux. This is also advantageous because it does not make practical sense to exactly fit $f_\text{SF} = 0$; there will inevitably be error fields from a host of real-world sources that will lead to deviations from the desired magnetic fields. Some precisely-quasisymmetric stellarator designs, like the Landreman\&Paul QA/QH \cite{LandremanPaul} configurations, have fairly rapid deterioration of the degree of quasisymmetry as $f_\text{SF}$ deviates from zero. Nonetheless, these configurations, which have been optimized only for aspect ratio, rotational transform, and quasisymmetry, will not have extremely high degrees of quasisymmetry once they are turned into reactor-designs. In other words, when finite-$\beta$ effects, ideal MHD stability, turbulence reduction, etc. are factored into the optimization problem, the degree of quasisymmetry will decrease and subsequently we expect the sensitivity with respect to errors in $f_\text{SF}$ will decrease. \textit{We believe this capability to impose $f_\text{SF}$ as an inequality constraint is a substantial advantage of our approach.}

\subsection{Engineering-informed coil constraints}
Regarding the definitions of the usual constraints ($c_i(\bm x)$) on the optimization, we use standard metrics for coil complexity, as given in the literature~\cite{zhu2017new,Kaptanoglu_2025}:
\medmuskip=0mu
\thickmuskip=2mu
\begin{subequations}
	\begin{gather}
        g_{\text{cs}} = \sum_{i = 1}^{N}\int_{\Gamma_i} \int_{S} \max(0, d_{0}^{cs} - \| \mathbf{r}_i - \mathbf{s} \|_2)^2 ~\text{d}l_i ~\text{d}S,\\
        g_{\text{cc}} = \sum_{i = 1}^{N} \sum_{j = 1}^{i-1} \int_{\Gamma_i} \int_{\Gamma_j} \max(0, d_{0}^{cc} - \| \mathbf{r}_i - \mathbf{r}_j \|_2)^2 ~\text{d}l_j ~\text{d}l_i, \\
	g_{\text{l}} = \frac{1}{2}\left(\max\left(\sum_{i=1}^{N} L_i - L_{0},0\right)\right)^2 \label{eq:length_penalty},\\
	g_\text{curv} = \sum_{i=1}^{N}\frac{1}{2}\int_{\Gamma_i} \max(\kappa - \kappa_0,0)^2 \text{d}l, \\
    g_\text{msc} = \frac{1}{2}\left(\max \left(\sum_{i=1}^{N}\frac{1}{L_i}\int_{\Gamma_i} \kappa^2 \text{d}l - K_0,0\right)\right)^2, \\
    	c_\text{link}(\Gamma_i, \Gamma_j) = \frac{1}{4\pi} \left| \oint_{\Gamma_i}\oint_{\Gamma_j}\frac{\textbf{r}_i - \textbf{r}_j}{|\textbf{r}_i - \textbf{r}_j|^3} (\text{d}\textbf{r}_i \times \text{d}\textbf{r}_j) \right|,	\\ 
        g_{f} = \frac{1}{2}\sum_{i=1}^N\left(\int \text{max}(|\bm{F}_i(l)| - F_0, 0)^2 \text{d}\ell\right).
        \end{gather}
\label{eq:all_constraints}
\end{subequations}
\medmuskip=4mu
\thickmuskip=4mu
\begin{table*}[ht!]
	\begin{tabular}{|l|l|l|}
		\hline
		\textbf{Objective}       & \textbf{Bound}         & \textbf{Motivation}                                                                                                                                                                                                                                                                    \\ \hline
		Normalized $f_\text{SF}$ & $\lesssim 10^{-5}$     & \begin{tabular}[c]{@{}l@{}}Sufficient accuracy to preserve important plasma properties.\end{tabular}                                                                                                                                                                                 \\ \hline
		Minimum CS distance      & $\gtrsim 1.3$m         & \begin{tabular}[c]{@{}l@{}}Sufficient distance for neutron-absorbing blanket.\end{tabular}                                                                                                                                                                                          \\ \hline
		Minimum CC distance      & $\gtrsim 0.7$m         & \begin{tabular}[c]{@{}l@{}}Sufficient distance for avoiding coil \\ intersections and assembly difficulties\\  from finite coil build.\end{tabular}                                                                                                                                    \\ \hline
		Total coil length        & $\lesssim 150-200$m    & \begin{tabular}[c]{@{}l@{}}As small as possible to decrease the costs of purchasing superconducting material. \\But there must be sufficient coil length for coils to be far enough from the plasma\\ to reduce coil-surface distance and improve squared flux.
		Scales with the major radius. \end{tabular} \\ \hline
		Max curvature $\kappa$   & $\lesssim 1$ m$^{-1}$  & \begin{tabular}[c]{@{}l@{}}Low enough local curvature to simplify geometry and manufacturing costs.\end{tabular}                                                                                                                                                                     \\ \hline
		Max MSC                  & $\lesssim 0.5$m$^{-1}$ & \begin{tabular}[c]{@{}l@{}}Similar to the curvature $\kappa$, the average curvature must remain low so that the coils \\can be manufactured precisely without extensive cost.\end{tabular}                                                                                      \\ \hline
		Coil linking             & 0                      & \begin{tabular}[c]{@{}l@{}}Interlinking coils could greatly increase the complexity of assembly.\end{tabular}                                                                                                                                                                       \\ \hline
		Max force                & $\lesssim 0.5$ MN/m    & \begin{tabular}[c]{@{}l@{}}Max forces must stay within the force tolerances of the HTS material, estimated for\\ industrial REBCO-based cables ~\cite{hartwig2020viper,zhao2022structural,riva2023development}.\end{tabular}                                                      \\ \hline
	\end{tabular}
\caption{Rough estimates of upper $\uparrow$ and lower $\downarrow$ bounds for engineering metrics of a reactor-scale stellarator design.}
\label{tab:bounds_table}
\end{table*}
$g_{\text{cs}}$ is a function penalizing the minimum distance between the last closed flux surface of the plasma and the coils, preventing it from being smaller than $d_{0}^{cs}$. This can allow enough space for the positioning of a tritium-breeding blanket and shielding wall to keep the integrity of the coils and fusion device that can be damaged by high energy neutrons. $g_{\text{cc}}$ restrains the minimum coil-to-coil distance from being smaller than $d_{0}^{cc}$, accounting for finite dimensions of the coils as well as coil assembly and diagnostics. The length of the coils is set by a quadratic penalty $g_{\text{l}}$ that penalizes the sum of length of all the coils if it goes above the threshold $L_0$. Longer coils means more expensive superconducting material is required to construct them. The cost function $g_\text{curv}$ calculates the $L_2$ norm of the curvature of a coil and then penalizes the local curvature of the coil if it exceeds the threshold $\kappa_{0}$. Similarly, $g_\text{msc}$ penalizes the mean squared curvature of a curve when it exceeds the threshold $K_{0}$, this yields smoother curves. Higher curvature coils tend to be more difficult to manufacture and assemble. In order to prevent configurations with interlinked coils, which would be very challenging to build in practice, the Gauss linking number $c_\text{link}$ is calculated for each pair and becomes an integer $\geq 1$ if they are interlinked. Finally, $g_f$ minimizes the pointwise self and mutual forces that are above some $F_0$ threshold representing the tolerances of the superconducting material in the coils. The details for the force estimation can be found in Hurwitz et al.~\cite{hurwitz_2024}. It  has become an increasing necessity to find configurations minimizing the coil-to-coil forces because the required coil currents in a reactor-scale stellarator can reach tens of Mega-Ampere values~\cite{Robin_2022,Guinchard_2025,Fu_2025,Hurwitz_2025,Kaptanoglu_2025}. 
For the purpose of this work, the estimation of the maximum force per High Temperature Superconducting (HTS) tape turn is done by dividing the total force value by the number of HTS turns per coil. 

In order to determine the relative HTS material requirements for coil sets with varying number of coils per half-field period, for all the following results in this work we assume 200 HTS turns per coil for a reference coilset (e.g. 800 turns total in a four-coil solution) and recalculate the number of turns assuming fixed total current per half-field period (e.g. a three coil solution will be assumed to have $800/3$ HTS turns in each coil to keep the total current fixed). This is probably a pessimistic lower bound on the amount of HTS required for solutions with fewer coils, but it also helps to mitigate the forces in these designs. These calculations are also valid if using Low Temperature Superconductors (LTS). 

The interest behind designing stellarators with fewer coils comes from the increased plasma access that is made available on the outboard side. Fewer coils means more space in between coils for placing additional ports for plasma diagnostics, increasing the heating power, and the easing remote maintenance of the device. We find in practice that often the minimum coil-coil and coil-surface distances are higher, meaning less stringent assembly tolerances and more space for a neutron-absorbing blanket. However the price is paid on the quality of the magnetic field and a potentially higher amount of required HTS material. 

Lastly, note that while we have avoided the use of manually determined \textit{weights}, we do not get rid of appropriate upper and lower \textit{bounds} on these objectives, as this is typically critical for avoiding undesirable coil solutions such as coils which are all topologically interlinked. Furthermore, the objectives all have natural upper and lower bounds ($d_0^{cs}$, $d_0^{cc}$, ...) for  feasibility that are well-motivated by physical and engineering considerations. A summary of natural upper and lower bounds for a reactor-scale stellarator, and their motivation, is provided in Table~\ref{tab:bounds_table}.

\subsection{Comparison with standard methods}

\begin{figure*}[ht!]
    \centering\includegraphics[width=0.8\linewidth]{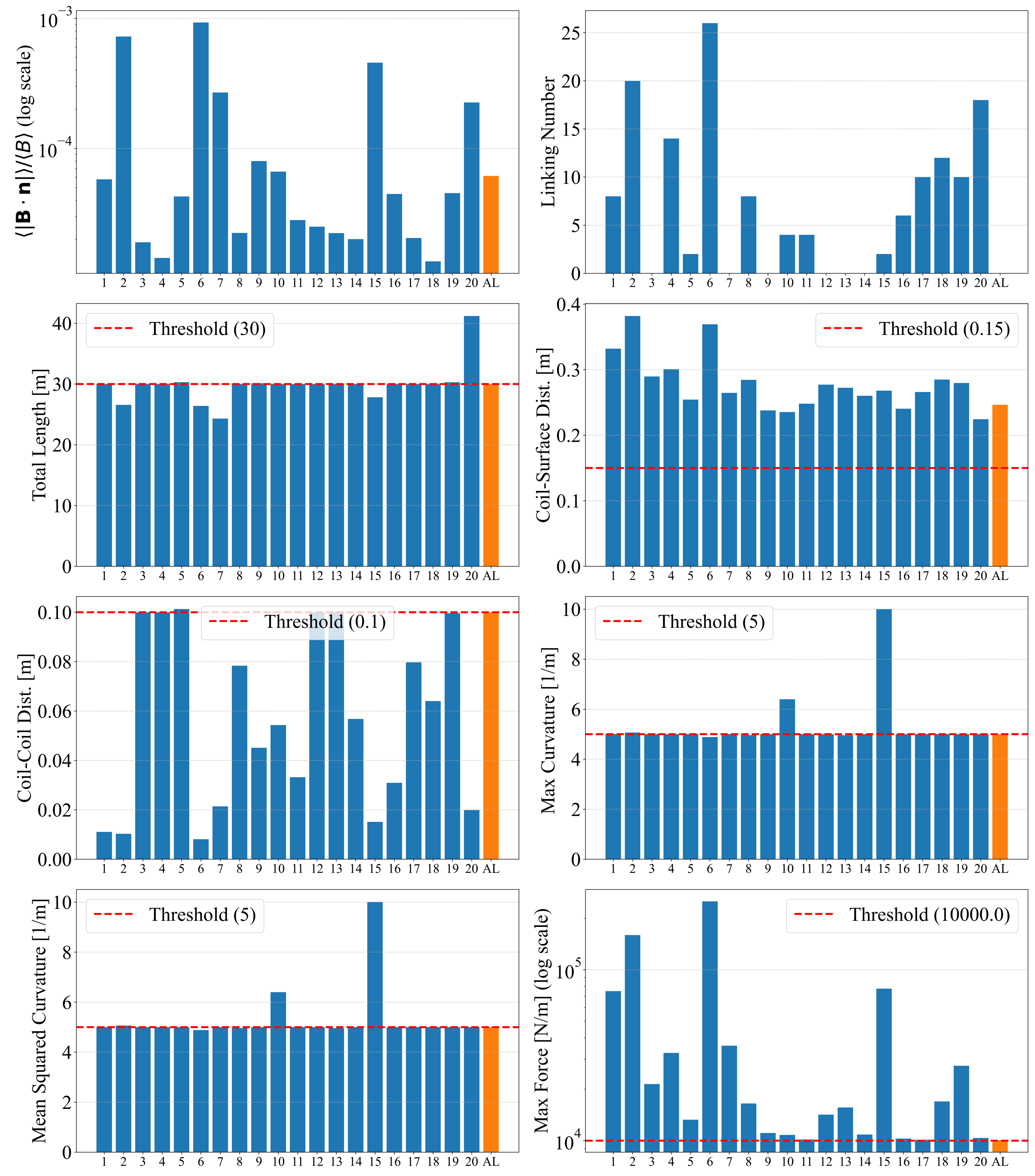}%
    \caption{Performance comparison of the augmented Lagrangian (AL) method against 20 standard coil optimization runs. The eight plots evaluate multiple targeted metrics in the optimization, the red dashed line in each plot indicates a design constraint as expressed in Equation \ref{eq:all_constraints}. While each of the 20 initial runs (blue bars) violates at least one constraint, the proposed augmented Lagrangian method (orange bar) is the only solution that successfully satisfies all criteria simultaneously.}
    \label{fig:20_runs}
\end{figure*}

In this section, we compare current filamentary methods implemented in SIMSOPT with the augmented Lagrangian approach. For this we use Fourier continuation, where the coils are optimized with a small number of degrees of freedom, and that solution is used to initialize subsequent optimizations with more and more degrees of freedom. We start with order $M = 3$ coils and solve successive optimization problems until $M = 19$. One run therefore corresponds to 16 successive optimizations. In this example we use the Landreman\&Paul quasi-axisymmetric plasma
equilibrium scaled at 1 meter major radius found in \cite{LandremanPaul} and include all the constraints as specified in Equation \ref{eq:all_constraints}. For the standard method, as the weights are traditionally user-determined, we choose to perform 20 runs with randomly assigned weights in a small range centered around known close-to-optimal weights from previous work. We then compare these 20 runs to an identical set-up but optimized using the augmented Lagrangian formalism as described in Section \ref{sec:auglag_formalism}. The results are displayed in Figure \ref{fig:20_runs}. We also repeated this comparison without continuation, i.e. direct minimization using $M = 19$, and found that almost all runs produced worse results, and the improvements using augmented Lagrangian were similar to those shown in Fig.~\ref{fig:20_runs}.

The figure illustrates all of the resulting optimization metrics that are penalized in the constraints. These plots show that standard methods are generally unable to balance all the constraints, while the augmented Lagrangian method is able to find a solution within all the upper and lower bounds. Out of the 20 runs, the runs 3, 12, and 13 violate the least number of constraints (only one) but significantly fail this constraint (a prescribed force tolerance). In contrast, the augmented Lagrangian run successfully passes all tests, without needing any weight tuning. We emphasize the strength of this result; without significant manual weight tuning, the standard methods cannot regularly produce solutions with a prescribed list of constraints like the augmented Lagrangian results.

\section{Results}
\subsection{Quasi-axisymmetric stellarator}

The first goal for this paper is to demonstrate a novel computational framework capable of efficiently reaching, and in fact improving on previously Pareto-optimal solutions. 
We first consider quasi-axisymmetric stellarators, which are thought to combine the benefits of both stellarators and tokamaks into one device. Like tokamaks, the coils exploit a form of toroidal invariance of the magnetic field strength and allow for the inherent steady-state operation of stellarators. A QA optimized stellarator is in principle also capable of showing neo-classical transport levels comparable to axisymmetric devices as mentioned previously. The start-up company Thea Energy is currently pursuing this concept using dipole coils for their prototype reactor EOS, which is planned to be at first a neutron source~\cite{Gates_2025}. 

The Landreman\&Paul quasi-axisymmetric plasma equilibrium scaled at 1 m major radius found in ~\cite{LandremanPaul} was chosen in order to be able to compare to the dataset found in ~\cite{Hurwitz_2025}, and therefore the same optimization set-up is used. This is a plasma configuration with 2-field periods, aspect ratio 6 and rotational transform of 0.42, exceptional confinement properties with a quasisymmetric field error comparable to Earth's magnetic field amplitude. 

Kaptanoglu et al.~\cite{Kaptanoglu_2025} performed 8,500 stellarator coil optimizations with five coils per half-field period (hfp) in order to find trade-offs between the magnetic field accuracy that can be achieved with the coils and the Lorentz forces that act on the coils. A similar scan was performed in Hurwitz et al.~\cite{Hurwitz_2025}. From the 8,500 minimization runs on a supercomputer, about 41 defined a Pareto-front between the coil forces and the field accuracy which is measured using the surface integrated average of the normal component of the magnetic field to the target surface normalized by its amplitude: $\langle \bm B\cdot \bm n\rangle / \langle B \rangle$. 

In order to filter suitable candidates from the 8,500 configurations, multiple thresholds are applied a-posteriori to be able to determine a valid Pareto-front and remove any undesirable coil solutions that should not be considered (e.g. interlinked coils). The thresholds include a maximum allowed curvature of 12 $\text{m}^{-1}$, maximum allowed coil length: 6.5 m, minimum distance between coils and plasma surface: 16.6 cm, minimum distance between coils: 8.3 cm. The number of coil optimizations that were performed in the scans is a testament to the computing power needed in previous methods to achieve optimal devices. 

\begin{figure}[ht!]
    \centering
    \includegraphics[width=\linewidth]{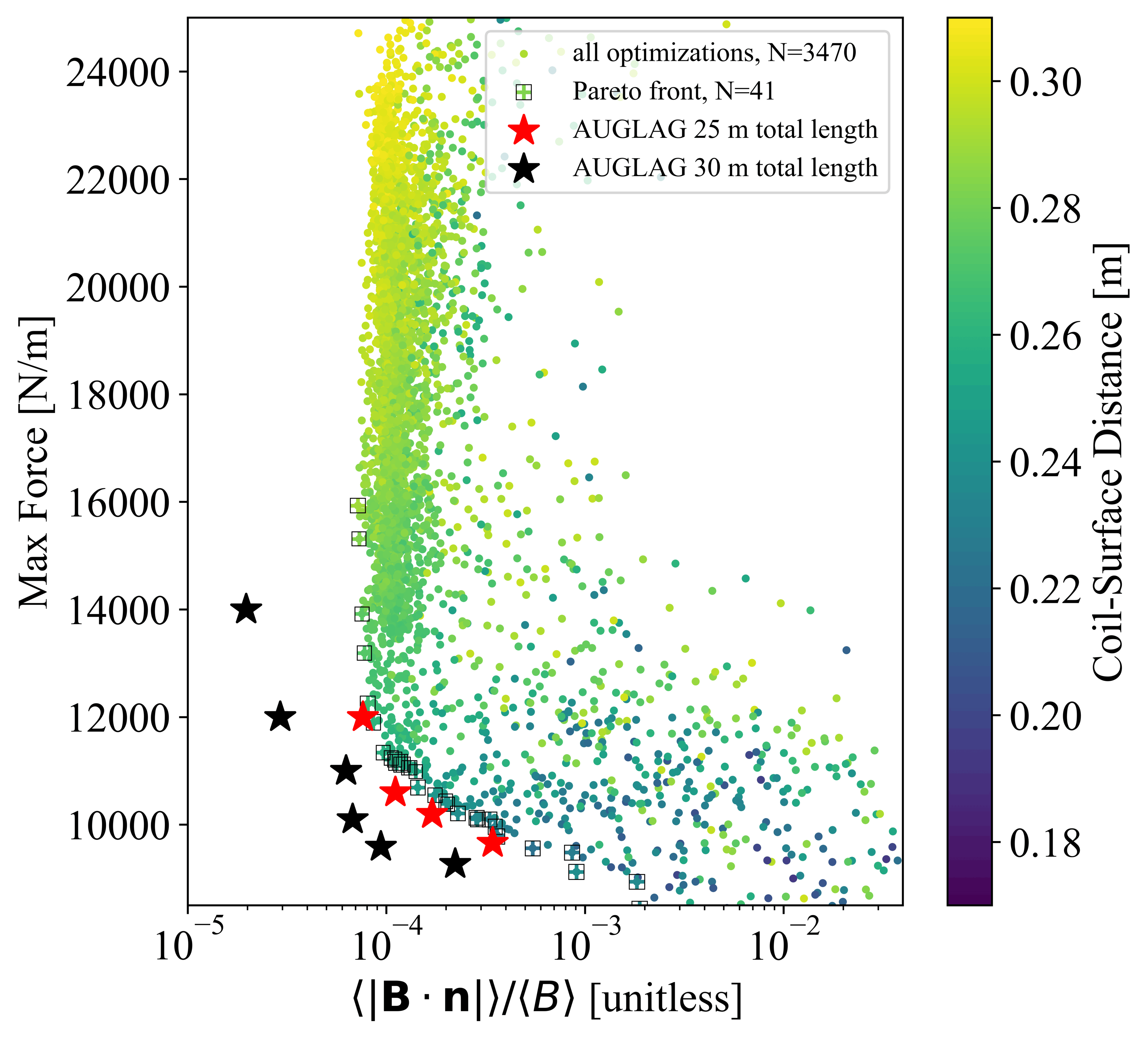}%
    \caption{\label{fig:QA_Pareto} Scans of QA stellarator configurations from~\cite{Kaptanoglu_2025} showing trade-offs between coil forces and field accuracy, with red stars placing the new configurations at the found Pareto-front and black ones outlining a new Pareto-front for configurations with larger coil length.}
\end{figure}

The initial goal of this work is to be able to map this existing Pareto-front. This is done by running our augmented Lagrangian scheme and adjusting the upper linear force threshold to 9, 10, 10.5 and 12 kN/m. Each run takes around 40 minutes on a single CPU core, and is stopped once the algorithm reaches the maximum number of iterations, set to 50 in this case. The quadratic flux also possesses an upper boundary which is on purpose set to low values ($10^{-15}$) to force the optimizer to reach the lowest error configurations. All four configurations are plotted on Figure \ref{fig:QA_Pareto} and they all sit at the Pareto-front. These results are calculated with a fixed total length of 25 m; we then try to define a new Pareto-front by relaxing this constraint and setting the total length threshold to 30 m. By running 6 different optimizations with the linear force thresholds set to 14, 12, 11, 10, 9.5 and 9 kN/m, it is possible to trace-out a new limit visible in Figure~\ref{fig:QA_Pareto} with the black stars. This is a similar class of stellarators as previously, with the main difference that they achieve up to one order of magnitude better field accuracy reaching a $\langle \bm B\cdot \bm n\rangle / \langle B \rangle = 1.97\times 10^{-5}$ at 14 kN/m.

\begin{figure*}[ht!]
    \centering
    \includegraphics[width=0.32\linewidth]{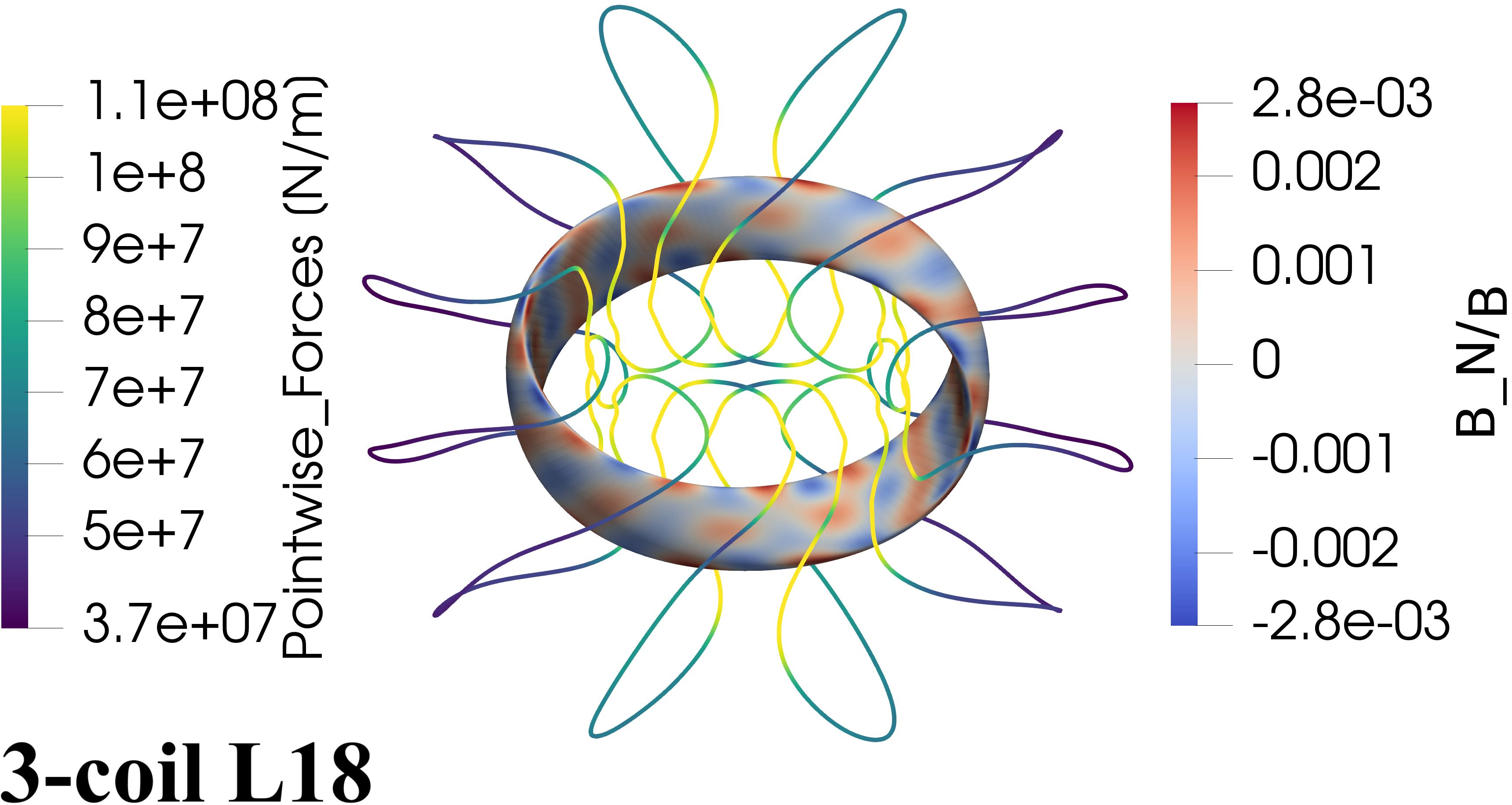}
    \includegraphics[width=0.32\linewidth]{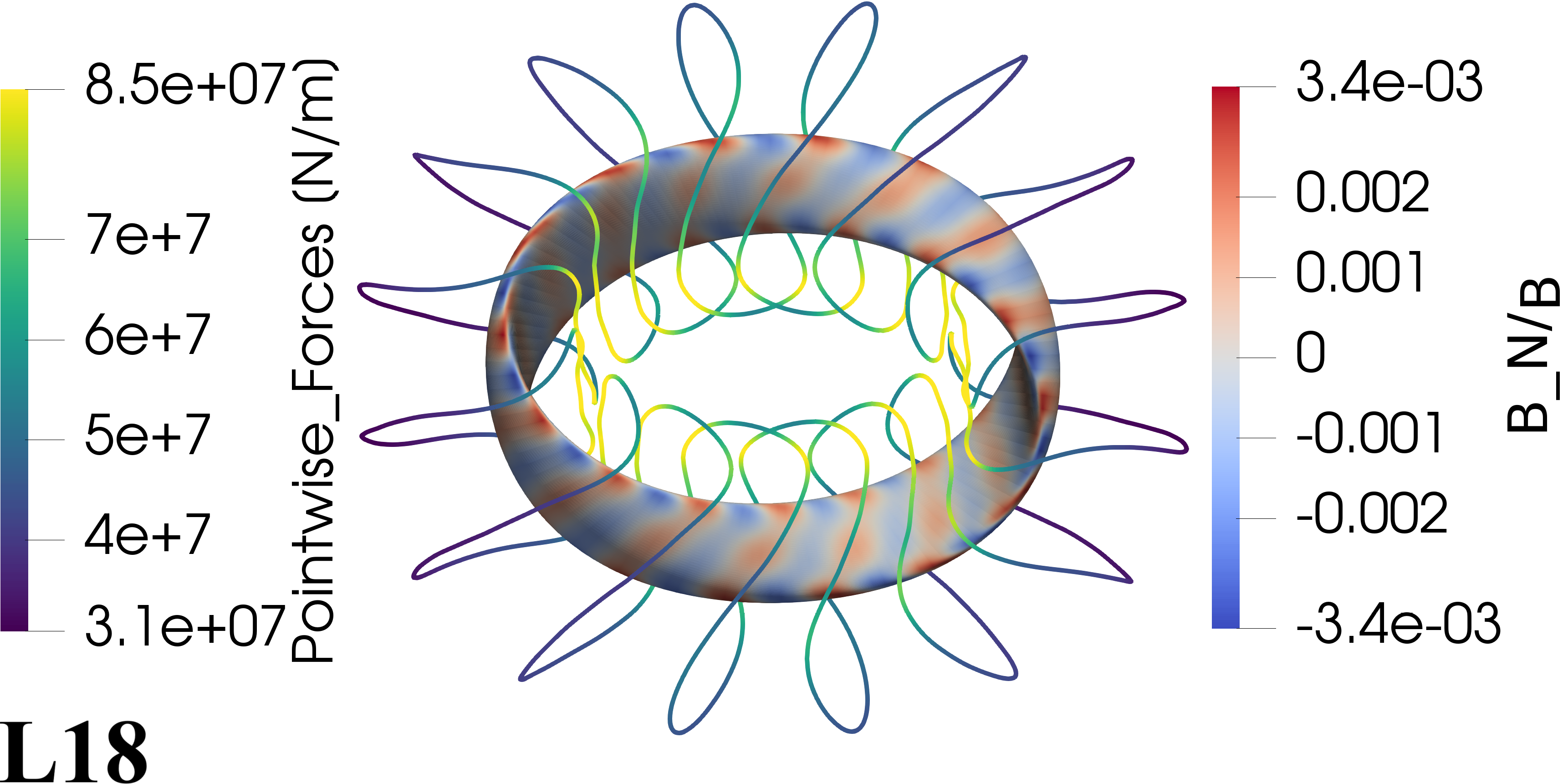}
    \includegraphics[width=0.32\linewidth]{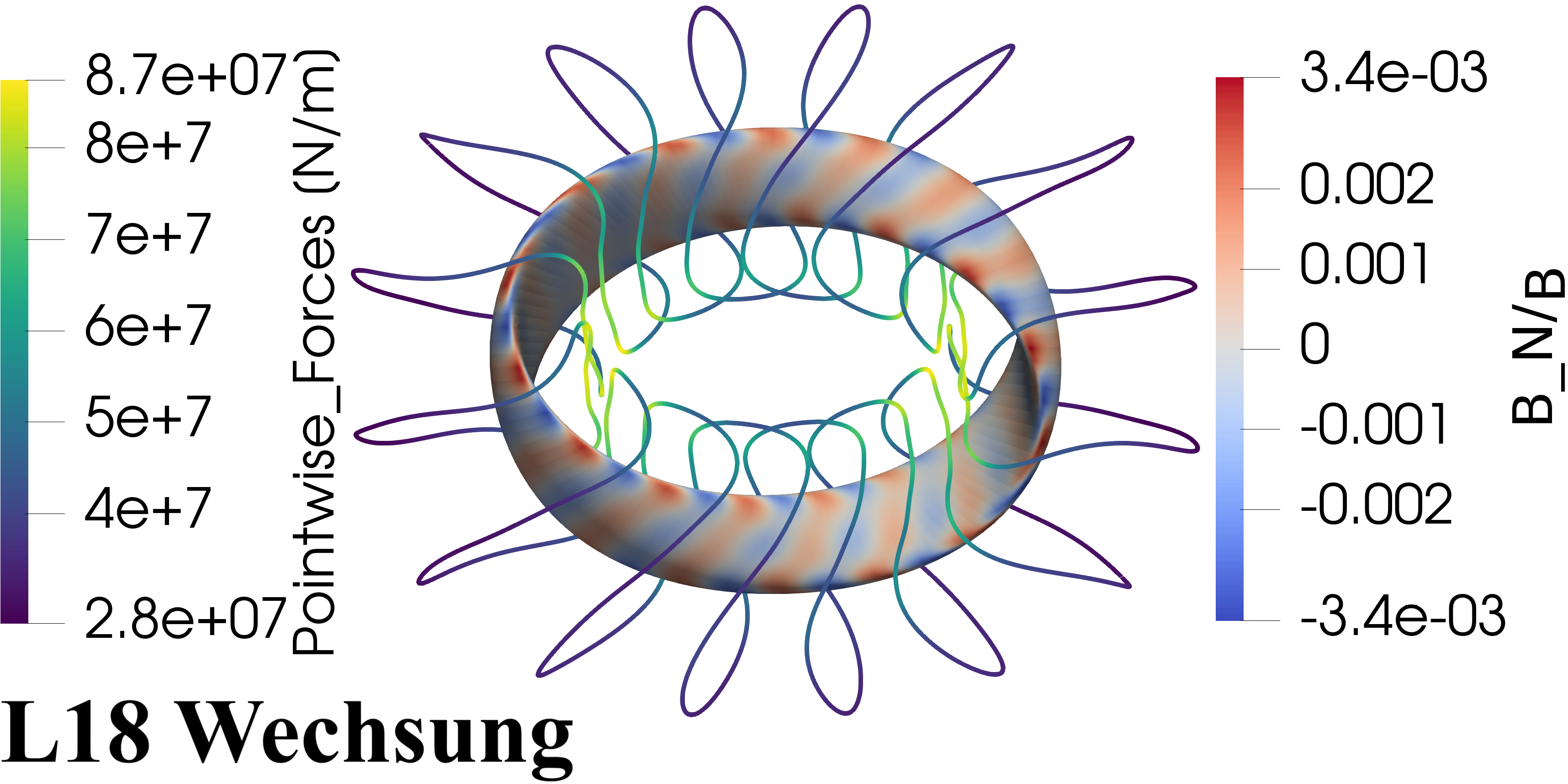}
    \includegraphics[width=0.49\linewidth]{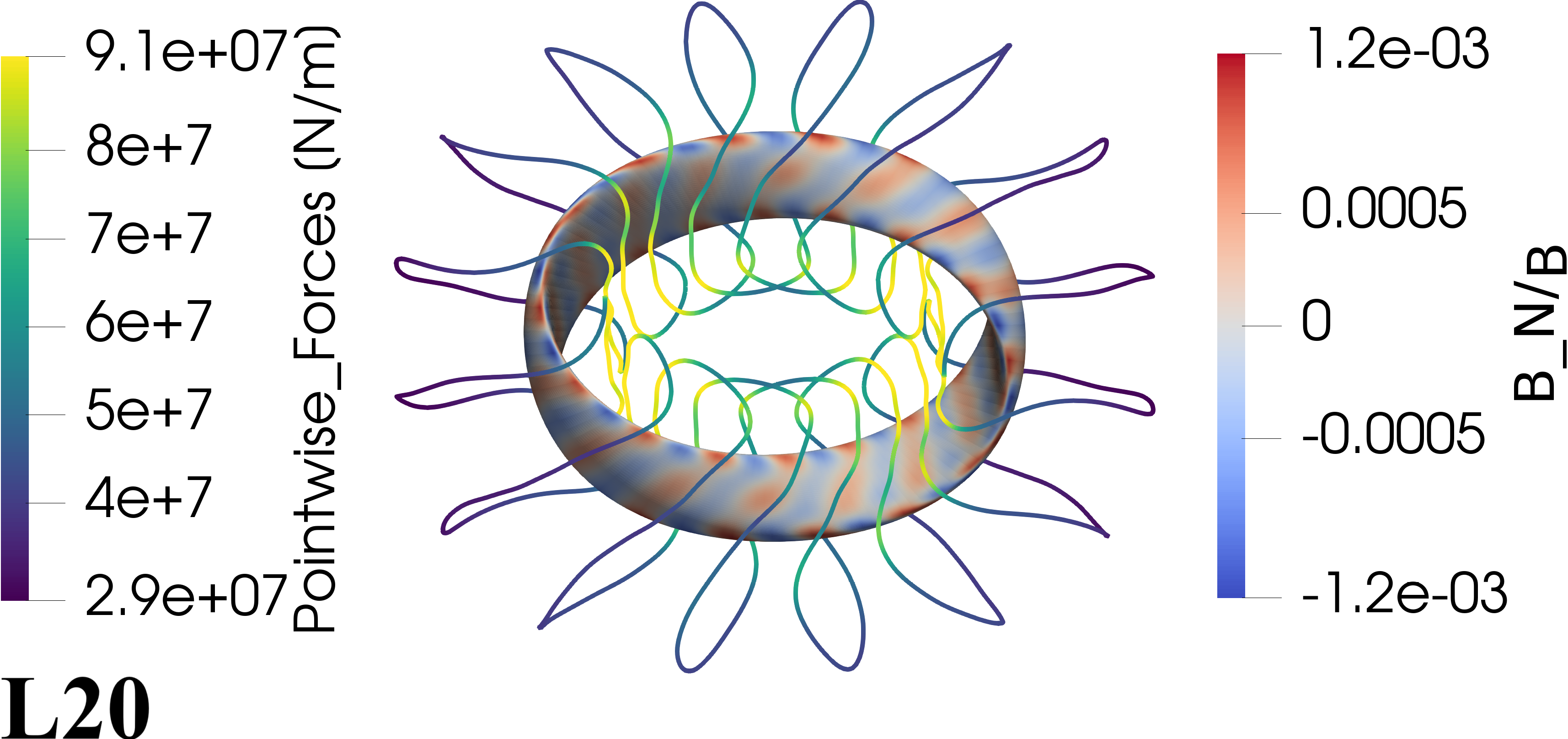}
    \includegraphics[width=0.49\linewidth]{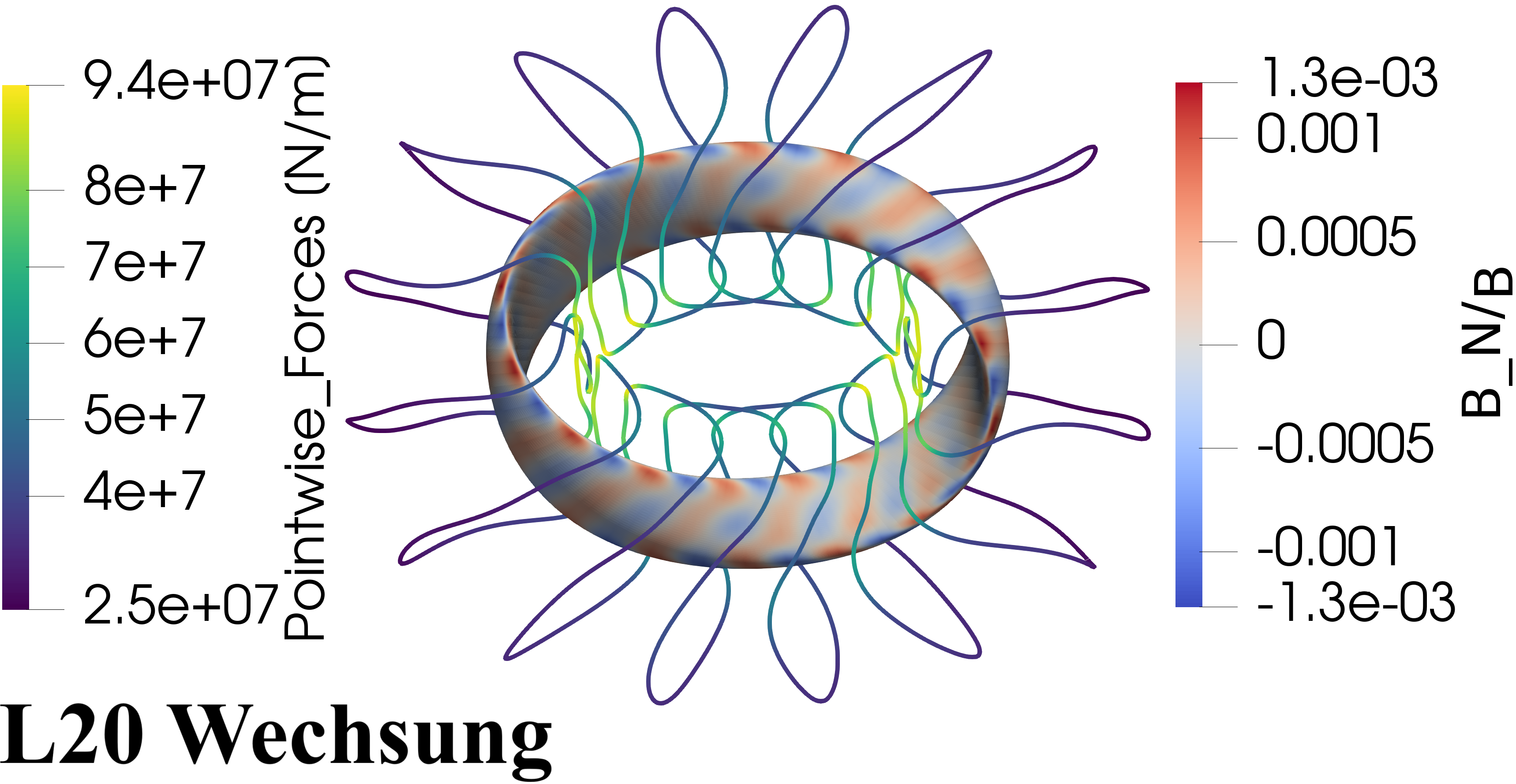}
    \includegraphics[width=0.49\linewidth]{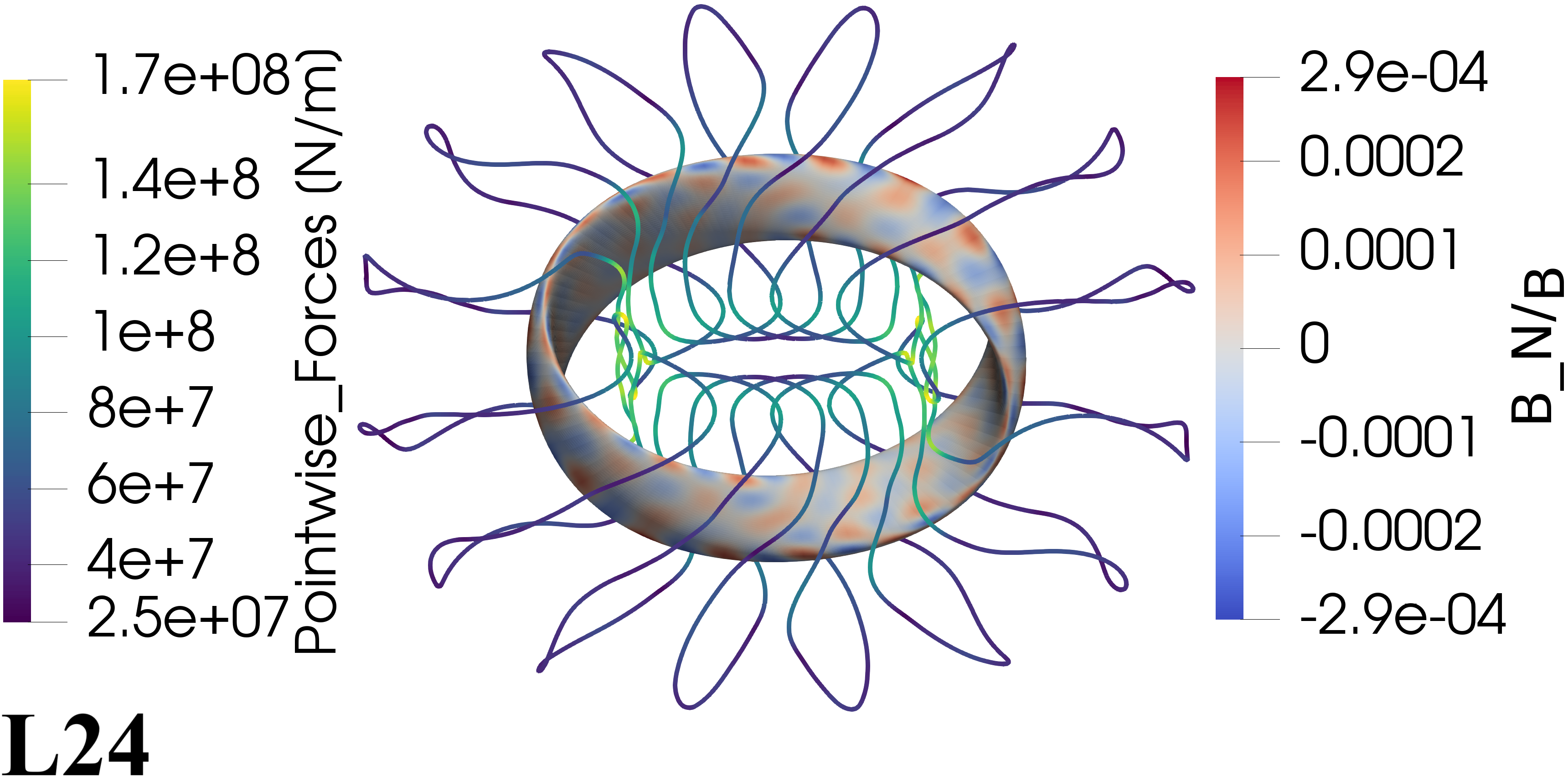}
    \includegraphics[width=0.49\linewidth]{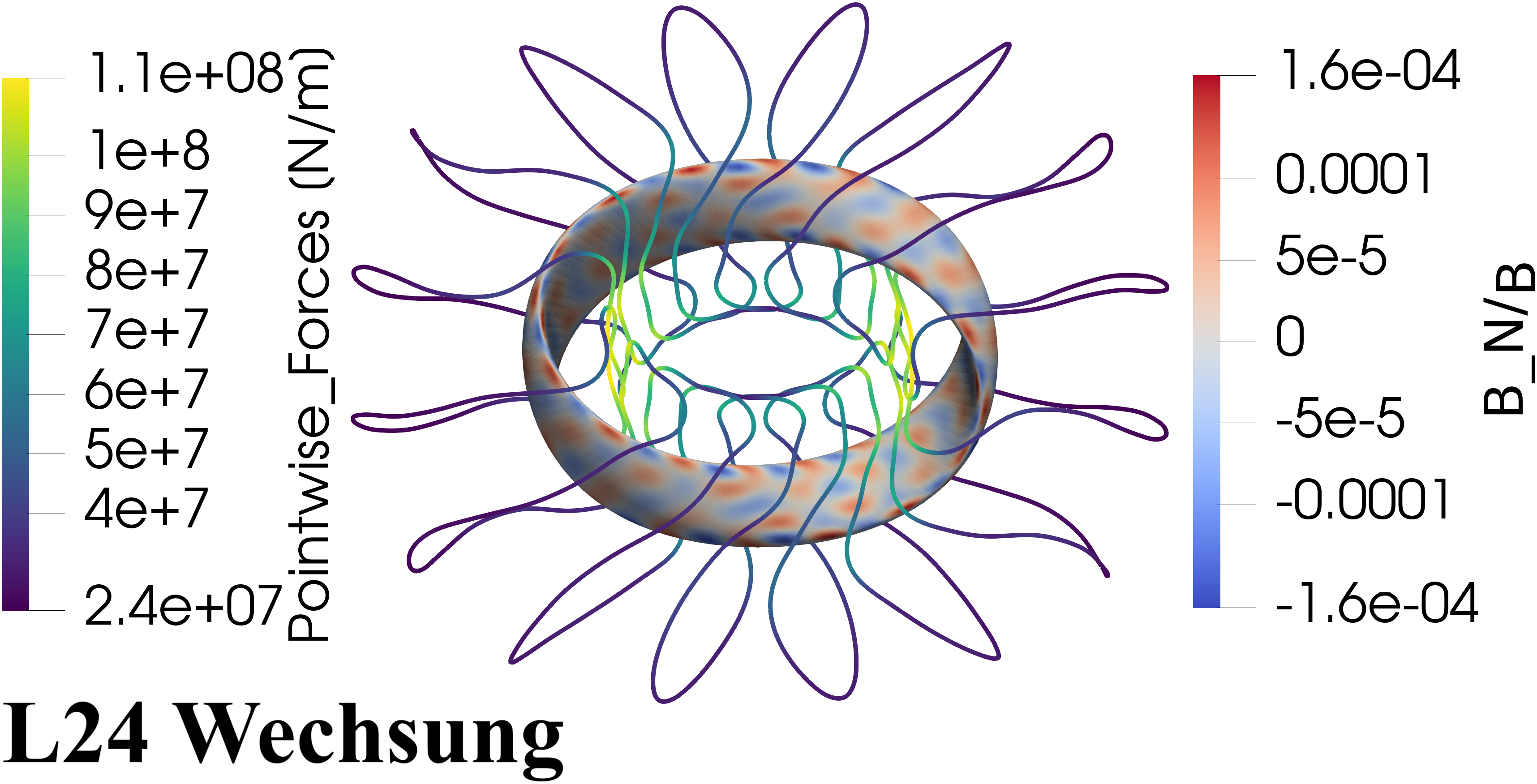}
    \caption{(Top, left to right) Three-coil solution, four-coil solution, and four-coil Wechsung solution, all with same total coil length. (Middle, left to right) Four coil solution with augmented Lagrangian versus the Wechsung solution obtained at same total coil length. (Bottom, left to right) Same as the middle panel, but with increased total coil length. Forces appear very large because they are reported here before dividing by the number of turns.}\label{fig:12kNmStellarator}
\end{figure*}

\begin{table*}[ht!]
\begin{tabular}{|l|c|c|c|c|c|c|c|c|}
\hline
Property    & Goal                        
& 3-coil L18
&Wechsung L18 & L18 & Wechsung L20 & L20 & Wechsung L24 & L24 \\ 
\hline
$\#$ of coils per hfp & $\downarrow$                                   
& $\bm{3}$ & $4$ &$4$ &$4$   &$4$ &$4$  & $4$     \\ 
$\langle \bm B\cdot \bm n\rangle / \langle B \rangle \times 10^{-5}$  & $\downarrow$ 
& 85.5 & 91.1 & 93.7 &  32 &  34  &$\bm{4.35}$ & 6.0   \\ 
Max $(\bm B\cdot \bm n / B)\times 10^{-4}$  & $\downarrow$                          
& 28& 34& 34& 13  & 12  &$\bm{1.6}$ & 2.9 \\ 
Max Force [MN/m]  & $\downarrow$              
&$\bm{0.41}$&  $0.44$&   $0.43$  &$0.47$ &$0.46$ & $0.55$ & $0.85$    \\ 
Coil Length per hfp [m]   & $\downarrow$                       
& $\bm{182}$ & $\bm{182}$ &  $\bm{182}$ & 203  & 203 & 243   & 240   \\ 
Min CC Distance [m] &   $\uparrow$                       
& $\bm{1.49}$& 1.34& $1.44$& 1.14  & 1.31  &$1.0$ & $1.0$      \\ 
Min CS Distance [m] & $\uparrow$                 
&2.74&2.75 &  2.57 & 2.9 & 2.71   &$3.1$  & $\bm{3.14}$   \\ 
Max $\kappa$ [$\text{m}^{-1}$] & $\downarrow$       
&  0.53& $\bm{0.44}$ &  0.5 & 0.49    & 0.51  &$0.49$ & $0.5$  
\\ 
Max MSC [$\text{m}^{-1}$] &    $\downarrow$                   
&  0.05& 0.05 & 0.05 & 0.05 & 0.05 &0.05   & 0.05    \\
Quasisym. Error $\times 10^{-6}$ & $\downarrow$ & 2000 & 2100 & 1900 & 190 & 260 & 2.4 & 10.7 \\
HTS length [km]    &  $\downarrow$                  
& $194$& $\bm{146}$ & $\bm{146}$ & 162  & 162 & 194 & 192     \\
\hline
\end{tabular}
\caption{Comparison of magnetic field and coil properties for our solutions and the corresponding Wechsung solutions scaled to the ARIES-CS \cite{Najmabadi01102008} reactor with 5.7T on axis and 1.7m minor radius. Properties that should ideally be lowered (increased) in the design are indicated by $\downarrow$ ($\uparrow$). Here, we assume a fixed total current per half-field period, taking 200 HTS turns per coil for the 4 coil solution as reference, explaining the difference in the max force between the table values and Figure~\ref{fig:12kNmStellarator}. 
}
\label{tab:qa_table}
\end{table*}

For the purposes of comparison with the Wechsung et al.~\cite{wechsung_precise_qs_2022} solutions in Table~\ref{tab:qa_table}, four-coil configurations are now scaled to reactor size with 5.7T on-axis magnetic field strength and 1.7m minor radius, following ARIES-CS~\cite{Najmabadi01102008}. The total length upper bounds are set to the lengths used for the coils sets in Wechsung's solutions and our solutions are systematically compared. For instance, the $18$m total length coils for a 1m major radius design are indicated by L18. Scaled to a reactor, the total coil length is $182$m with a $10.13$m major radius. As far as we can tell, the Wechsung solutions are essentially Pareto-optimal. Nonetheless, \textit{without tuning the weights}, we can produce a L18 solution that is essentially equivalent to the Wechsung L18 solution; the only significant difference is that the coil-coil distance and max curvature are slightly larger in our solution. Similarly our L20 solution is essentially identical, although arguably improved by reducing the unnecessarily large coil-surface distance and increasing the minimum coil-coil distance by $0.17$m. For the L24 solution, we purposely avoid penalizing the force in order to see that only marginal improvements in the field accuracy and other metrics are available. At the same time, the max force increases by almost $50\%$ compared to the Wechsung L24 solution. We show this result to illustrate the diminishing returns to ``over-optimizing'' on the field accuracy. 

Lastly, at a given fixed total length, we attempted some three-coil and five-coil solutions. Five-coil solutions were found with comparable or marginally improved performance to the four-coil solutions, so we concluded that the extra coil per half-field period was not worth the inevitable reduction in plasma access. However, assuming that fairly long modular coils can be constructed in practice, an excellent solution was found with $182$m total length coil solution with only three coils per hfp. It replicates most of the equivalent (i.e. same total coil length) four-coil solution complexity metrics while exhibiting better field accuracy and higher coil-coil distances. It can be seen visually in Fig.~\ref{fig:12kNmStellarator} to provide very good plasma access. More remarkably, although it was expected that a low number of coils per hfp would reduce the degree of quasisymmetry because of increased coil ripple, the 3-coil solution still manages to outperform the Wechsung L18 solution. Throughout this work, we use the two-term formulation defined in the Appendix~\ref{sec:qs_metric} for a quantitative measure of the quasisymmetric error displayed in Table \ref{tab:qa_table}. Moreover, a decrease of about 3 orders of magnitude in the quasisymmetric error is visible as the total coil length threshold is increased.
These benefits will have to be weighed against the additional coil turns in the three-coil solution that will be necessary to continue avoiding the critical current limits in the superconducting material. One plausible explanation as to why coil length impacts field accuracy is that 2 field-period QA configurations allow for considerable variations in coil length. This is especially true on the inboard side, while other configurations do not (e.g. a 4 field-period QH has relatively low space inside the corners), and increasing the length can effectively help to compensate coil ripple. This is however only valid at comparable plasma aspect ratios.

 For completeness, the Boozer plot which displays qualitatively the departure from quasisymmetry together with the Poincaré plots showing the nested flux surface structure of the magnetic equilibrium are shown in the Appendix~\ref{sec:extra_plots}. Straight lines indicate that very low levels of quasisymmetric error are achieved. We also show in Appendix~\ref{sec:fp} that the 3-coil solution found here keeps fast-particle losses to roughly $6\%$; this is excellent performance, especially given that this solution primarily addresses improving engineering criteria such as plasma access, forces, total length, and coil-coil distances.

Furthermore, the reason why, out of the 8,500 runs performed in previous work only 41 defined a Pareto-front, lies in the fact that the adjustment of the weights but also the arbitrary truncation of thresholds modify the stellarator optimization landscape in such a way that it prevents convergence towards good configurations. In contrast, our proposed method takes care of the weight adjustment dynamically throughout the optimization, and informs their update based on the penalty violation of each of the constraints.

\subsection{Quasi-helical symmetric Stellarator}
\label{sec:lp_qh}
In a second optimization demonstration, we target the plasma boundary of the Landreman\&Paul quasi-helically symmetric (QH) configuration~\cite{LandremanPaul}. Like quasi-axisymmetric plasmas, QH equilibria are designed to confine collisionless trapped particles while having also reduced alpha-particle
losses. The QH configuration has only been effectively built once with the Helically Symmetric eXperiment (HSX)\cite{anderson_almagri_anderson_matthews_talmadge_shohet_1995}, which is explored in Section \ref{sec:hsx}. One of the reasons QH symmetry and more generally quasisymmetric stellarators are of interest for a potential fusion reactor lies in the prediction of shear flows in the plasma, allowing for the suppression of microturbulence \cite{Spong_2007}. However, a four-field-period QH configuration is considered to be more demanding from a coil optimization standpoint since the plasma contains more shaping than the QA configuration shown previously. The objective is to reproduce particle confinement properties of this configuration to high enough accuracy. Therefore, the coils obtained here are compared and optimized using the same set-up as in~\cite{Wiedman_Buller_Landreman_2024}, where the main purpose was to obtain a suitable coil configuration at reactor-scale. The major radius of this configurations is approximately 13 m, the aspect ratio is approximately 8 and the average rotational transform is 1.24. 

\begin{figure}[ht!]
    \centering
    \includegraphics[width=\linewidth]{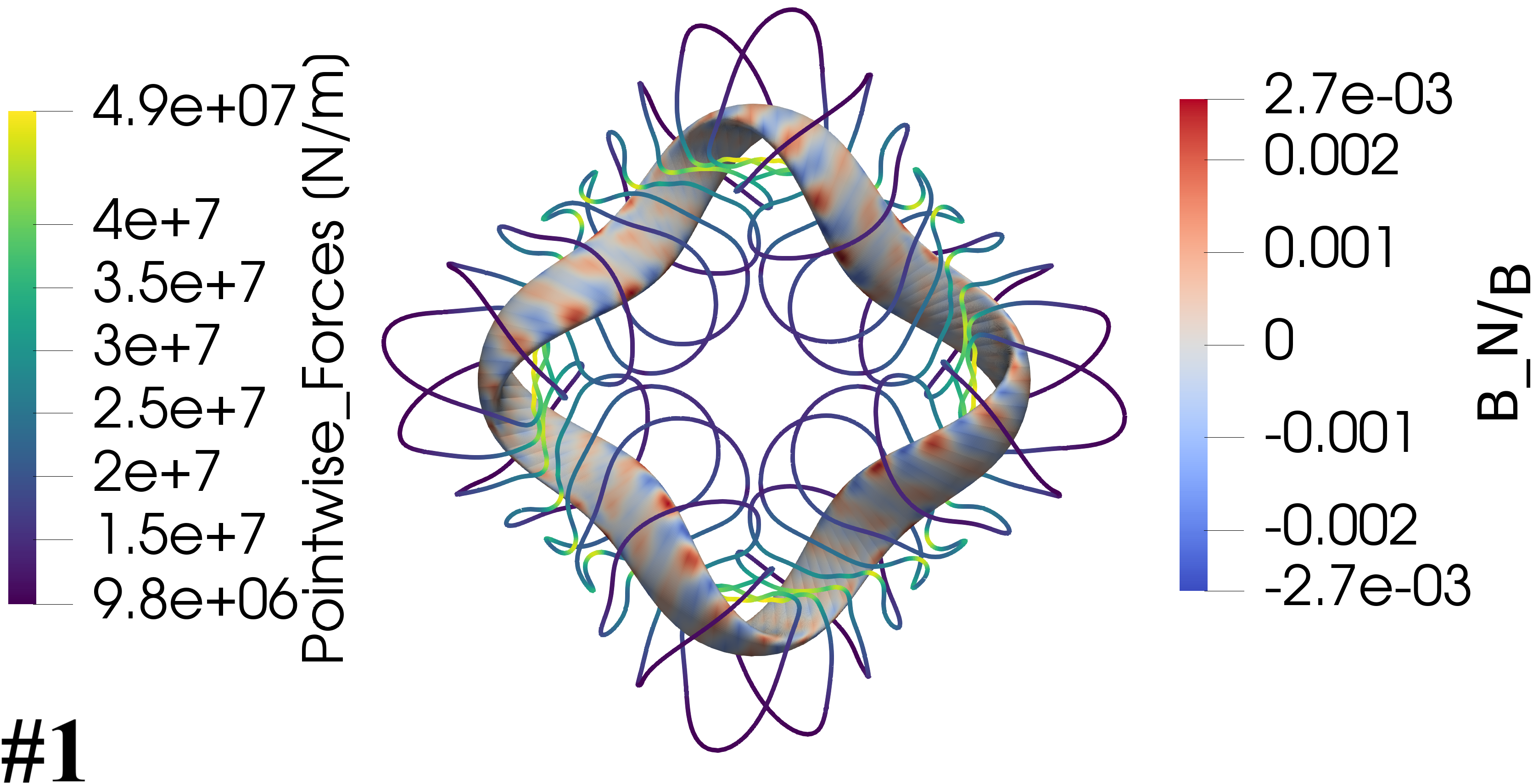}

    \includegraphics[width=\linewidth]{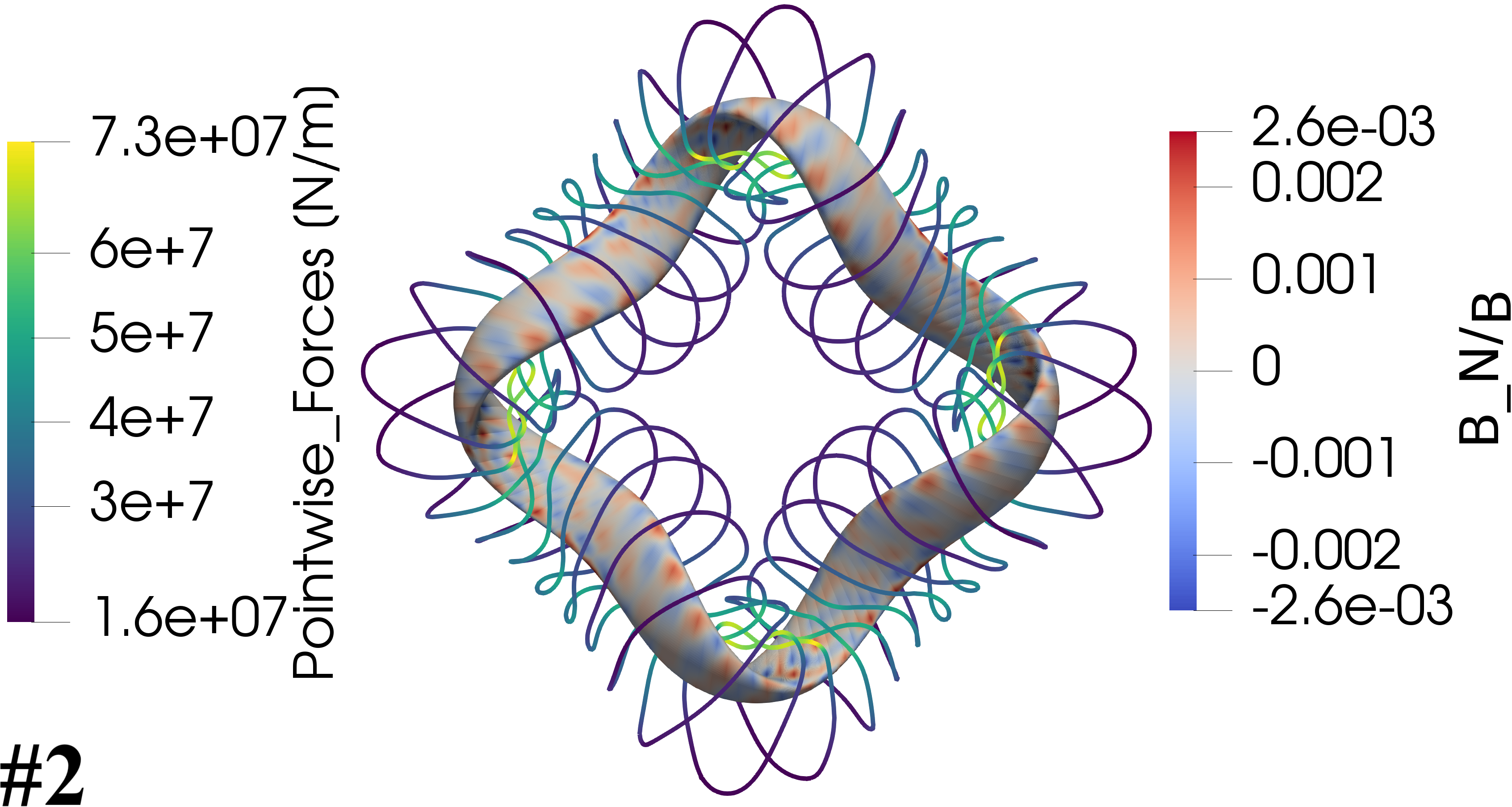}

    \includegraphics[width=\linewidth]{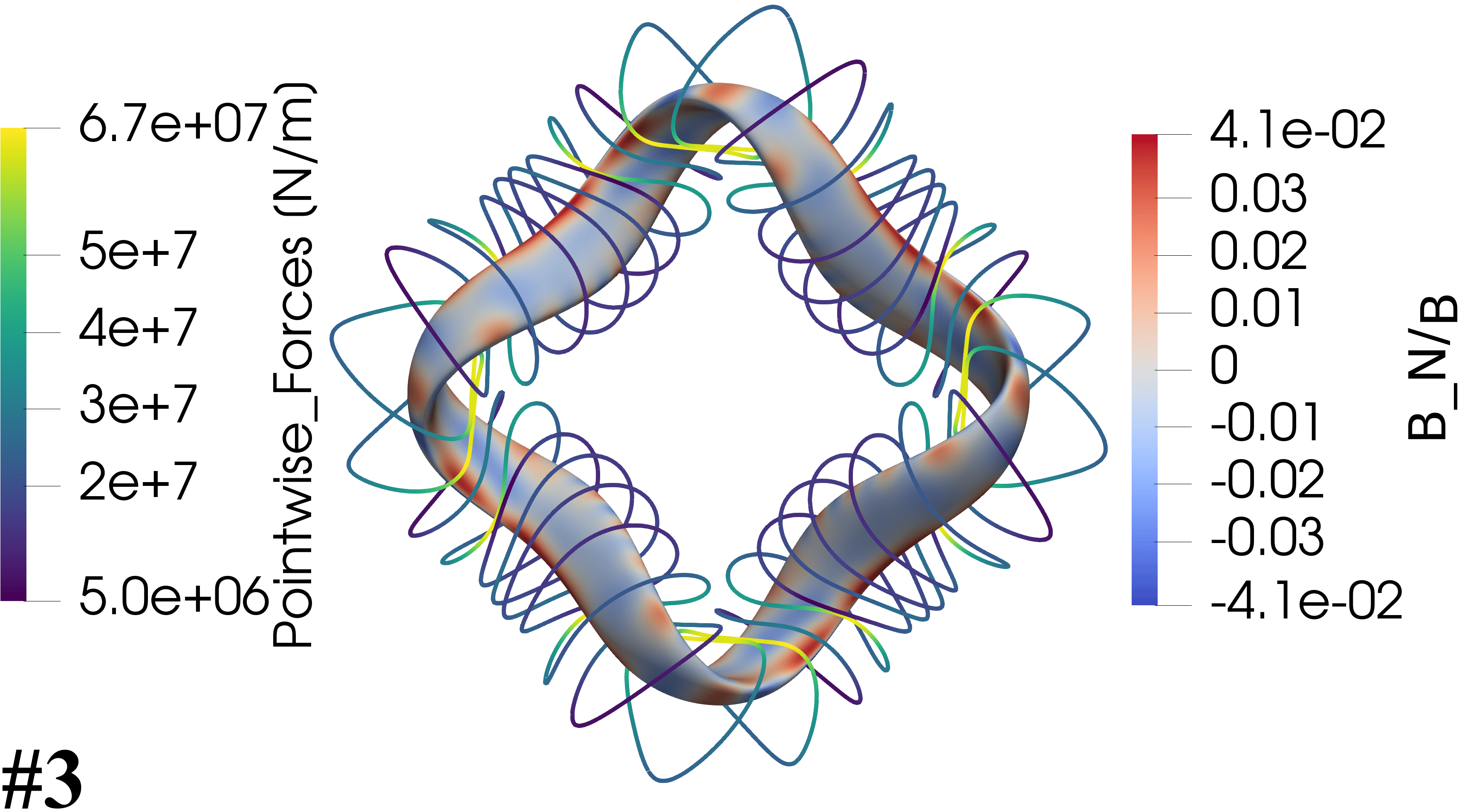}
    
    \includegraphics[width=\linewidth]{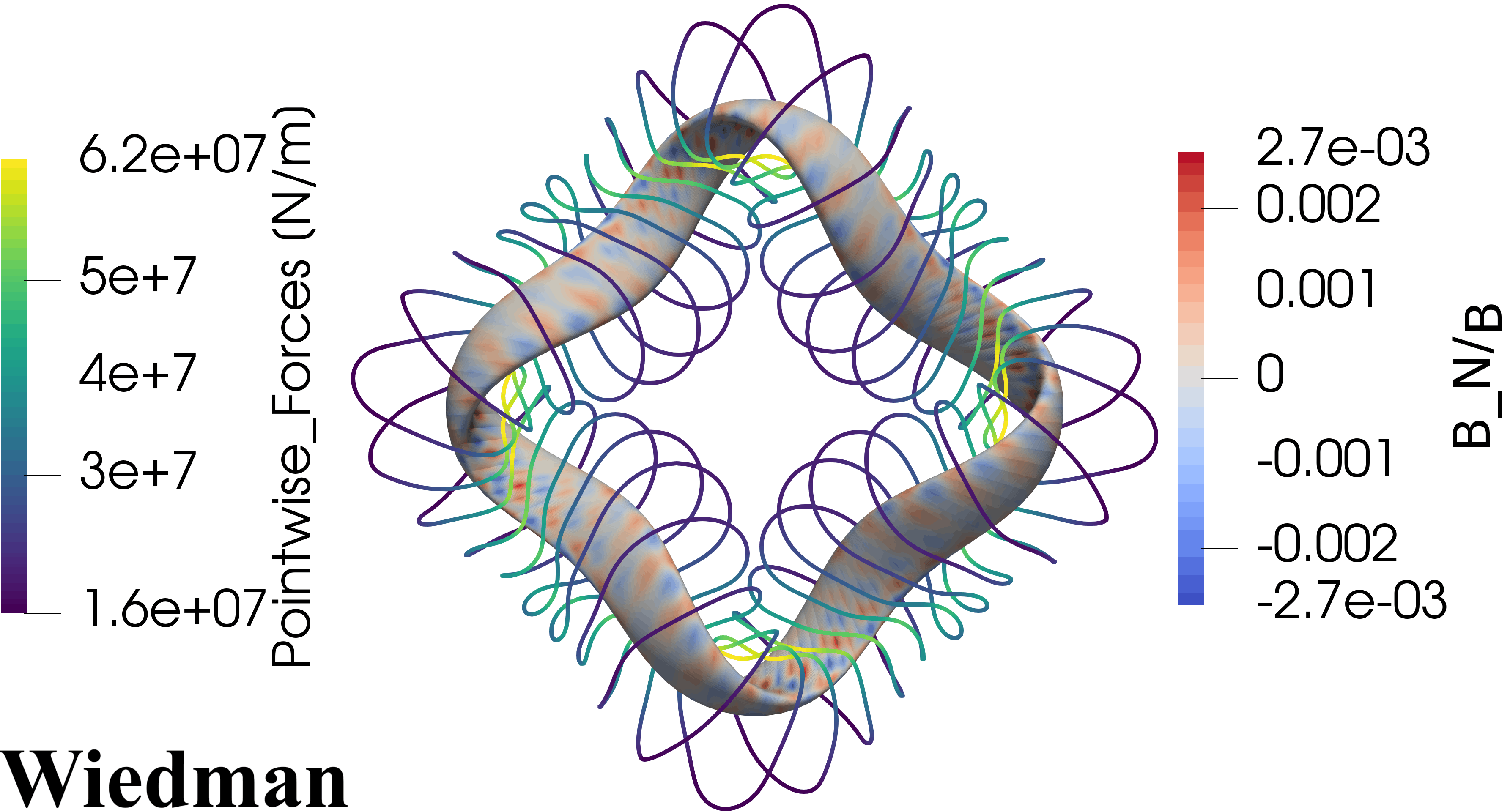}
    \caption{\label{fig:qh_stellarator} QH configurations offering new trade-offs between physics and engineering. (\#1) The stellarator coil configuration for the 4-coil case. (\#2) The stellarator coil configuration for the 5-coil case. (\#3) The stellarator coil configuration designed for optimal engineering properties. (Wiedman) Original 5 coil per half-field period QH coilset optimized by Wiedman \textit{et al.} \cite{Wiedman_Buller_Landreman_2024}. Forces appear very large because they are reported here before dividing by the number of turns.}
\end{figure}

In the original work in Wiedman et al.~\cite{Wiedman_Buller_Landreman_2024}, five coils per hfp were chosen as it was considered that fewer coils did not yield high enough accuracy to reproduce the plasma equilibrium at the coil-to-coil and coil-to-surface distances required. Here, we show that not only is it possible to reproduce a similar configuration as in~\cite{Wiedman_Buller_Landreman_2024}, but it is also possible to both improve the engineering metrics and reduce the number of coils to four per half field period. The total coil length is also reduced by slightly relaxing other physical requirements. The results for all three optimizations are displayed on Table \ref{tab:qh_table}. Both the average field accuracy and maximum field error are comparable to the original work for both the \#1 and \#2 configurations. The stellarators are shown in Figure \ref{fig:qh_stellarator}. Except for $0.2$m-reduced minimum coil-coil distance and more HTS material, the four-coil solution is as good, or better, in every engineering metric as the Wiedman solution. 

To further verify that these coils generate magnetic fields that fulfill physics requirements, we plot the Boozer plot of the outmost closed magnetic surface together with the Poincaré cross-sections of the equilibrium, shown in Appendix~\ref{sec:extra_plots}. As expected the magnetic field amplitude contours appear straight with the low-field regions displaying higher levels of coil ripple than as reported by Wiedman et al., however this is consistent with the fact that the configuration has one less coil per half-field period than previously. It is also possible to reveal the nested structure of the QH equilibrium without any apparent magnetic island chains, capturing the high degree of shaping of the equilibrium.
\begin{table}[ht!]
\begin{tabular}{|l|c|c|c|c|c|}
\hline
\textrm{Property} & Goal & \#1 & \#2 & \#3 &\textrm{Wiedman} \\
\hline
$\#$ of coils per hfp & $\downarrow$ & \textbf{4} & 5 & 5 & 5 \\
$\langle \bm B\cdot \bm n\rangle / \langle B \rangle\times 10^{-4}$ & $\downarrow$ & $6.2$ & $6.2$ & 122 & $\bm{6.1}$ \\
Max $(\bm B\cdot \bm n / B)\times 10^{-3}$ & $\downarrow$ & $2.7$ & $\bm{2.6}$ & 41.2 & $3.1$ \\ 
Max Force [MN/m] & $\downarrow$ & $\bm{0.2}$ & 0.36 & 0.33 &0.31 \\
Coil Length per hfp [m] & $\downarrow$ & \textbf{160} & 177.8 & \textbf{160} & 177.8 \\
Min CS Distance [m] & $\uparrow$ & 1.91 & 1.59 & \textbf{2.0} &1.62 \\
Min CC Distance [m] & $\uparrow$ & 0.8 & 1.09 & \textbf{1.20} &1.09 \\
Max $\kappa$ [$\text{m}^{-1}$] & $\downarrow$ & 1 & 0.77 & \textbf{0.39} &0.81 \\
Max MSC [$\text{m}^{-1}$] & $\downarrow$ & 0.1 & 0.08 & \textbf{0.05} &0.079 \\
Quasisym. Error $\times 10^{-3}$ &  $\downarrow$ & $5.2$ & $2.3$ & 1602 &$\bm{2.2}$ \\ 
HTS length [km] & $\downarrow$                    & $320$        & 284.5            & \textbf{256.0} & 284.5     \\
\hline
\end{tabular}
\caption{Comparison of coil properties for the three different QH configurations, assuming 200 turns per coil for the max force calculations. Properties that should ideally be lowered (increased) in the design are indicated by $\downarrow$ ($\uparrow$).
}
\label{tab:qh_table}
\end{table}

Confinement of collisionless trapped particles is also strongly suggested by the low quasisymmetric error present in configurations \#1 and \#2. Note that this is achieved without necessarily breaching the thresholds and activating the constraints in the optimization. In particular, this is visible in table \ref{tab:qh_table} for the coil-surface distance (lower bound set to $1.5$m) and the maximum curvature (upper bound set to $1$m$^{-1}$) for the \#1 and \#2 configurations respectively. This shows that \textit{it is possible to set loose upper and lower bounds and still obtain feasible solutions that do not necessarily achieve the bounds.} 

Finally, the goal of configuration \#3 is to show that by relaxing the field accuracy target our approach is able to improve every engineering metric that is optimized. As shown in table \ref{tab:qh_table}, while the field accuracy is in the order of 1\%, configuration \#3 achieves all the best engineering properties, except for the forces, which in this case were not directly targeted during optimization. Notably there is a 38cm increase of coil-to-surface distance, a 17.8m decrease in the coil length per hfp, and a 11cm increase in the coil-to-coil distance. This translates into a 10\% cost reduction in HTS tape, while having considerably simpler coils with 52\% percent less local curvature and 37\% percent less mean squared curvature. The simplicity of the coils is visible in Figure \ref{fig:qh_stellarator}. The maximum force also remains comparable to the Wiedman case. However, the quasisymmetric error reaches values that are three orders of magnitude higher than the other configurations. Although the field accuracy is within reasonable bounds ($\langle \bm B\cdot \bm n\rangle / \langle B \rangle \approx 1\%$), it is larger by a factor of 20 than the Wiedman configuration and the rapid deterioration of quasisymmetry comes from high sensitivity of the equilibrium to field errors from the coils. As was mentioned previously, the Landreman\&Paul configuration was targeted to reach minimal levels of quasisymmetric errors. Complementary, single-stage methods can be employed to improve the compatibility between equilibrium and coils.
In Section \ref{sec:hsx}, we present results in the same spirit as for this configuration but for the fully manufactured and assembled QH HSX stellarator. 


\subsection{Stellaris: Quasi-Isodynamic Stellarator}

A third demonstration of the improvement brought by our method is shown on a Quasi-Isodynamic (QI) stellarator configuration ~\cite{goodman_prx} that was recently published by the fusion start-up Proxima Fusion~\cite{LION2025114868} to serve as prototype for a fusion power plant. This builds upon the promising experimental results of the Wendelstein 7-X experiment, an optimized near-QI device \cite{Beidler2021}. The interest behind this class of stellarators lies in their physics properties and recent optimization reported in~\cite{goodman_prx}, where they appear MHD stable, as a consequence of the maximum-$\mathcal{J}$ property described in~\cite{Helander_2009, Helander_2014}. This property is also predicted to mitigate turbulence from the trapped electron mode (TEM)~\cite{Helander_2012, mynick_prl, Proll_Plunk_Faber_Görler_Helander_McKinney_Pueschel_Smith_Xanthopoulos_2022}. The same class of stellarators is being pursued by other companies such as Type One Energy \cite{Anderson_Canik_Hegna_Mowry_2025}. Moreover, stellarators such as Stellaris display low small collisional thermal transport, bootstrap current, and fast-ion losses, which can have a great impact on plasma heating, stability and divertor operation~\cite{Helander_2012, helander_bootstrap, Helander_2014}. Finally, there is also evidence suggesting that ITG turbulent heat fluxes of stellarators of this class are below that of the W7-X stellarator \cite{goodman_prx}. However, for past QI configurations it has been challenging to find buildable and accurate coils such as W7-X.
On top of that, fine-tailored field line curvature imposing the maximum-$\mathcal{J}$ property or MHD stability render the task even more complex. Proxima Fusion published a four field-period QI configuration of aspect ratio 9.8 with major radius 12.7m, minor radius 1.3m, a plasma volume of 425$\text{m}^3$ (about half the volume of ITER), an axis averaged magnetic strength of 9T, and is supposed to deliver 2700MW of peak fusion power. 
The coilset found by Proxima consists of six coils for half-field period. While this allows for more field accuracy, it also comes at the cost that by accounting for the finite-dimensions of the winding pack, the minimum casing-to-casing distance reaches 11.4 mm and assuming 200 turns per coil, peak HTS turn linear force on the filamentary coils is of 0.9 MN/m. This value is at the peak material force tolerances and limits the possibility for structural supports. These very tight engineering constraints are typical for reactor-scale stellarators. 

\begin{figure}[ht!]
    \centering
    \includegraphics[width=\linewidth]{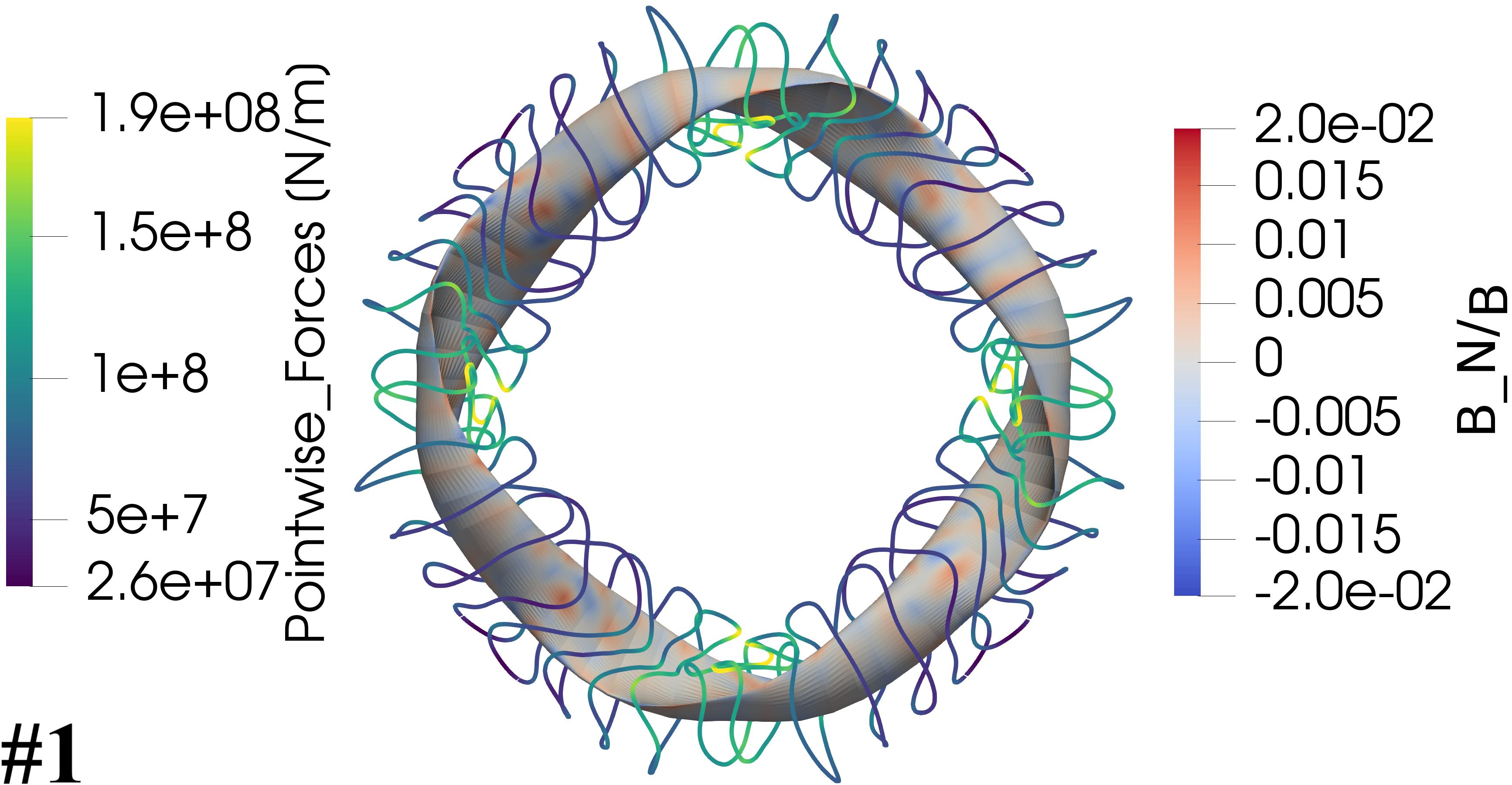}
    
    \includegraphics[width=\linewidth]{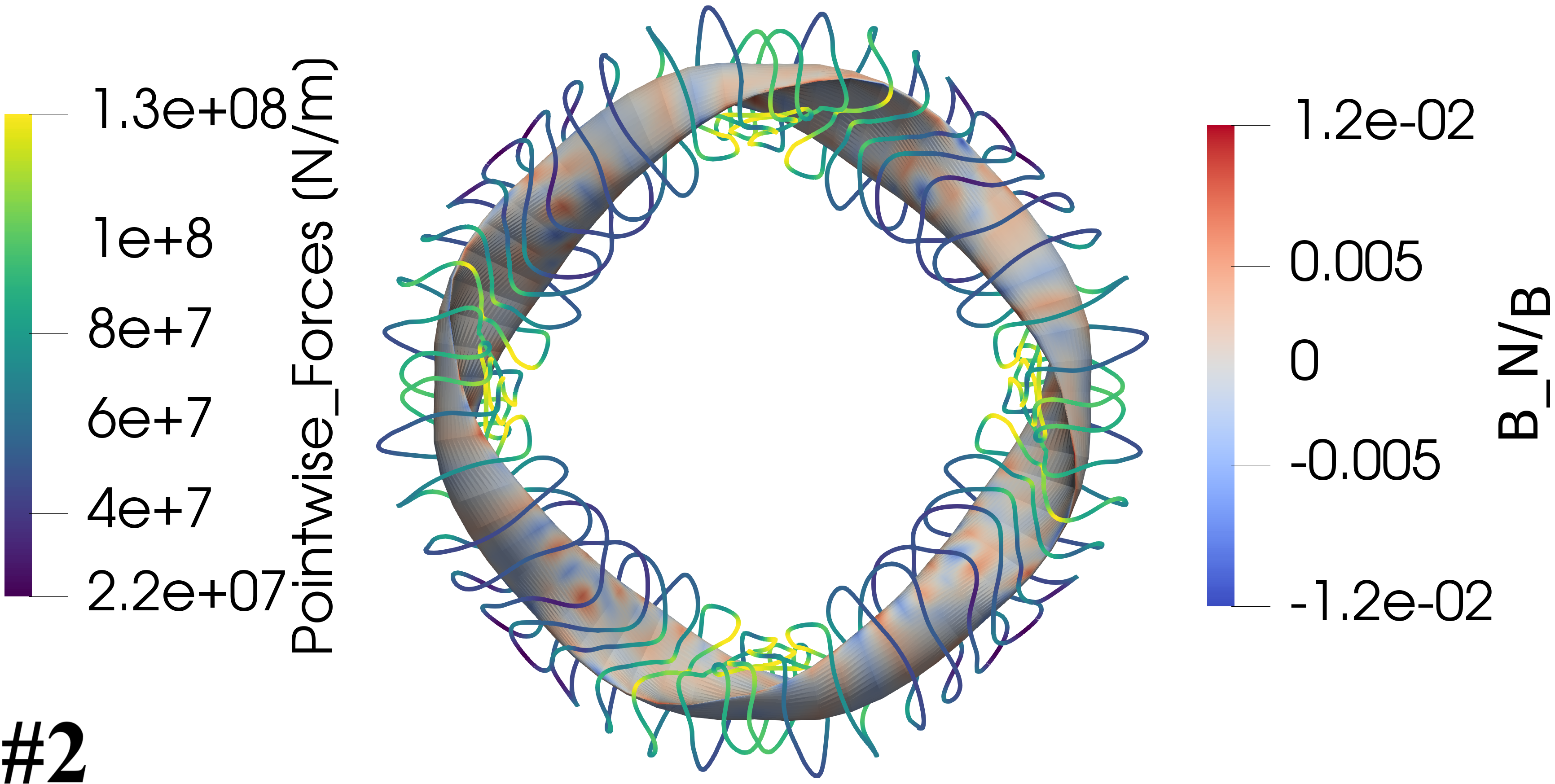}

    \includegraphics[width=\linewidth]{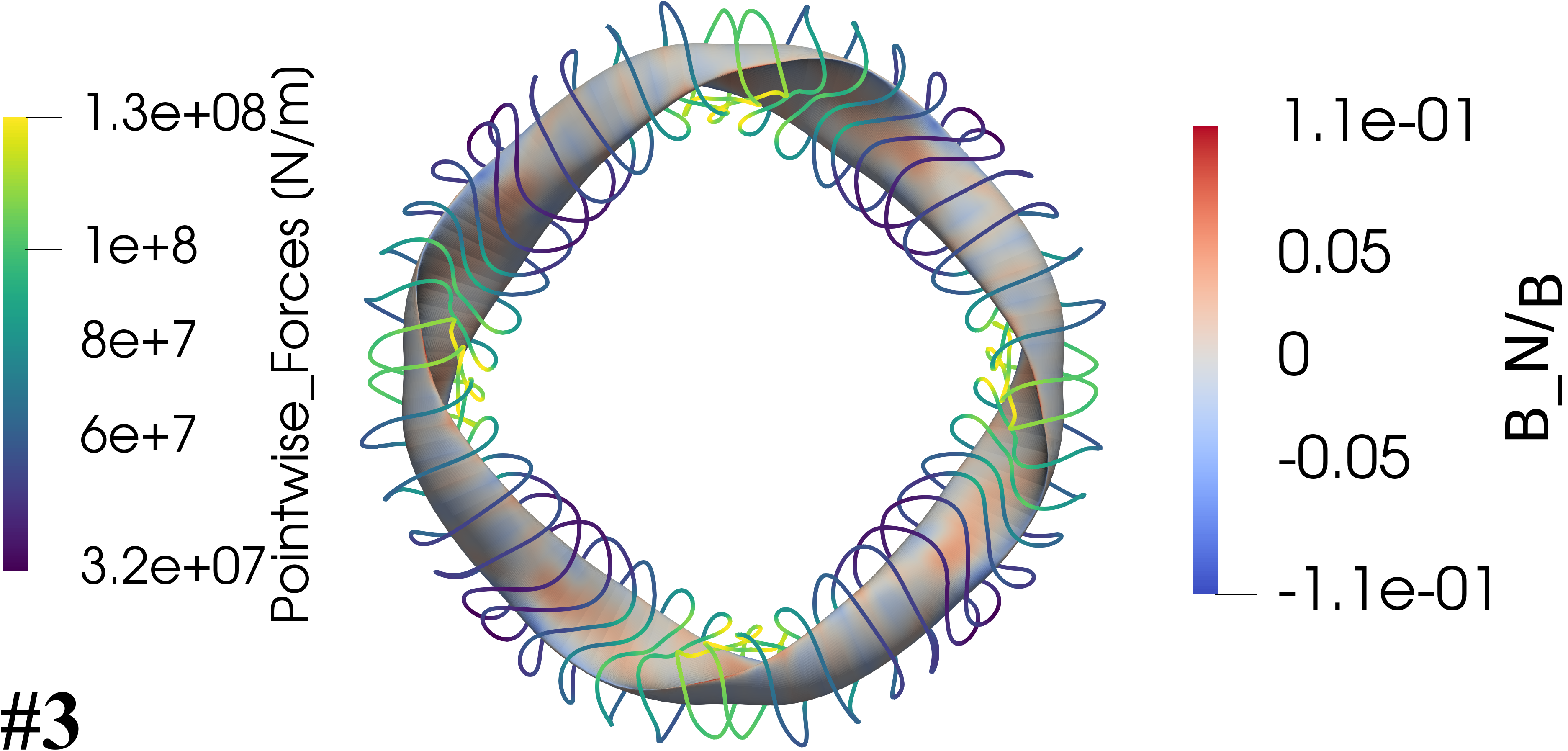}

    \includegraphics[width=\linewidth]{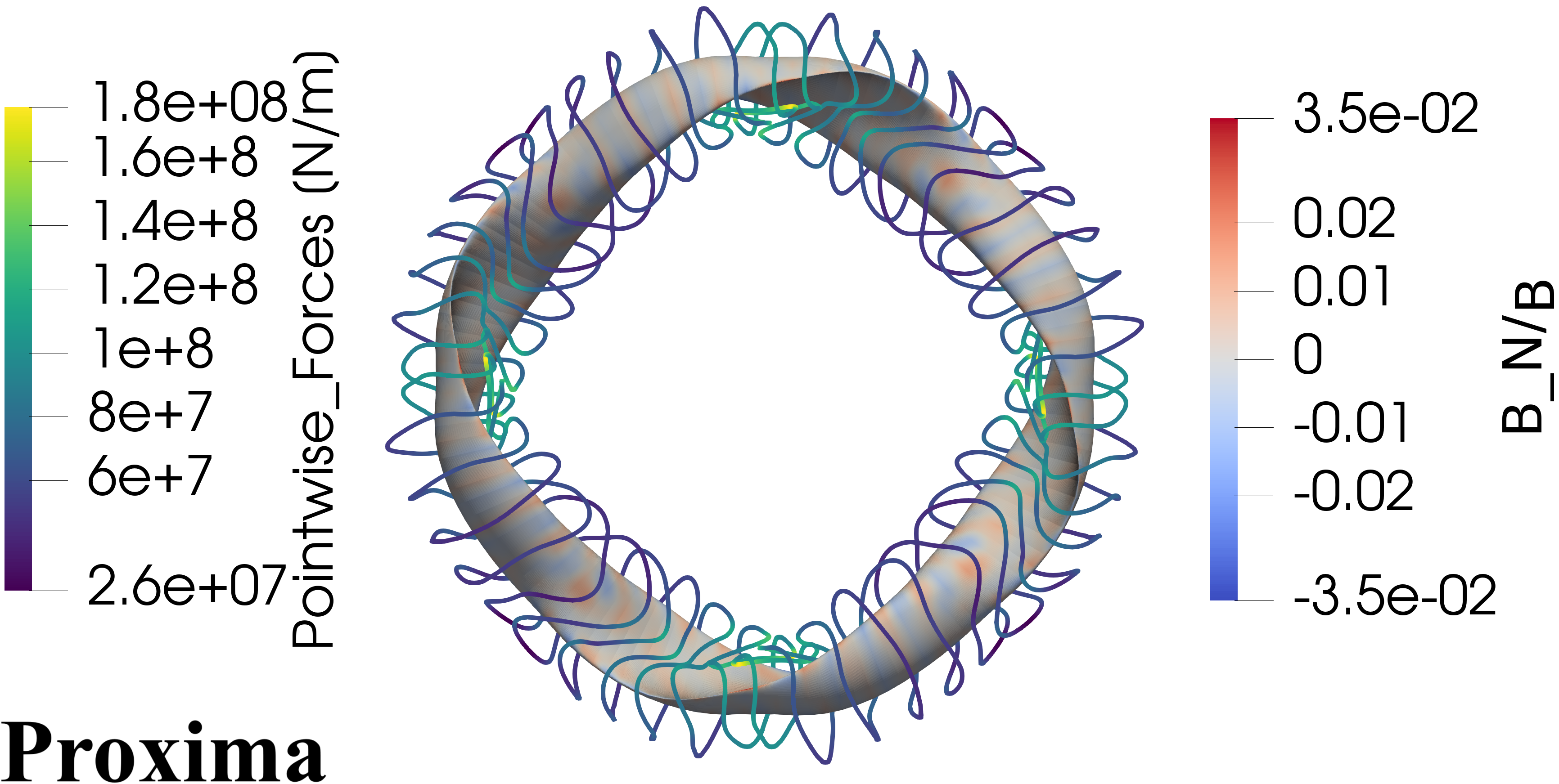}
    \caption{New alternative coil configurations found for the Stellaris fusion power plant concept. (\#1) First possible configuration for the Stellaris configuration with 5 coils per half-field period, same field errors. (\#2) Second possible configuration with similar parameters but forces are reduced by $29\%$. (\#3) Third possible configuration designed for optimal engineering properties. (Proxima) Proxima's configuration with 6 coils per hfp, recently published in \cite{LION2025114868}. Forces appear very large because they are reported here before dividing by the number of turns.}
    \label{fig:stellaris_coils}
\end{figure}

Here three alternative configurations are presented, one with 5 coils, two with 6 coils per half-field period. In the former case, setting similar thresholds as the characteristics of Proxima's coilset except for the intercoil distance, yields a configuration that improves on the magnetic field error average by 25\%, as well as giving an additional 13cm of space between coils which allows for better handling of eventual inaccuracies from manufacturing, assembly, thermal contraction or magnetic forces which appear 5\% larger. The coilset is displayed on Figure \ref{fig:stellaris_coils}. The main parameters are summarized in Table \ref{tab:stellaris_table}. Taking 200 turns per coil in the 6-coil solutions as a reference, and fixed total current per half-field period, it is possible to extrapolate the amount of HTS length required for the 2 solutions. While the two solutions come at a slightly higher cost of HTS tape, these configurations can enable extra access to the plasma, and facilitate reactor maintenance, which is a cost driver. Moreover, the 6 coil case \#2 improves on both the 5 coil configuration and Proxima's coils with a field error that is cut by around half, linear forces on the HTS turns that are 29\% lower, similar distances between coils and between the plasma surface and the conductor as Proxima. Note that this is however a conservative estimate of the amount of HTS required as it assumes that the tape is always operated at the limit of arbitrarily set engineering requirements for the critical current.
In order to achieve the reduced force configuration \#2, the curvature and length curvatures constraints were relaxed, allowing the total length of the conductors to reach 145m.
The corresponding Boozer plots which serve as a verification for the departure from the target isodynamicity are shown in Appendix~\ref{sec:extra_plots}. 

Furthermore, configuration \#3 takes advantage of the new threshold formulation for the quadratic flux by relaxing it while making every other constraint stricter. This yields a configuration with field errors four times higher than Stellaris but improves essentially every engineering metric. This is of special interest since it is unclear if achieving accuracies of $\langle \bm B\cdot \bm n\rangle / \langle B \rangle \sim 10^{-3}$ is at all possible at a reactor scale without auxiliary coils. Therefore, we choose to optimize for a coilset with higher errors but with for example more 22cm of spacing between coils leaving space for error correction coils. This is achieved while having simpler geometries with a 39\% decrease in local curvature, a 35\% decrease in mean squared curvature and 8m/turn shorter coils giving a total 6\% decrease in HTS cost. Similarly to configuration \#2, forces are also reduced by 29\%. 
Most importantly, despite the substantial engineering improvements of configuration \#3, fast-particle tracing comparisons in Appendix~\ref{sec:fp} show that total losses remain below $4\%$. 

%
%
\begin{table}[ht!]
\begin{tabular}{|l|c|c|c|c|c|}
\hline
Property  & Goal                     & \#1     & \#2 & \#3 &Proxima \\ \hline
\# of coils per hfp & $\downarrow$                             & \textbf{5}      & 6  & 6 & 6              \\ 
$\langle \bm B\cdot \bm n\rangle / \langle B \rangle\times 10^{-3}$ & $\downarrow$ & $3.0$   & $\bm{2.3}$ & 14.7   & $4.0$             \\ 
Max $(\bm B\cdot \bm n / B)\times 10^{-2}$ &  $\downarrow$                       & $2.2$   & $\bm{1.6}$   & 11.4 &$3.5$         \\ 
Max Force [MN/m] & $\downarrow$              & 0.79    & \textbf{0.64}    & \textbf{0.64} & 0.9            \\ 
Coil Length per hfp [m] & $\downarrow$                        & \textbf{130}    & 145   & \textbf{130} & 138            \\ 
Min CC Distance [m] & $\uparrow$                       & 0.8    & 0.7 & \textbf{0.89} & 0.67            \\ 
Min CS Distance [m] &  $\uparrow$                & 1.38   & 1.3 & \textbf{1.40} & 1.37           \\ 
Max $\kappa$ [$\text{m}^{-1}$] & $\downarrow$       & 1.57   & 1.6  & \textbf{0.96} & 1.57           \\ 
Max MSC [$\text{m}^{-1}$] & $\downarrow$                    & 0.3    & 0.35  & \textbf{0.22} & 0.34          \\ 
HTS Length [km] & $\downarrow$  & 250 & 232 & \textbf{208} & 221
\\ \hline
\end{tabular}
\caption{Comparison of magnetic field and coil properties for the three different Stellaris configurations. \#1, \#2 and \#3 are results from this work, the Proxima configuration was optimized by the start-up Proxima Fusion. Forces were scaled by assuming 200 turns per coil in the six-coil solution and fixed total coil current per half-field period.}
\label{tab:stellaris_table}
\end{table}
As shown in Table \ref{tab:stellaris_table}, the Stellaris configuration is improved in essentially all aspects by the 3 different coil solutions found here. Notably, configuration \#3 is able to improve all the engineering metrics, including 29\% less forces and a 6\% reduction in estimated HTS cost. Nonetheless, final decisions about coil and HTS stack design require full finite-element modeling of the coils and structural supports beyond the scope of this work.   
\subsection{Wendelstein 7-X (W7-X)}

So far the configurations described correspond to devices that have been conceptualized but not yet built. In this and the next sections, we show that it is possible to improve the coil designs of stellarators that have been fully manufactured and assembled and already delivered experimental results. The predecessor of Wendelstein 7-X, Wendelstein 7-AS, was the first stellarator to successfully  demonstrate that the two stage approach to stellarator design, as previously described, is able to generate a feasible reactor \cite{stellarator_intro}. In order to bring forward the stellarator program the main goals of W7-X were to demonstrate the generation of nested magnetic surfaces with modular superconducting coils, enhanced fast-particle confinement, reduced parallel currents, minimal neoclassical transport at low collisionality, and MHD stability up to an average plasma beta of 5\% \cite{w7x_beidler}. Wendelstein 7-X is currently the largest operating optimized stellarator in the world, with a major radius of 5.5m and an aspect ratio of 10, being at the forefront of stellarator research and fusion science generally. The first experimental campaign was performed in 2015, using NbTi superconducting coils to generate a quasi-isodynamic magnetic field with 2.5T on axis. In addition to its non-planar coils, W7-X has a set of planar coils necessary for flexibility in the magnetic configuration. As mentioned in the introduction, the W7-X manufacturing and assembly process presented various delays due to tight coil tolerances. In practice this meant that for example ``six magnets were returned to the manufacturers three times for repair, and six others were returned twice for repair before passing the tests successfully" \cite{cea}. This resulted in additional costs. 

\begin{figure}[ht!]
    \centering
    \includegraphics[width=\linewidth]{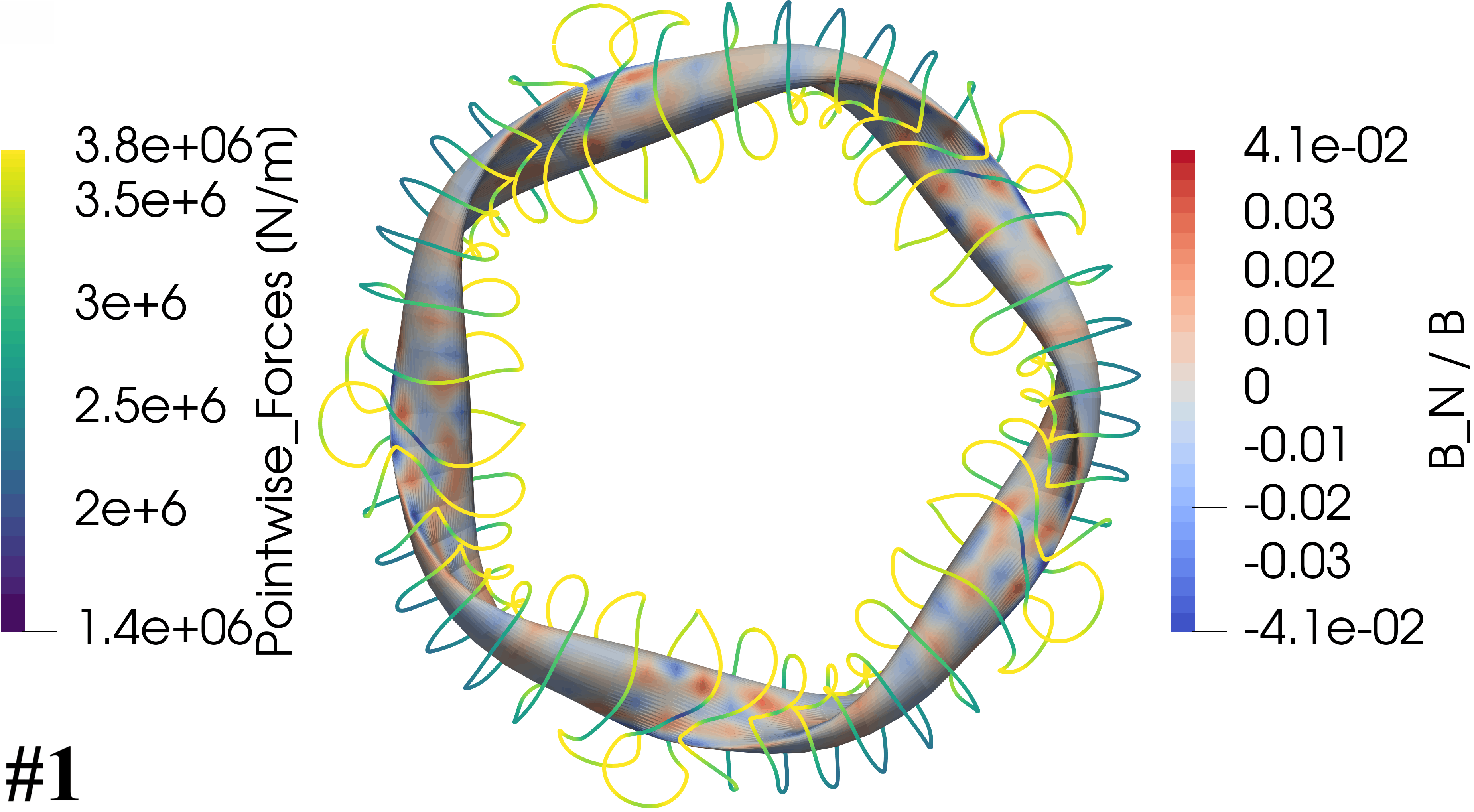}
    
    \includegraphics[width=\linewidth]{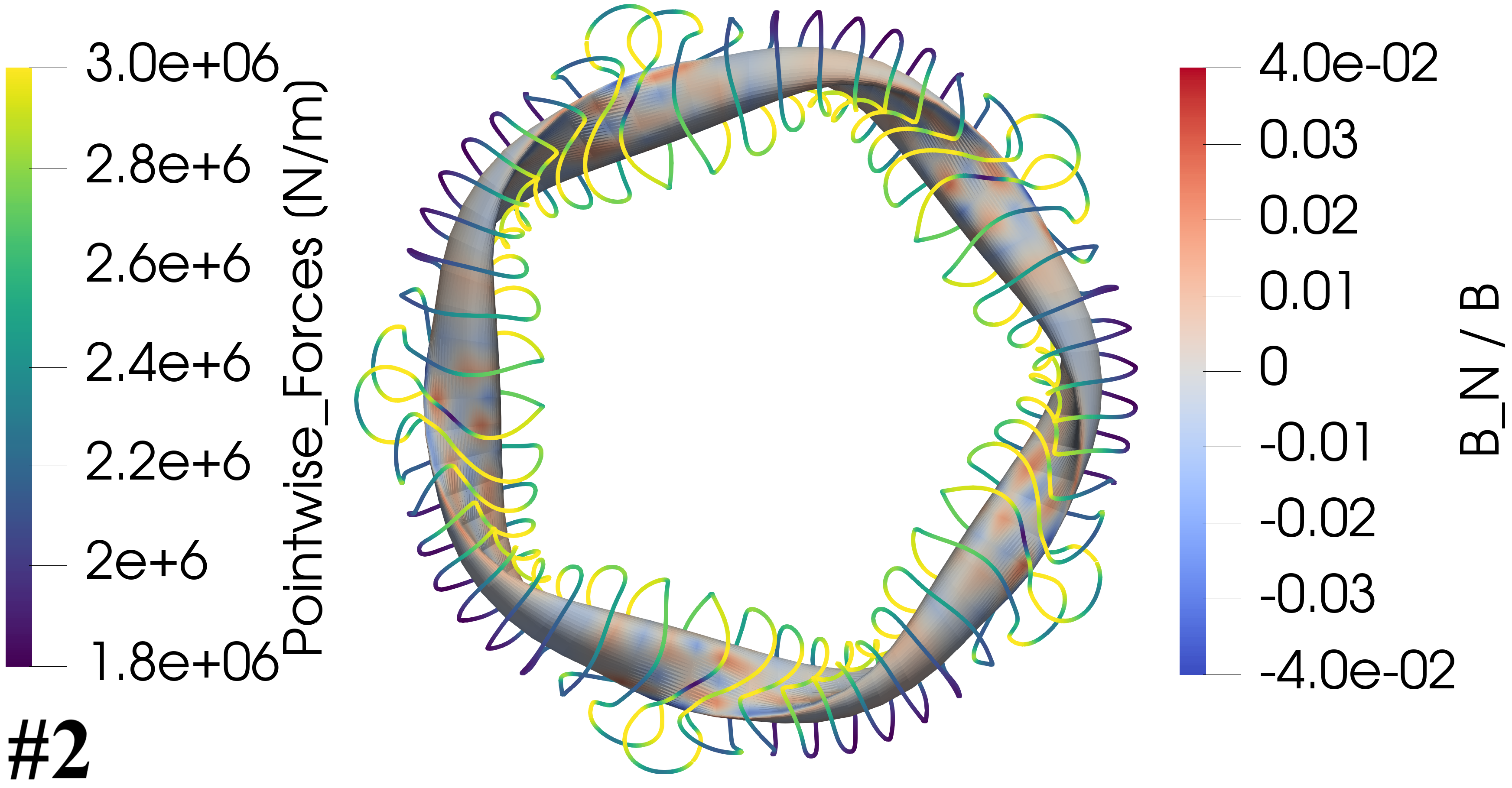}

    \includegraphics[width=\linewidth]{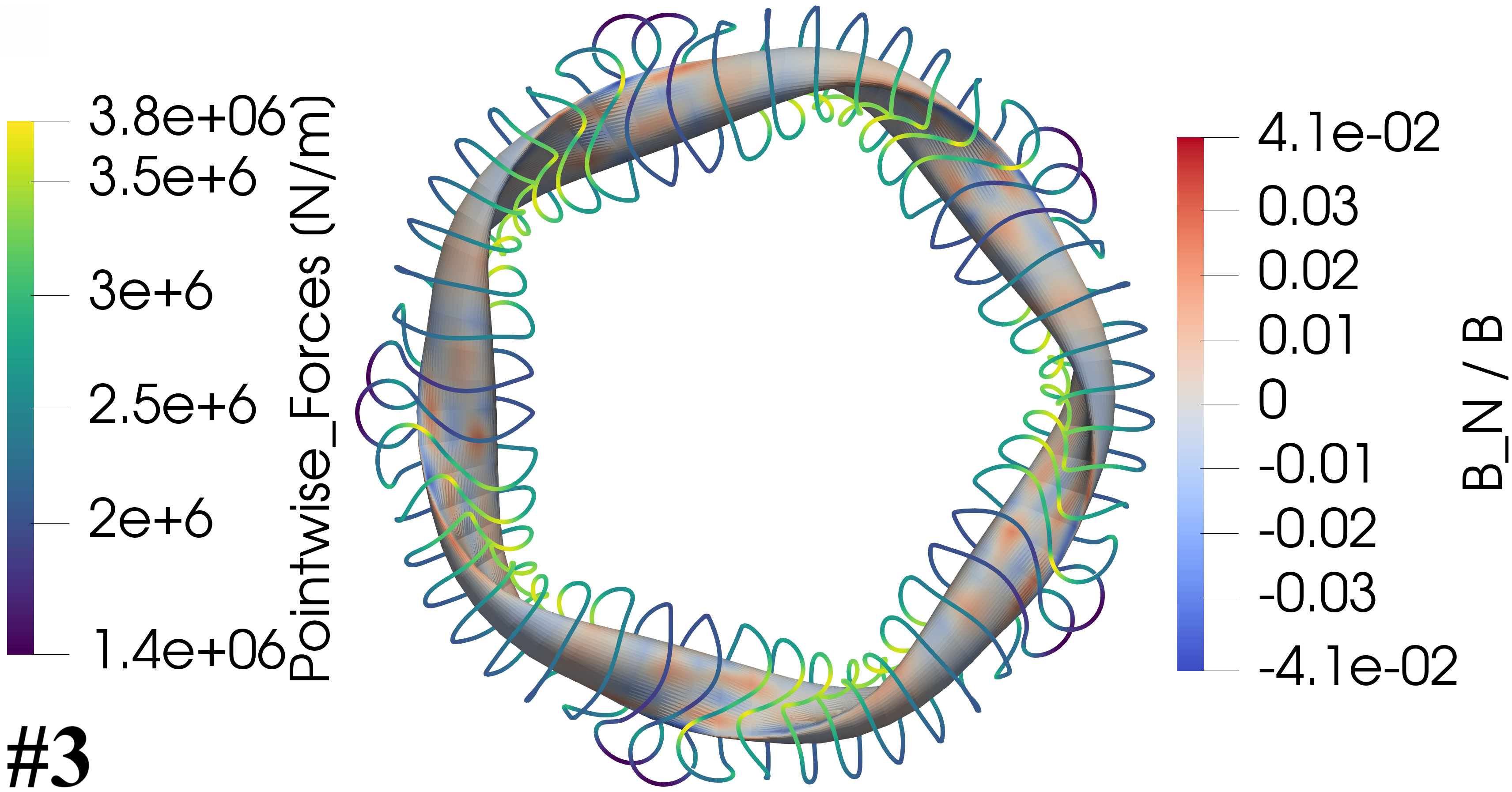}

    \includegraphics[width=\linewidth]{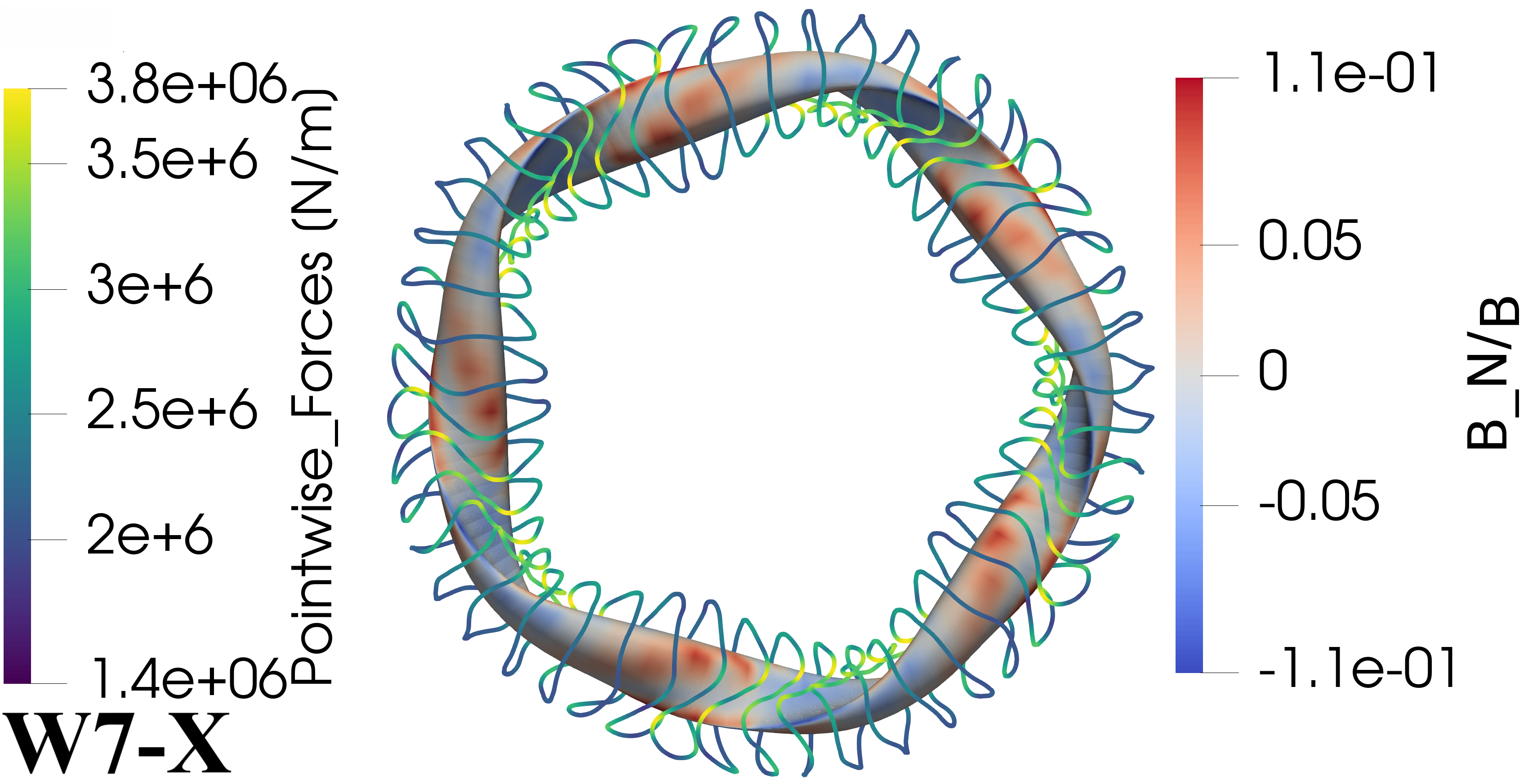}
    \caption{Newly optimized W7-X coil sets with the augmented Lagrangian formalism. (\#1) First alternative configuration for the W7-X stellarator with 4 coils per half-field period, improved field errors, simpler coils. (\#2) Second alternative coilset for the W7-X stellarator with improved field accuracy, 20\% less forces, longer coils, but lower curvature.
    (\#3) Third alternative coilset for W7-X with improved field error, similar coil length, same forces and simpler coils.
    (\#4) Original W7-X coilset extracted from the CAD design with 5 coils per half-field period. Forces appear very large because they are reported here before dividing by the number of turns.}
    \label{fig:w7x_coils}
\end{figure}

Three different configurations are presented as options for W7-X in Figure~\ref{fig:w7x_coils}. 
%
%
\begin{table}[ht!]
\begin{tabular}{|l|c|c|c|c|c|}
\hline
Property  & Goal                          & \#1        & \#2            & \#3     & W7-X \\ 
\hline
\# of coils per hfp & $\downarrow$                              & \textbf{4} & 5              & 5       & 5             \\ 
$\langle \bm B\cdot \bm n\rangle / \langle B \rangle\times 10^{-3}$ & $\downarrow$ & $10$   & $6.7$ & $\bm{6.6}$ &$29$          \\ 
Max $(\bm B\cdot \bm n / B)\times 10^{-2}$  & $\downarrow$                        & $5.5$   & $\bm{4.1}$& $4.2$    &$11$         \\ 
Max Force [kN/m] & $\downarrow$                      & \textbf{15}      & \textbf{15}           & 19    &  19      \\ 
Coil Length per hfp [m]  & $\downarrow$                       & $\bm{38}$ & 45             & 43  &  42.6        \\ 
Min CC Distance [m]  & $\uparrow$                      & \textbf{0.3}        & 0.27           & 0.26  &  0.28       \\ 
Min CS Distance [m]  & $\uparrow$                & 0.32       & 0.3            & 0.34  &   \textbf{0.37}      \\ 
Max $\kappa$ [$\text{m}^{-1}$] & $\downarrow$     & 2.51       & \textbf{2.0}              & \textbf{2.0} &  2.56         \\ 
Max MSC [$\text{m}^{-1}$] & $\downarrow$                    & \textbf{1.5}        & \textbf{1.5}            & \textbf{1.5} &   1.96      \\
HTS Length [km] & $\downarrow$                    & 95        & 90            & 86 &   $\bm{85.2}$      \\
\hline
\end{tabular}
\caption{Comparison of magnetic field and coil properties for the four different W7-X configurations. \#1, \#2 and \#3 are results from this work, the W7-X configuration was taken from the original CAD files and adapted as single filaments for the W7-X standard configuration (without additional planar coils). Here the max force and HTS length are computed with 200 HTS turns per coil using the five-coil solution and fixed total current per half-field period.}
\label{tab:w7x_table}
\end{table}
As is illustrated in Table \ref{tab:w7x_table}, all the configurations are optimized to possess lower curvatures than the original W7-X coilset. Here, the last closed flux surface of W7-X that does not contain coil ripple is used as target. The results are visible in the coils shown in Figure \ref{fig:w7x_coils}, as all the newly optimized coils appear smoother and with less local curvature than the W7-X ones. Note that all configurations possess comparable lengths to the W7-X coils. This freedom is purposely imposed for this study, as the goal is to explore options to the current W7-X reactor coils and not to replicate it. 

Configuration \#1 presents substantial advantages over the original W7-X coilset; the full reactor requires 40 coils instead of 50, but still provides better field accuracy, reduced total coil length, increased minimum coil-coil distance, and reduced curvatures. In every metric it is improved over the original design, except for a $5$cm reduced minimum plasma-coil distance. Access to the plasma is substantially increased, as can be seen visually in Fig.~\ref{fig:w7x_coils}. The reduced number of coils can help to drive costs down by reducing material, reducing assembly difficulties, and increasing the amount of plasma access for diagnostics, which can become a bottleneck in the operation of W7-X. Configuration \#2 is targeted to reduce the electromagnetic forces compared to the original design. By allowing only an increase of 5\% of the total coil length, slightly closer coils to each other and to the plasma, it is possible to achieve a reduction of about 20\% of the force load on the support structure and coils of W7-X. Note that it has been reported by Pedersen et al. \cite{Pedersen2016} that the rotational transform profile and island chain structure are modified due to deformations in the coil shapes resulting from increasing electromagnetic forces. Additionally this new configuration comes with an improvement of about an order of magnitude in field accuracy. 
Configuration \#3 is also an exciting find. With very tiny changes of $0.4$m more total coil length, $0.02$m reduction to minimum coil-coil distance, and $0.03$m reduction to minimum coil-surface distance, the squared flux and curvatures are dramatically improved, without change to the max forces. 

The degree of omnigeneity for all coil sets displayed by the magnetic field strength contours is plotted in  Appendix~\ref{sec:extra_plots}.
All W7-X alternatives show contours following the target field.
As shown in Figure \ref{fig:w7x_coils}, the \#2 and \#3 alternatives possess four visibly simple coils out of five with only one coil type standing out as longer. It also appears that lower overall forces can be achieved by relaxing the length constraint on an individual coil type. To our current knowledge and given the current parameters of the W7-X coils displayed in Table \ref{tab:w7x_table}, any of these three alternatives of W7-X could constitute viable experimental devices.

\subsection{Helically Symmetric eXperiment (HSX)}
\label{sec:hsx}
\begin{figure}[ht!]
    \centering
    \includegraphics[width=\linewidth]{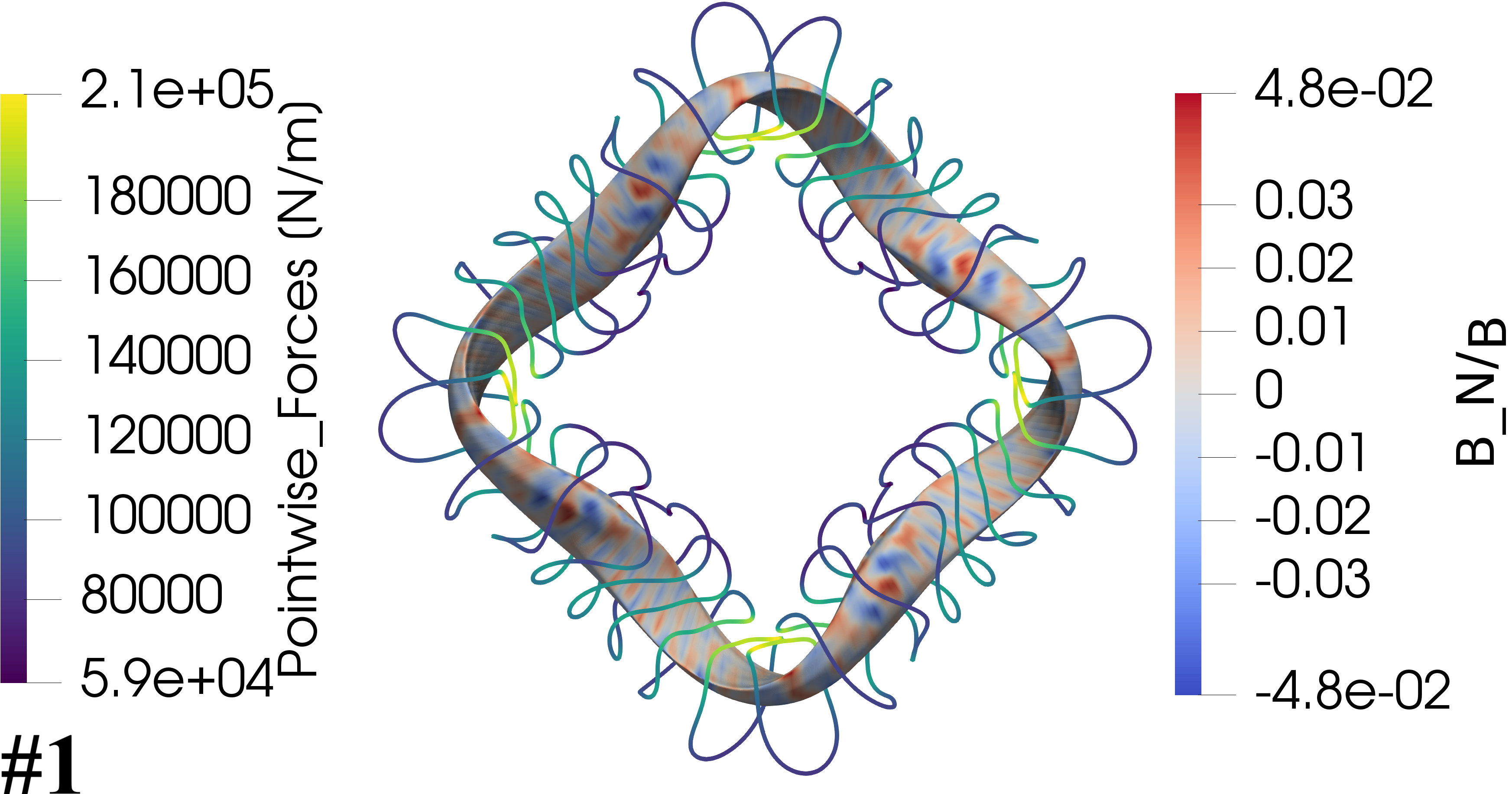}

    \includegraphics[width=\linewidth]{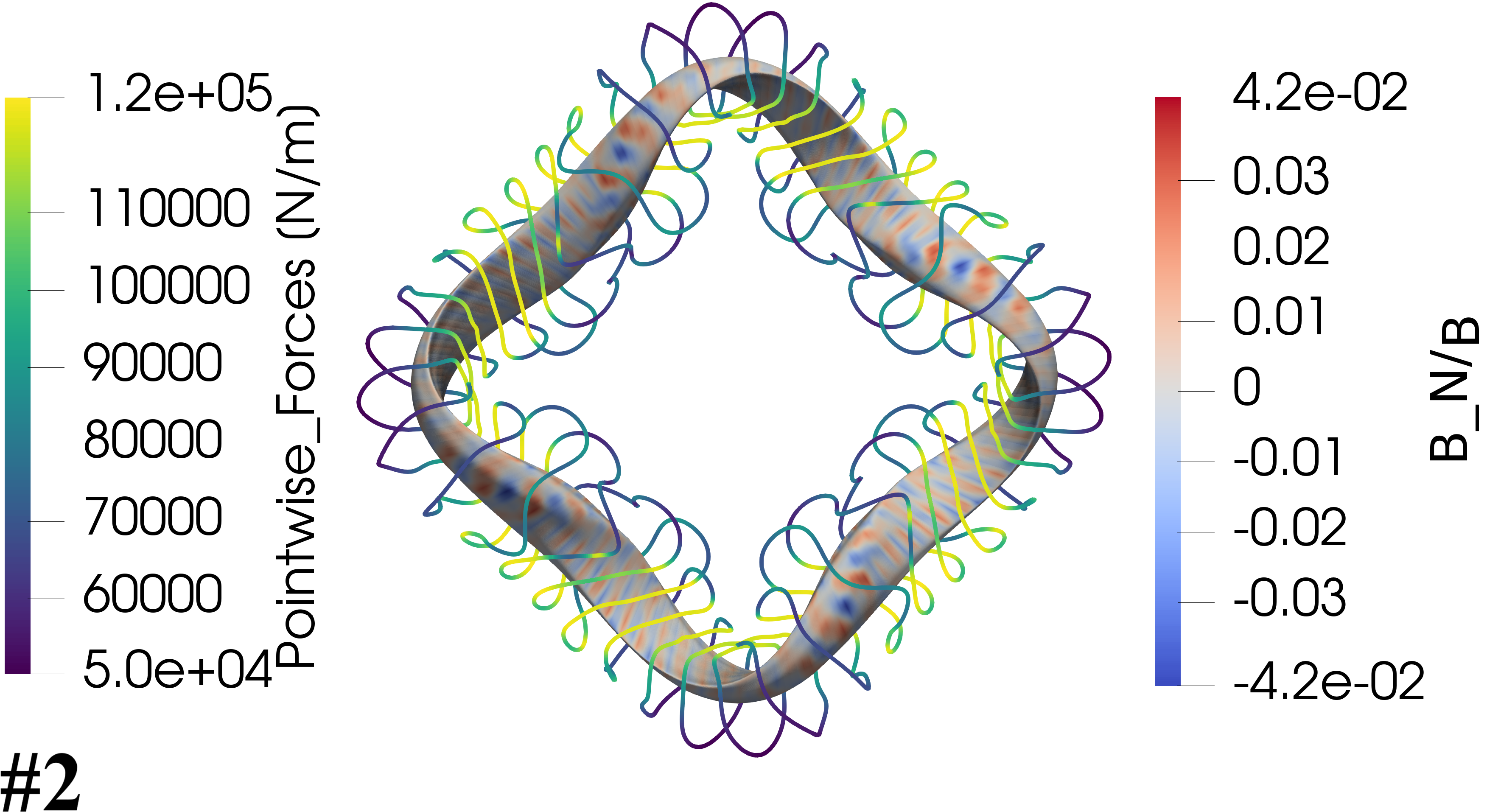}
    
    \includegraphics[width=\linewidth]{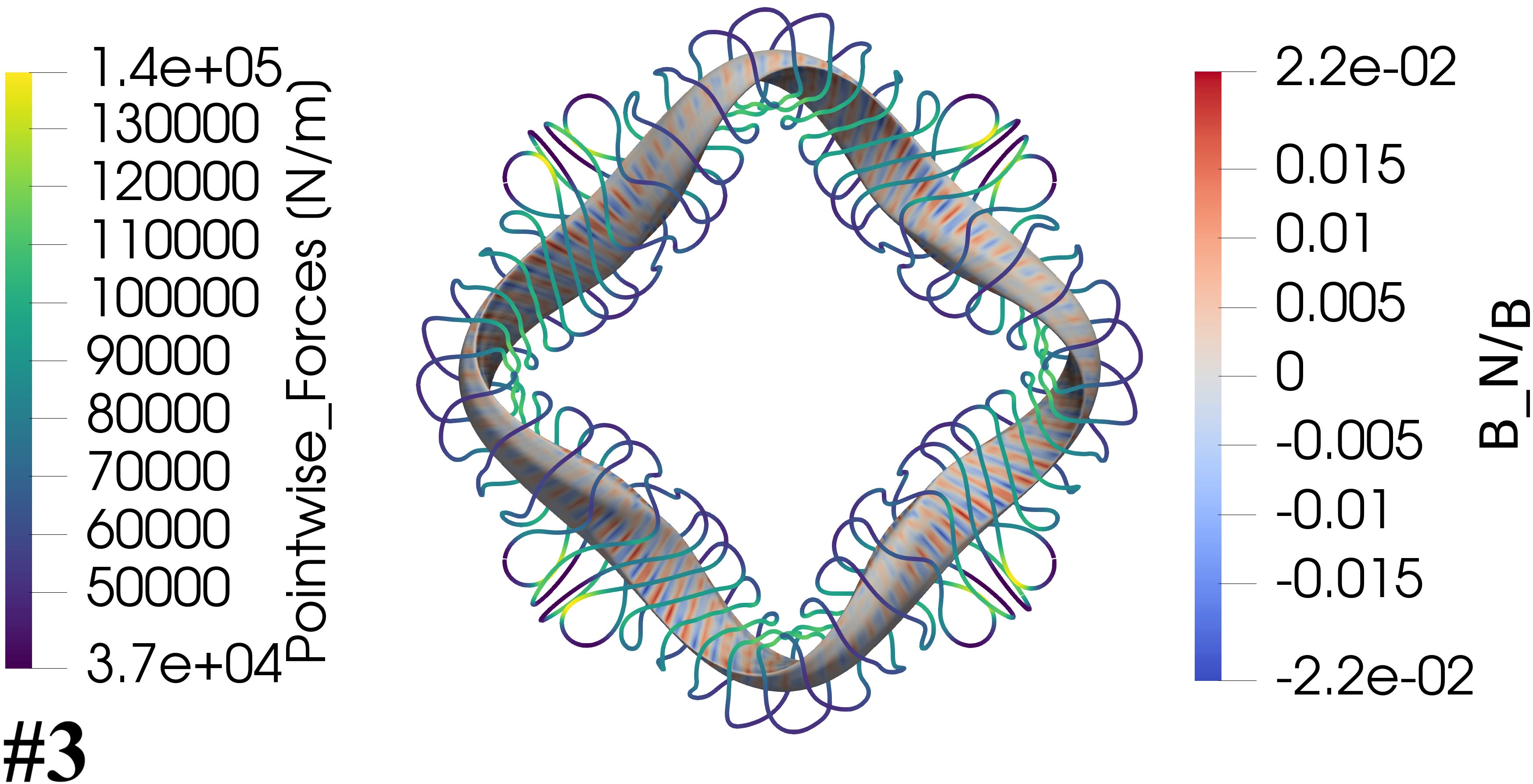}
    
    \includegraphics[width=\linewidth]{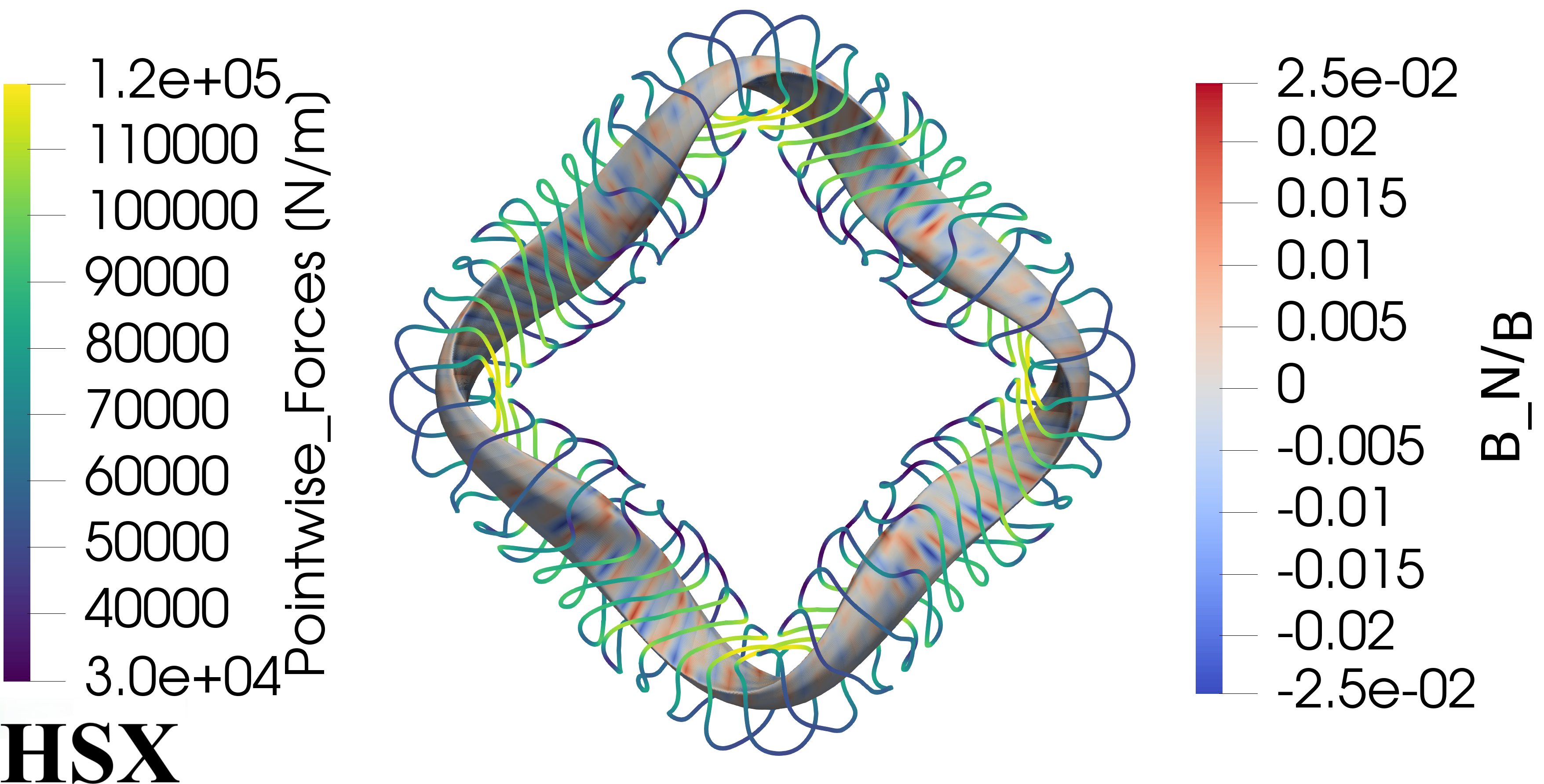}
    \caption{New coil configurations for the HSX reactor optimized with the augmented Lagrangian method. (\#1) First alternative configuration for the HSX stellarator with 4 coils per half-field period, simpler coils, improved coil-to-surface and intercoil distance. (\#2) Second alternative coilset for the HSX stellarator with 5 coils per half-field period and simpler coils.
    (\#3) Third alternative coilset for HSX with improved field error, similar coil length, and considerably simpler coils.
    (\#4) Original HSX coil design with 6 coils per half-field period. Forces appear very large because they are reported here before dividing by the number of turns.}
    \label{fig:hsx_coils}
\end{figure}

The Helically Symmetric eXperiment (HSX) is a stellarator located at the University of Wisconsin in Madison. The device began operation in 1999, and was and still is the only device in the world with a quasi-helical stellarator field. It is a four field period stellarator with 48 copper coils, around 1T on axis, major radius of 1.2m and aspect ratio 10. HSX was designed to demonstrate reduced direct orbit losses, Mercier stability, and reduced neoclassical transport. Due to its rotational transform of 1.05 on axis and quasi-helical field, its neoclassical confinement is similar to a $q=1/3$ tokamak \cite{anderson_almagri_anderson_matthews_talmadge_shohet_1995}. It has also managed to demonstrate reduced flow damping and reduced electron heat transport \cite{flow_damping_2005, canik_2007}. Unlike the QH stellarator equilibrium shown in Section \ref{sec:lp_qh} which was optimized for low quasisymmetric error throughout the volume, HSX was only optimized for quasisymmetry on the last closed flux surface because at the time no robust optimization methods had been developed yet. Similarly to W7-X, HSX coils were obtained using a winding surface method, which yielded six coils per half field period with high curvatures.
From the twisted filaments, a copper winding pack was constructed for the final device, these are shown in Figure \ref{fig:hsx_coils}. The procedure in this section does not aim at improving the forces, since the force requirements become much more stringent at reactor-scale. Given the smaller dimensions and relatively low magnetic fields of this device, we focus instead on finding simpler coils that are easier to manufacture. However, in HSX coils are made of copper, which means that unlike HTS coils curvature does not pose a risk regarding the integrity of the materials. We propose three different configurations with four, five, and six coils per half-field period, summarized in Table \ref{tab:hsx_table}. Note that the target configuration used in this section is a free-boundary surface, meaning that it already has coil ripple embedded into it, making it \textit{harder} to optimize for different number of coils per hfp and partially responsible for the high quasisymmetry errors observed in Table \ref{tab:hsx_table}. The reference coilset used here is obtained by taking the center spline of finite-built coils, which accounts for the field error being non-zero. Nevertheless, we report comparably low field errors, and as expected, the average field error progressively increases with decreasing number of coils. This is consistent with the results from the previous QH, Stellaris and W7-X configurations.

\begin{table}[ht!]
\begin{tabular}{|l|c|c|c|c|c|}
\hline
Property  & Goal                           & \#1        & \#2            & \#3     & HSX \\ 
\hline
\# of coils per hfp & $\downarrow$                               & \textbf{4} & 5              & 6      & 6            \\ 
$\langle \bm B\cdot \bm n\rangle / \langle B \rangle \times 10^{-3}$ & $\downarrow$ & $11$   & $7.5$ & $\bm{5.3}$ &$5.8$           \\ 
Max $(\bm B\cdot \bm n / B) \times 10^{-2}$   &  $\downarrow$                       & $4.7$   & $4.2$ & $\bm{2.2}$    & $2.5$      \\ 
Max Force [kN/m] & $\downarrow$                      & 0.7      & 0.88            & 0.7    &  $\bm{0.6}$      \\ 
Coil Length per hfp [m] &  $\downarrow$                        & $\bm{10.0}$ & 12.1             & 13.4  &  13.4        \\ 
Min CC Distance [m] & $\uparrow$                       & \textbf{0.10}        & \textbf{0.10}           & 0.09  &  0.09       \\ 
Min CS Distance [m] &  $\uparrow$                    & \textbf{0.15}       & 0.13            & 0.11  &   0.14      \\ 
Max $\kappa$ [$\text{m}^{-1}$] & $\downarrow$     &      9.7       & 12              & \textbf{8} & 12.34         \\ 
Max MSC [$\text{m}^{-1}$] & $\downarrow$                    & \textbf{25}        & 30            & 30.0 &   45.1      \\
Quasisym. Error $\times 10^{0}$ &  $\downarrow$ & 40.6 & \textbf{37.6} & 39.2 & \textbf{37.6}\\
HTS length [km] & $\downarrow$                    & 24        & 23.2            & \textbf{21.4} &   \textbf{21.4}      \\
\hline
\end{tabular}
\caption{Comparison of magnetic field and coil properties for the four different HSX configurations. \#1, \#2 and \#3 are results from this work, the HSX configuration was taken from the original CAD files and adapted as single filaments for the HSX standard Quasi-Helical Symmetry QHS configuration. Here the max force is the force on 200 turns of HTS per coil in the reference six-coil solution.}
\label{tab:hsx_table}
\end{table}

The three proposed configurations are mostly characterized by their reduction in curvature, which can be understood as a proxy for coil complexity. The reduction in mean squared curvature reaches 45\% in configuration \#1, and curvature is decreased by 35\% in configuration \#3. Despite configuration \#3 exhibiting the same number of coils per hfp, it manages to improve on the field accuracy while having simpler coils, the same clearance inbetween coils, and losing only 3 cm in the coil-to-surface distance. The forces are generally increased in all the optimized configurations, as they are not a target. It remains the sole aspect where the original HSX design is better. 

Configuration \#1 is an exemplary solution. Configuration \#1 has reduced by two the number of coils per half field period. This level of coil reduction has not been tried in previous sections, as it is often expected that field accuracy will not be precise enough with so few coils per hfp. Amazingly, and despite the fewer coils per hfp, configuration \#1 has reasonable field accuracy, and, compared with the original six-coil design, improved minimum coil-coil and coil-surface distances, lower curvatures, and achieves all of these improvements with 25\% less total coil length. Despite the factor of two increase in field error compared to the original design, the magnetic field contours plotted in \ref{fig:boozer_hsx} still display comparable accuracy to the other candidates. As previously mentioned, less coils are of particular interest in a research scale reactor where additional diagnostics are usually designed throughout the operation time and need space for new ports. The additional space in between coils on the outboard side is noticeable on Figure \ref{fig:hsx_coils}. Less coils also means less machining time and workforce required in the case of HSX as the coils are made out of copper and usually require the operation of CNC-milling or other heavy machinery devices. This can directly translate to reduced costs, which is often a limiting factor in university scale reactors and more generally in any type of reactor. 

\begin{figure*}[ht!]
    \centering
   \includegraphics[width=0.75\linewidth]{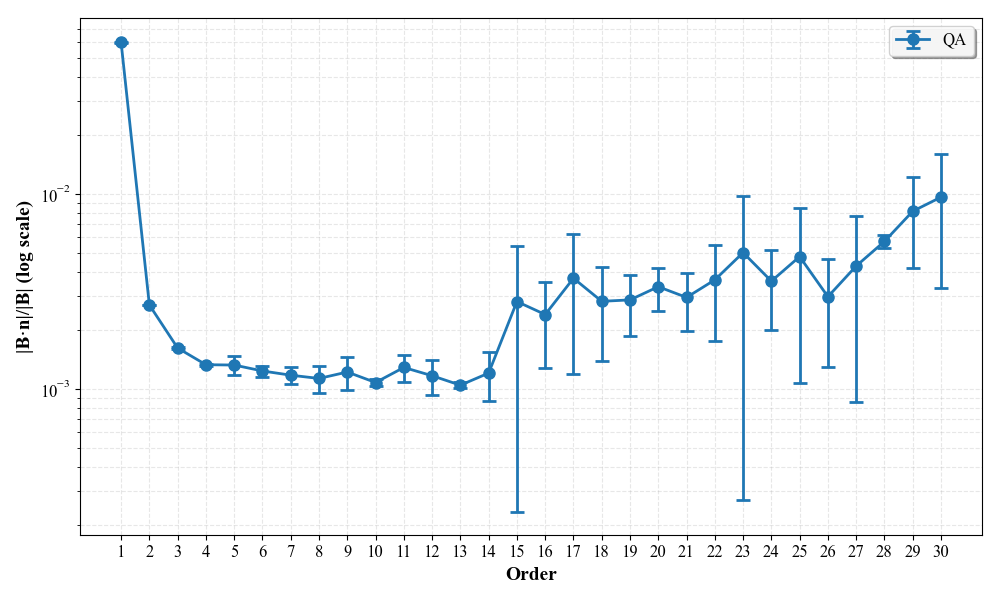}
    \caption{Log-scale error bar plot for normal field error vs.  order for the QA configuration after AugLag optimization.}
        \label{fig:OrderQA} 
\end{figure*}

\section{Conclusion}

In this work, we have introduced and validated a novel optimization methodology that fundamentally advances the state-of-the-art in stellarator coil design. The challenge of designing magnetic coils for a fusion reactor has long been characterized by difficult tradeoffs between achieving the precise magnetic field geometry required for plasma confinement and satisfying practical engineering constraints. Existing methods often lead to computationally intensive searches through vast parameter spaces, with no guarantee of arriving at a viable, let alone optimal, solution. Our approach confronts this challenge, establishing a robust and computationally efficient pathway for generating coil sets that are not only consistent with desired plasma equilibria but are also potentially manufacturable and meet the stringent requirements for reactor integration. Our methodology was successfully implemented and tested using the SIMSOPT framework. 

By applying an augmented Lagrangian approach to coil optimization, it was possible to address three critical issues: (1) the automated adjustment of weights for satisfying the constraints in a non-convex optimization problem, (2) the need for large computational resources and (3) optimally satisfying the tradeoff between low field errors and engineering constraints. In order to showcase the high performance of our method, five very different stellarator equilibria were considered, including quasi-axisymmetric, quasi-helically symmetric and quasi-isodynamic designs. New alternative coil sets were found for two  operating stellarators, Wendelstein 7-X and HSX. This demonstrates the versatility of this method in order to generate cutting-edge coils independently of the plasma boundary.
The quasi-axisymmetric design illustrated one of the main points of this work: the ability to perform single-optimization runs that achieve known Pareto-optimal solutions. By relaxing various constraints of the coil configurations, we were able to access new Pareto-boundaries. 
The automated weight tuning and reliance on upper and lower bounds allows one to specify loose parameter space bounds, increasing the space of possible solutions. 

All optimizations were performed on an Intel core i7-12650H laptop on 4 cores for Python multi-threading and on 1 core for the evaluations of the functions and gradients.
Future work includes accelerating the design of stellarator fusion reactors using modular coils, application to dipole coil arrays, incorporation into single-stage optimization efforts, and further algorithmic improvements.

\begin{acknowledgments}
This work was supported through grants from the Simons Foundation under award 560651 and from the Helmholtz Association Young Investigators Group program as project VH-NG-1430. 
We thank Jorrit Lion, Gabriel Plunk, Alan Goodman, Matt Landreman, and Benedikt Geiger for sharing plasma boundary configuration files, and Rory Conlin, Paul Huslage, David Bindel and Tom Simpson for useful discussions.
\end{acknowledgments}

\begin{figure*}[t]
    \centering
    \includegraphics[width=0.45\linewidth]{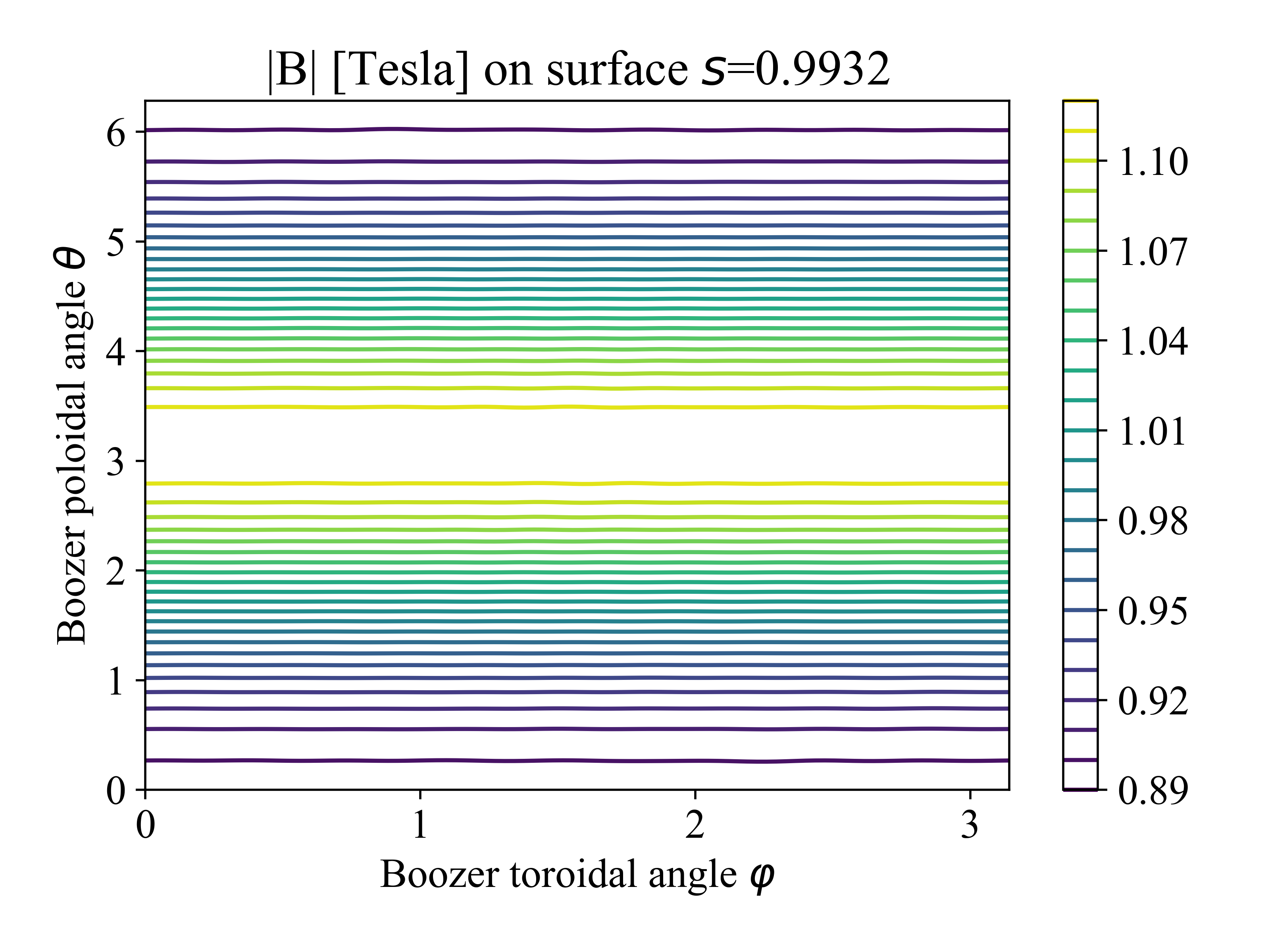}
    \includegraphics[width=0.45\linewidth]{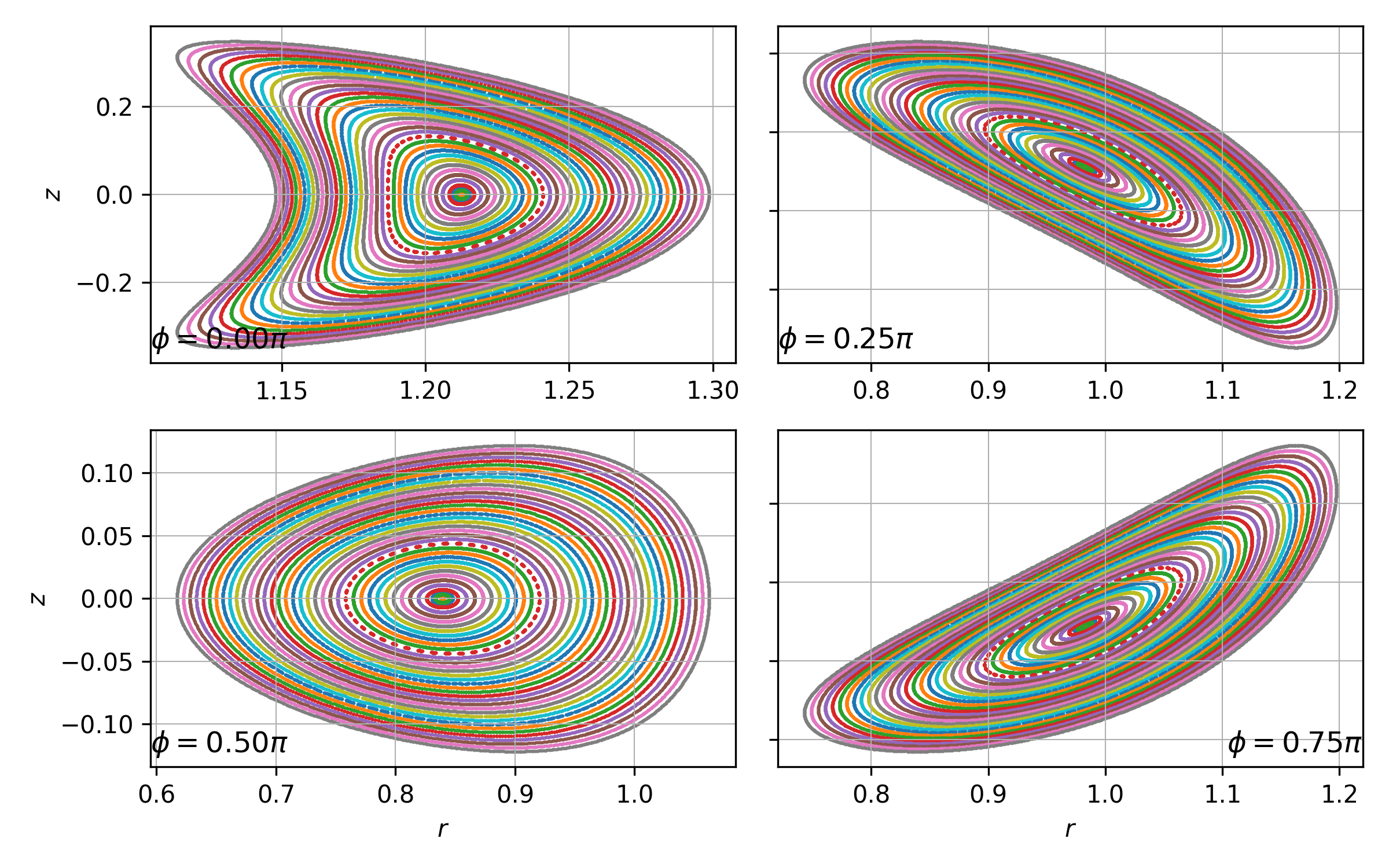}
    \caption{(Left) Boozer plot of one of the AUGLAG 25m coil legnth from Figure \ref{fig:QA_Pareto} configuration showing the quality of the quasi-symmetry of the last closed flux surface, which delimitates the plasma boundary. Perfectly straight lines mean that a very precise level of quasisymmetry is achieved. (Right) Poincaré plot showing a cross-section of the plasma where multiple nested flux surfaces are visible with no island chains. 
    }
    \label{fig:qa_boozer_poincare}
\end{figure*}
\appendix
\section{Order Investigation}
\label{sec:order_investigation}
This section provides a brief investigation into the specific effects of varying order $M$ on the quality of the optimized magnetic field via the Augmented Lagrangian method. For simplicity, we focus on the Landreman\&Paul QA reactor scale configuration. The quality of the coil solutions was quantified by $\langle\mathbf{B}\cdot \mathbf{n}\rangle/\langle\mathbf{B}\rangle$, as we found this was sufficient to draw conclusions. Orders 1 to 30 were investigated, and for each order, four optimization runs are performed with the Augmented Lagrangian method. The four optimizations were performed using four different initial conditions generated by adding Gaussian perturbations ($\sigma = 0.01$ and $L= 0.5$) to the initial set of coils. The results are summarized in Fig.~\ref{fig:OrderQA}. 

Surprisingly, $M = 2$ is already sufficient to get most of the benefits of higher order, although orders $8-14$ provide the most consistently low $\langle\mathbf{B}\cdot \mathbf{n}\rangle/\langle\mathbf{B}\rangle$ errors without adding overwhelming complexity. As $M \ge 15$ or so, the normalized field error begins to worsen (on average) and increase in variation from run-to-run, despite the extra degrees of freedom in the problem. Since the variation between runs is high, the problem may be increasingly nonconvex as $M$ increases. This result is consistent with the traditional explanation that high Fourier modes are complicating the optimization landscape by introducing many suboptimal local minima~\cite{jang2025exponential}. 

Lastly, we found a similar trend using traditional coil optimization techniques and with different stellarators (although the optimal $M$ choice can depend on the stellarator). In fact, with a traditional method it was easy to generate final solutions with infeasible coil sets, where the coils were interlinking or even intersecting with the plasma surface, if the weights were not carefully tuned. 

Initially, we found similar issues with the augmented Lagrangian approach. A critical consideration was that if the initial weight on the coil-surface distance was too small, the first sub-problem solved by augmented Lagrangian could generate a solution with coil-surface intersection, after which all subsequent optimizations struggled to recover from. This first sub-problem issue can be  permanently solved by starting the optimization with an aggressive \textit{initial} weight on the coil-surface distance penalty, after which the weight is tuned by the algorithm as before. We then manually checked in Fig.~\ref{fig:OrderQA} that all of the resulting coil optimizations were feasible. In other words, we were able to completely remove infeasible coil solutions from the scan, and yet the trend with $M$ remains.
In summary, this experiment generally suggests that orders in the $6-15$ range provide more consistent results for minimizing the normalized field error. Moreover, it suggests that if high order coils are desired, a Fourier continuation approach is likely to perform better than direct optimization starting with a large value of $M$. As far as we are aware, this is standard practice in stage-1 but not stage-2 optimizations, so continuation should be carefully considered by future researchers.

\begin{figure*}
    \centering
    \includegraphics[width=0.48\linewidth]{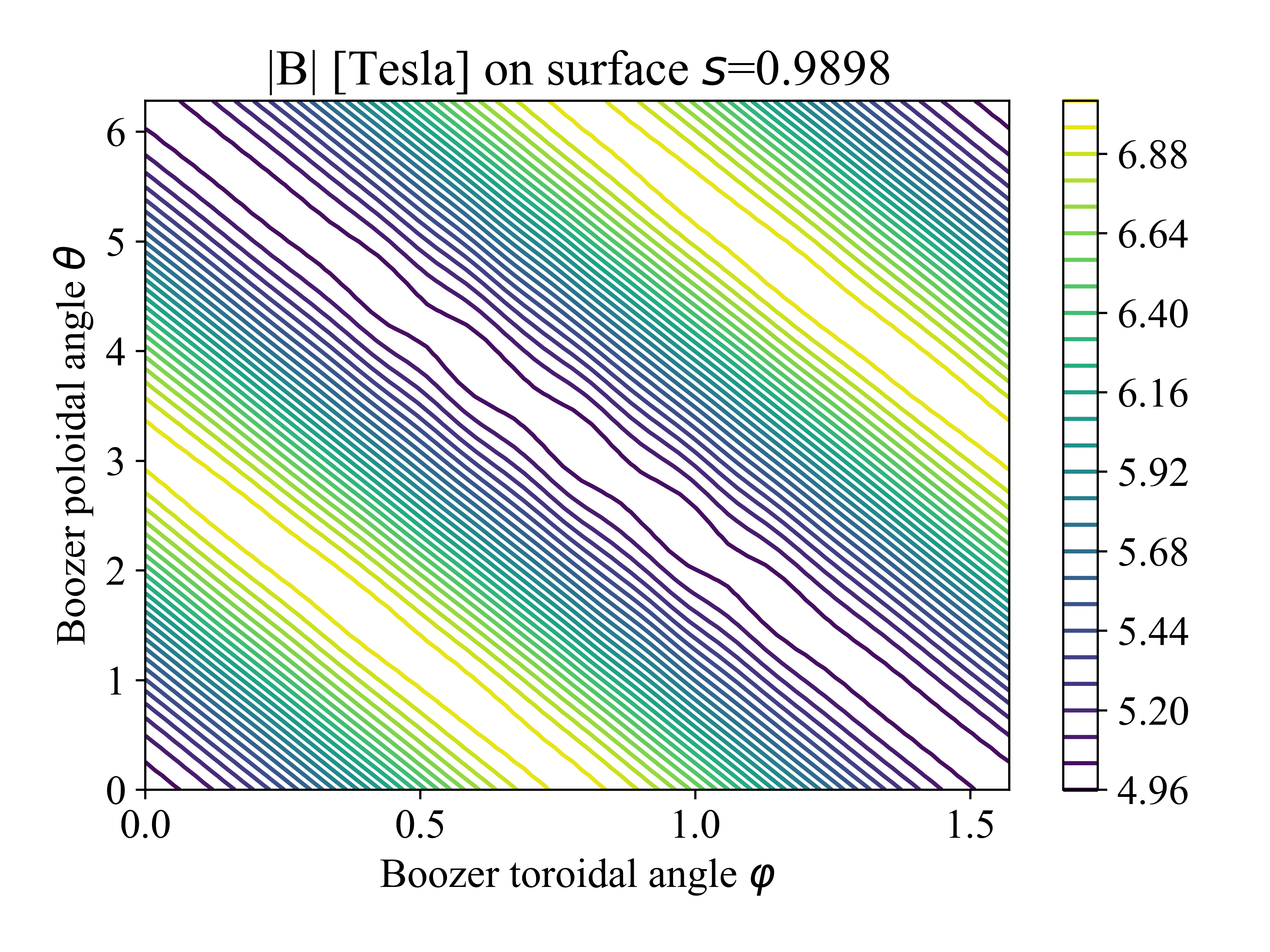}
    \includegraphics[width=0.48\linewidth]{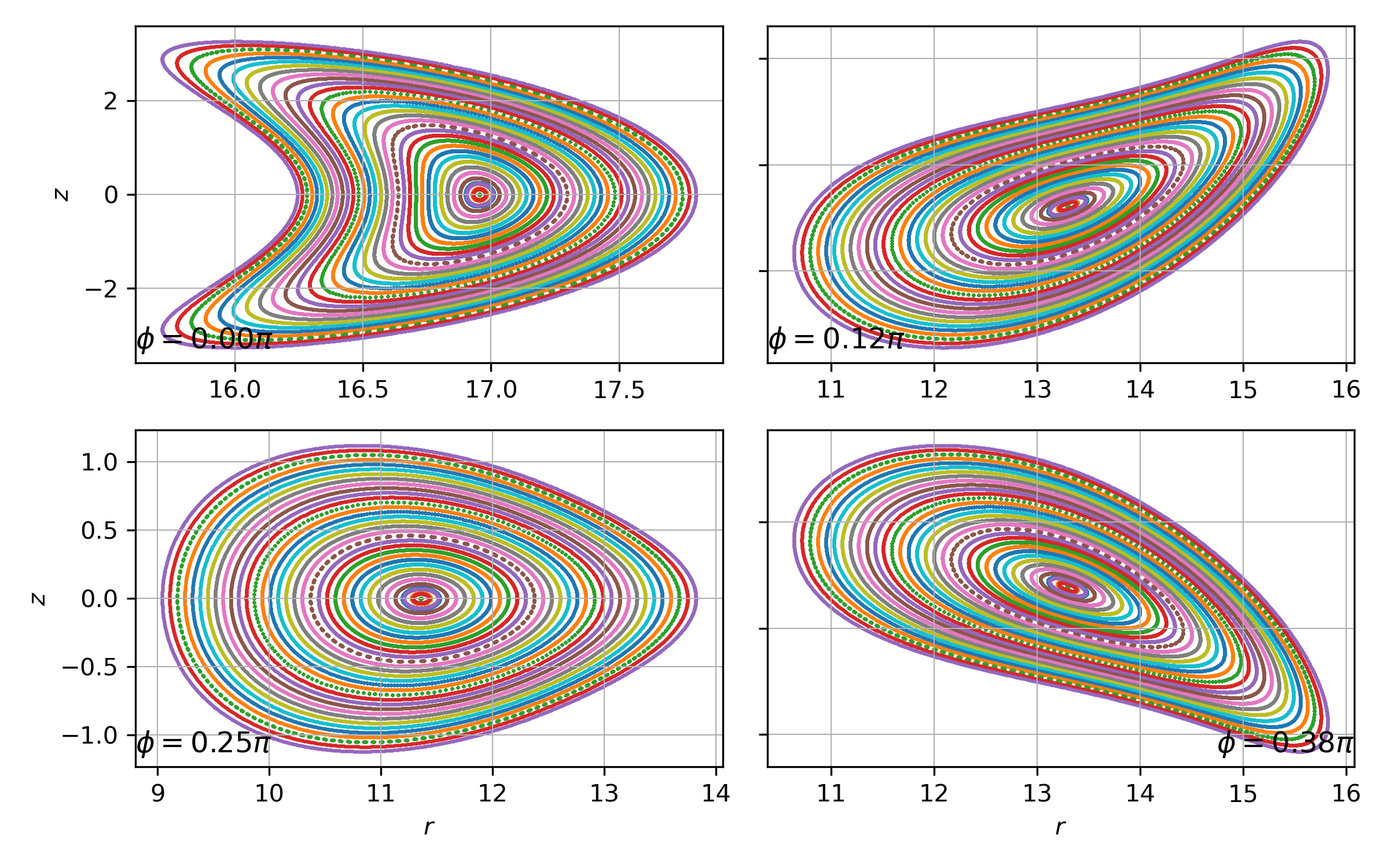}

    \includegraphics[width=0.48\linewidth]{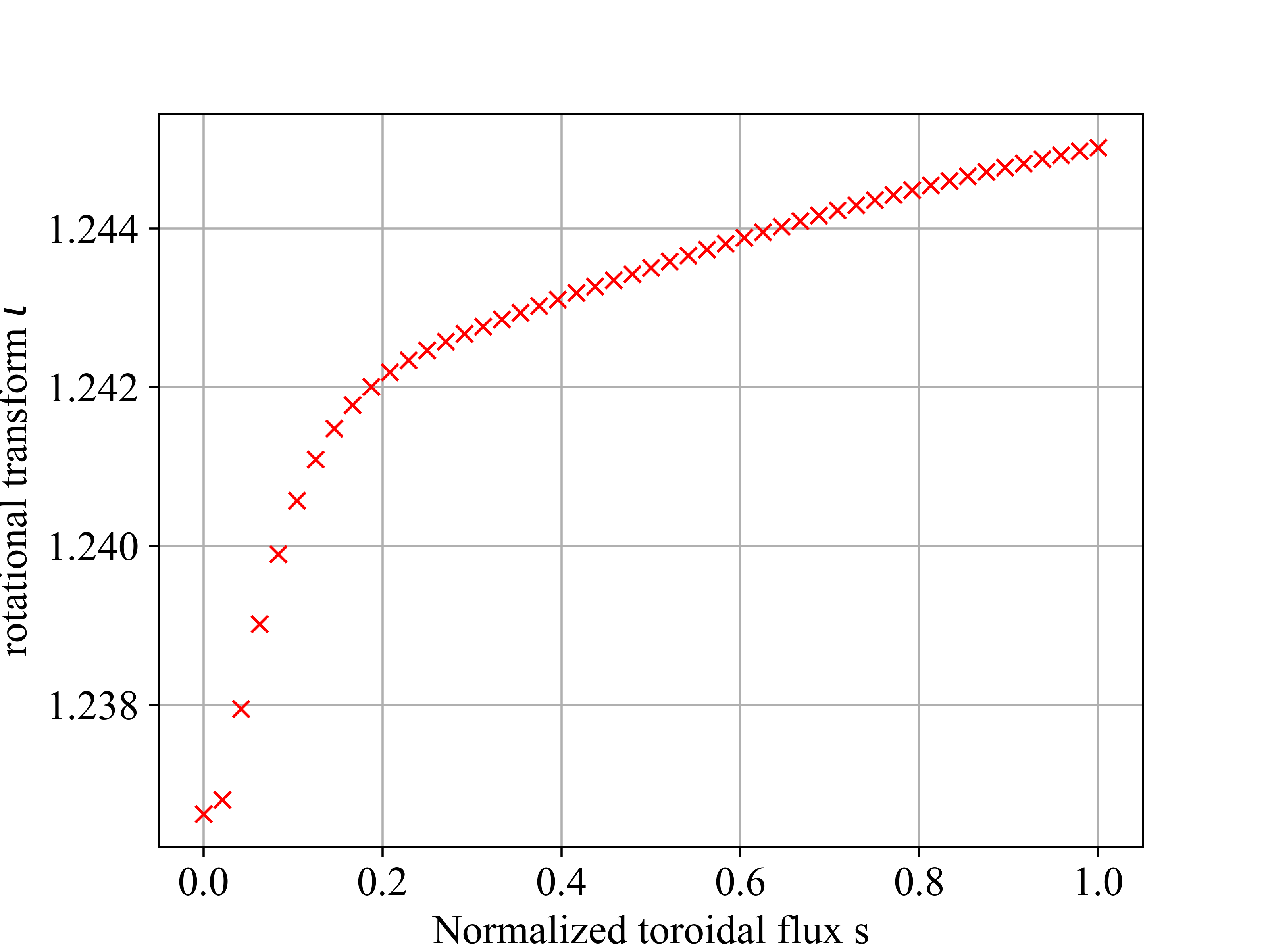}
    \includegraphics[width=0.48\linewidth]{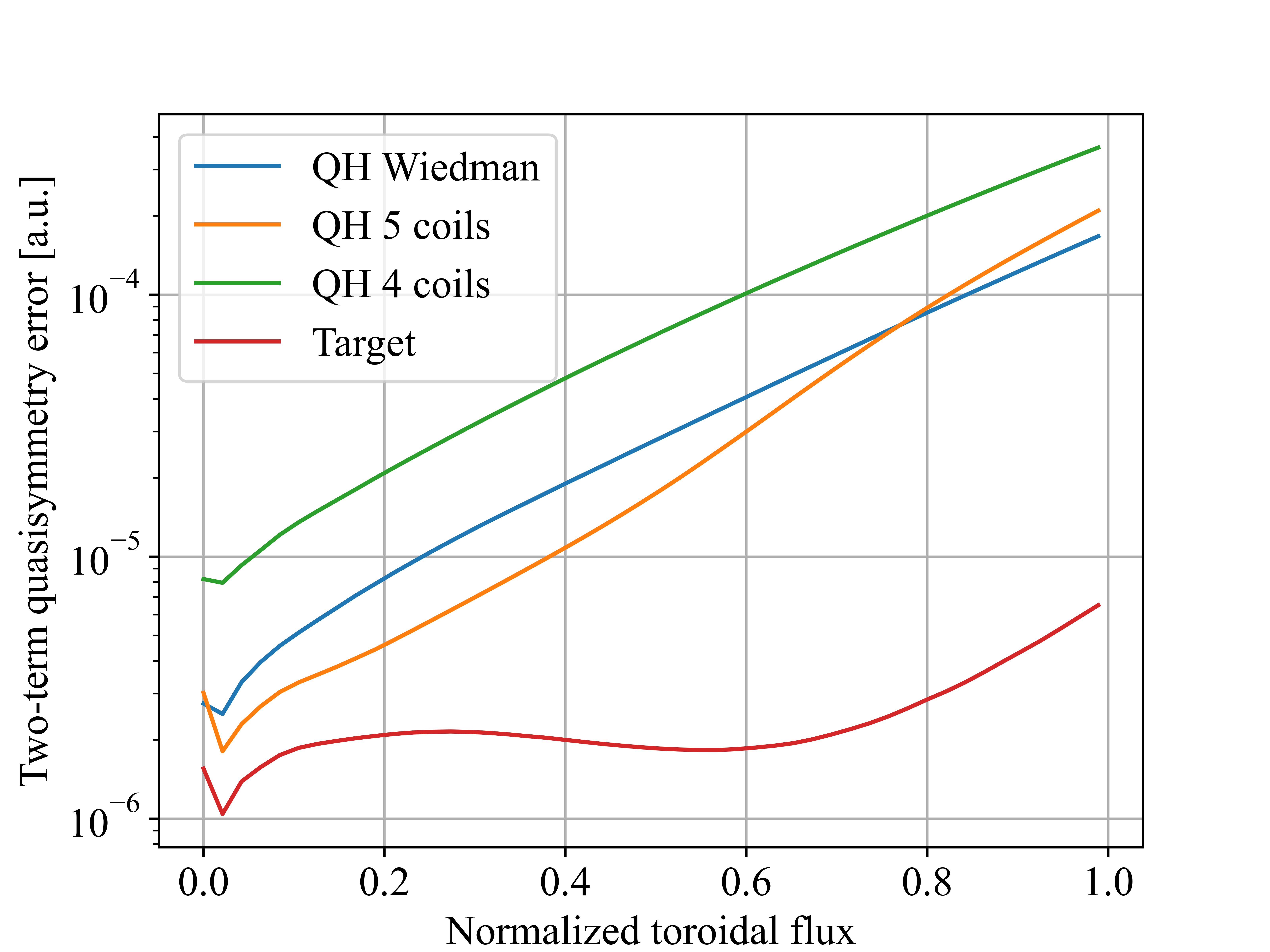}
    \caption{(Top Left) Boozer plot for configuration \#1 showing the quality of the quasi-helical symmetry of the last closed flux surface. Perfectly straight lines mean that a very precise level of quasisymmetry is achieved. (Top Right) Poincaré plot of configuration \#1 showing a cross-section of the plasma where multiple nested flux surfaces are visible with no island chains and rotational transform profile of the configuration which matches the ideal profile.(Bottom Left) Rotational transform profile of configuration \#1. (Bottom Right) QS profiles for the configurations \#1, \#2, Wiedman, and the original QH target plasma.}
    \label{fig:qh_boozer_poincare}
\end{figure*}

\section{Quasisymmetry Metric}
\label{sec:qs_metric}
In this work, the metric used to quantify the departure from quasisymmetry is the two-term metric \cite{LandremanPaul, Rodríguez_Paul_Bhattacharjee_2022}, which does not require a transformation into Boozer coordinates and can be expressed as
\medmuskip=2mu
\begin{equation}
\begin{split}
    f_{QS} = \sum_{s_i} w_i \biggr\langle \biggl( \frac{1}{B^3} \Bigl[ &(N-\iota M) \bm B \times \bm \nabla B \cdot \bm \nabla \psi \\
    &- (MG+NI) \bm B \cdot \bm \nabla B \Bigr] \biggr)^2 \biggr\rangle.
\end{split}
\end{equation}
\medmuskip=4mu
Here it appears as a weighted sum across all specified flux surfaces $s_i$, with the weights $w_i$ by default set to 1, $B$ is the magnetic field amplitude, $(N,M)$ are parameters specifying the class of quasisymmetry (either QH or QA), $\psi$ is the toroidal flux, $\iota$ is the rotational transform, and $G$ is the poloidal and $I$ the toroidal currents inside the surface.

\section{Equilibria verification}
\label{sec:extra_plots}
\begin{figure*}[t]
    \centering
    \includegraphics[width=0.47\linewidth]{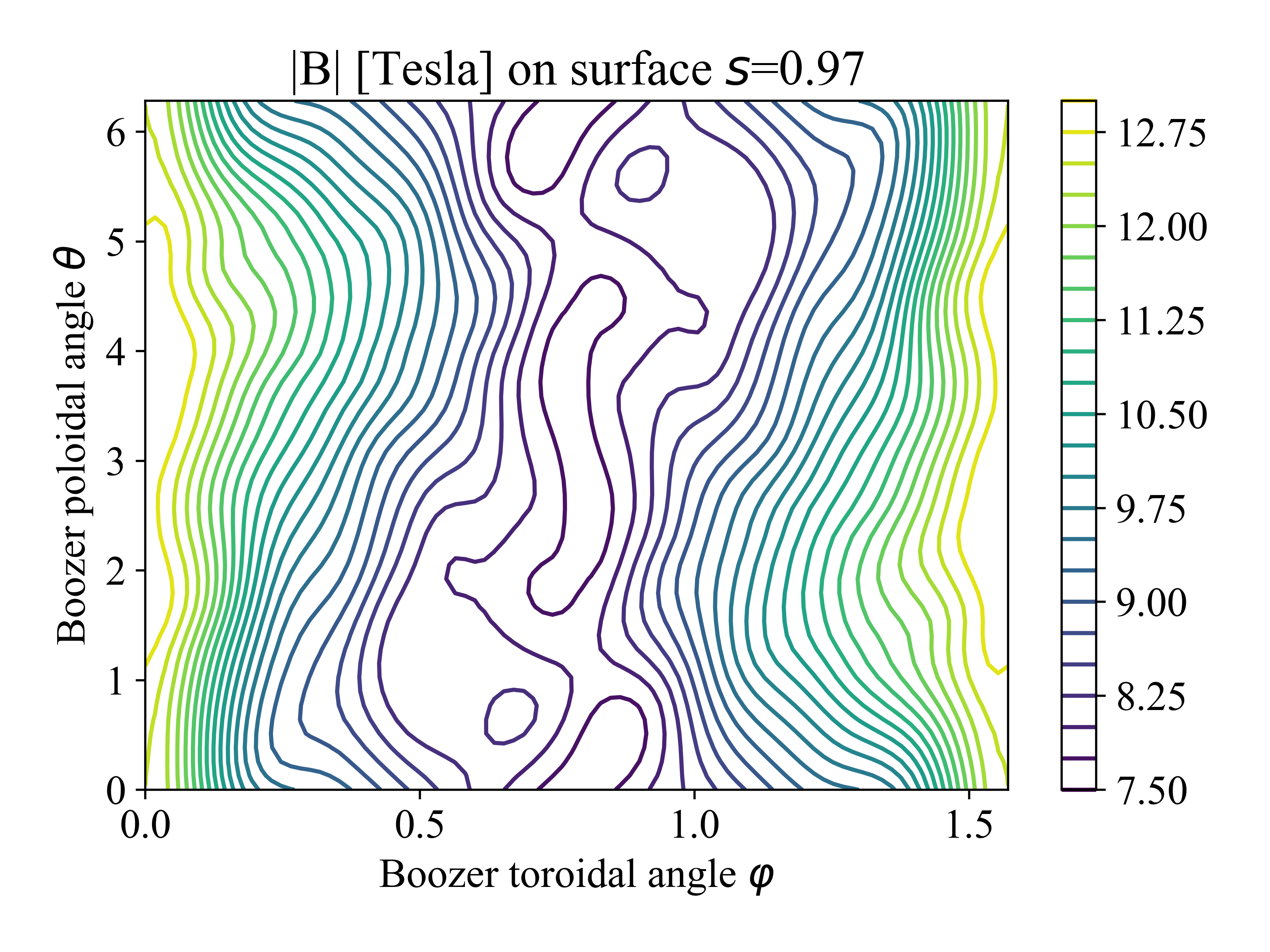}
    \includegraphics[width=0.47\linewidth]{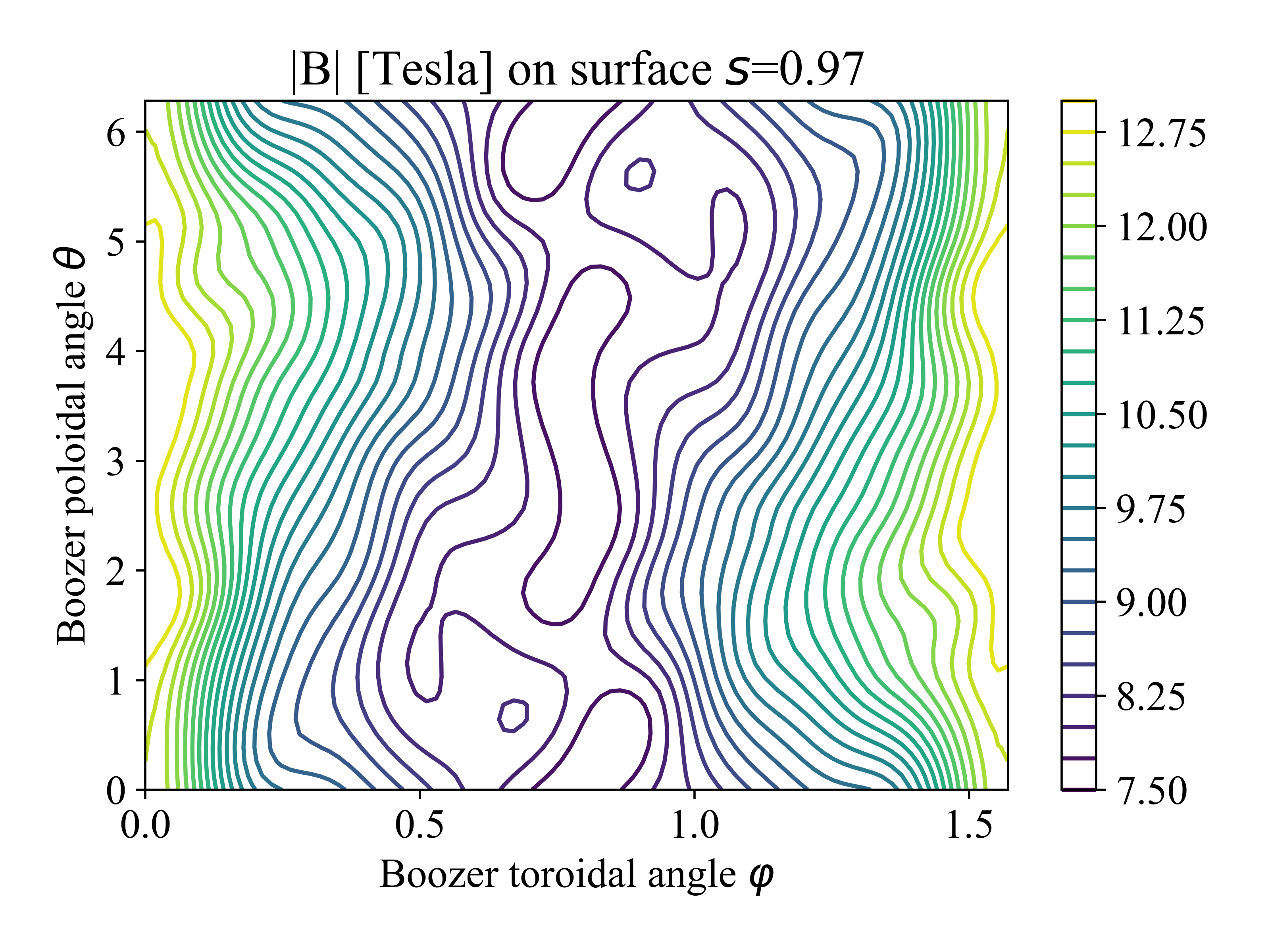}

    \includegraphics[width=0.47\linewidth]{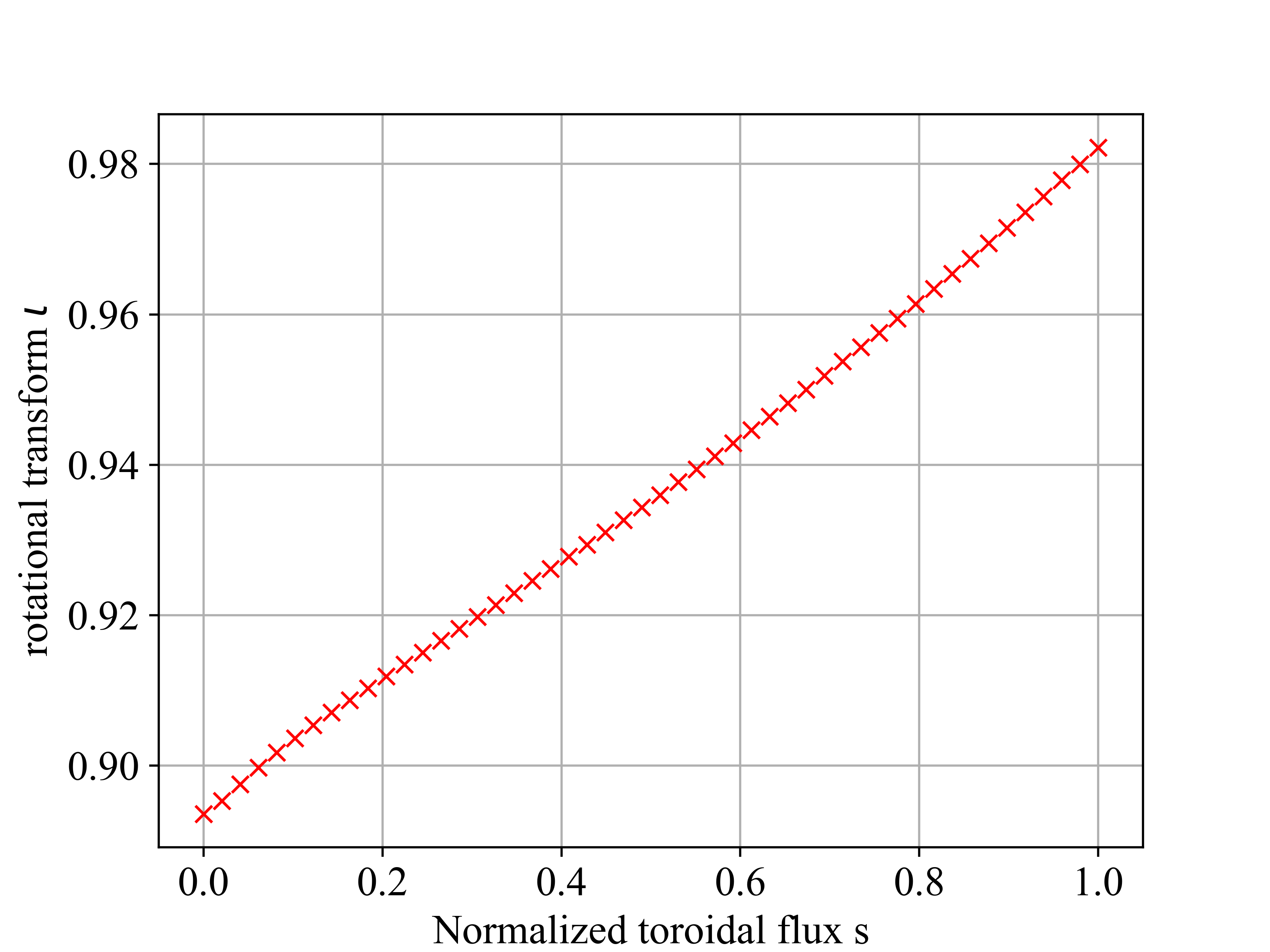}
    \includegraphics[width=0.47\linewidth]{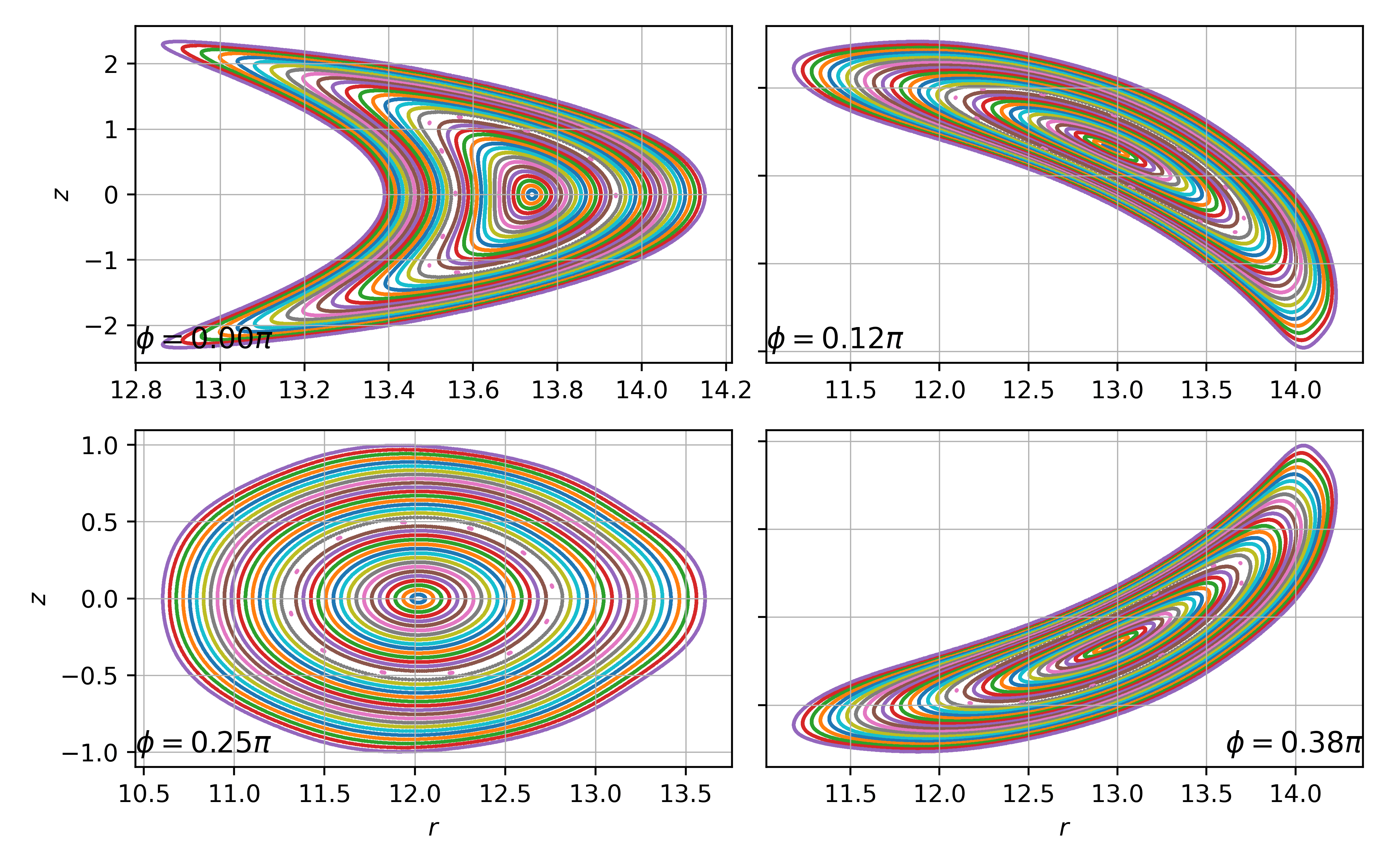}
    \caption{(Top Left) Boozer plot of the \#2 Stellaris configuration showing the quality of the quasi-isodynamic field generated by the coils (Top Right) Boozer plot of the target isodynamic-field of Stellaris. The coils appear to accurately capture the desired features of the boundary magnetic surface.(Bottom Left) Rotational transform profile of the \#2 Stellaris configuration. (Bottom Right) Poincaré plots of the \#2 configuration.}
    \label{fig:boozer_stellaris}
\end{figure*}

\begin{figure*}[t]
    \centering
    \includegraphics[width=0.47\linewidth]{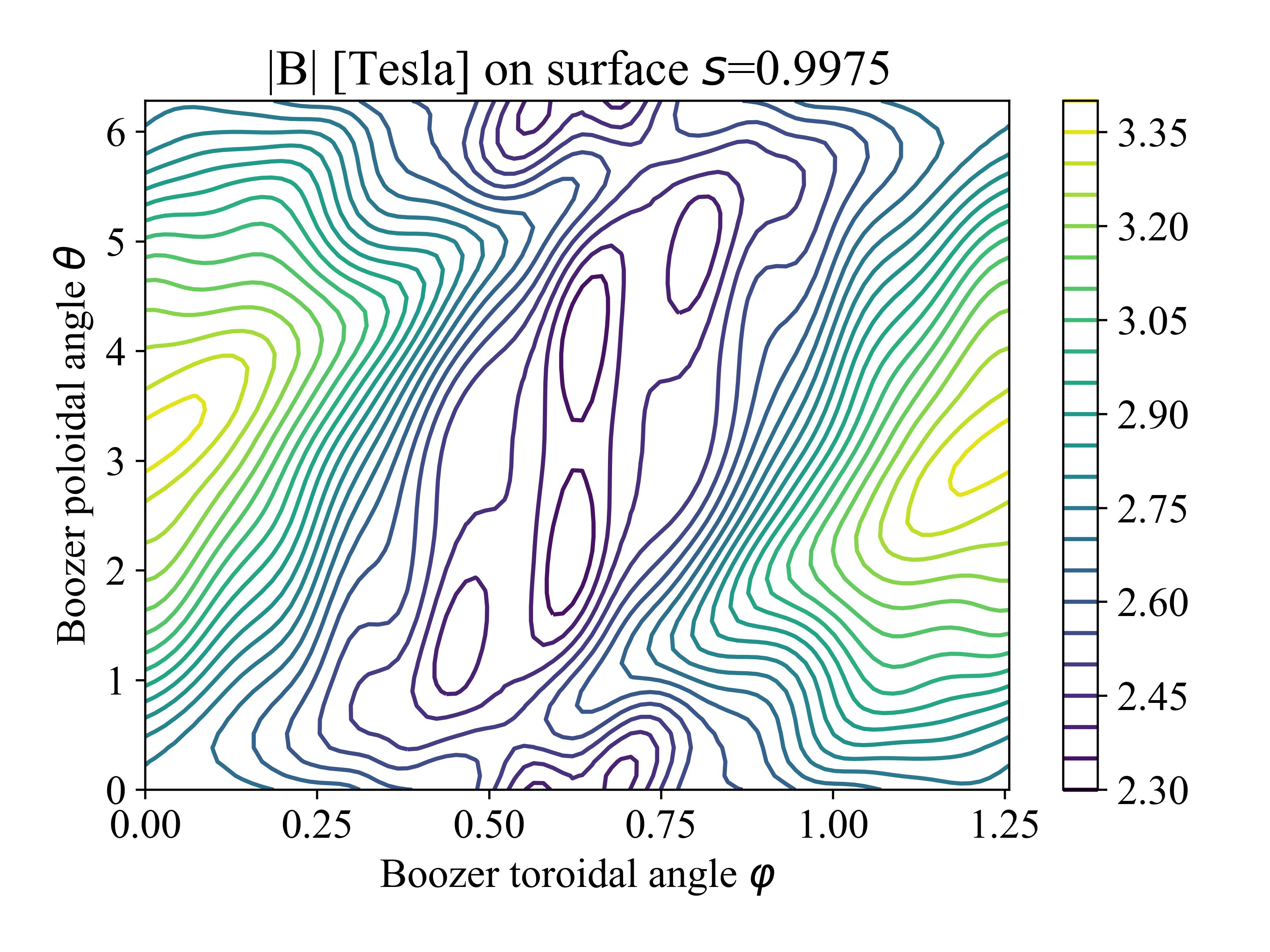}
    \includegraphics[width=0.47\linewidth]{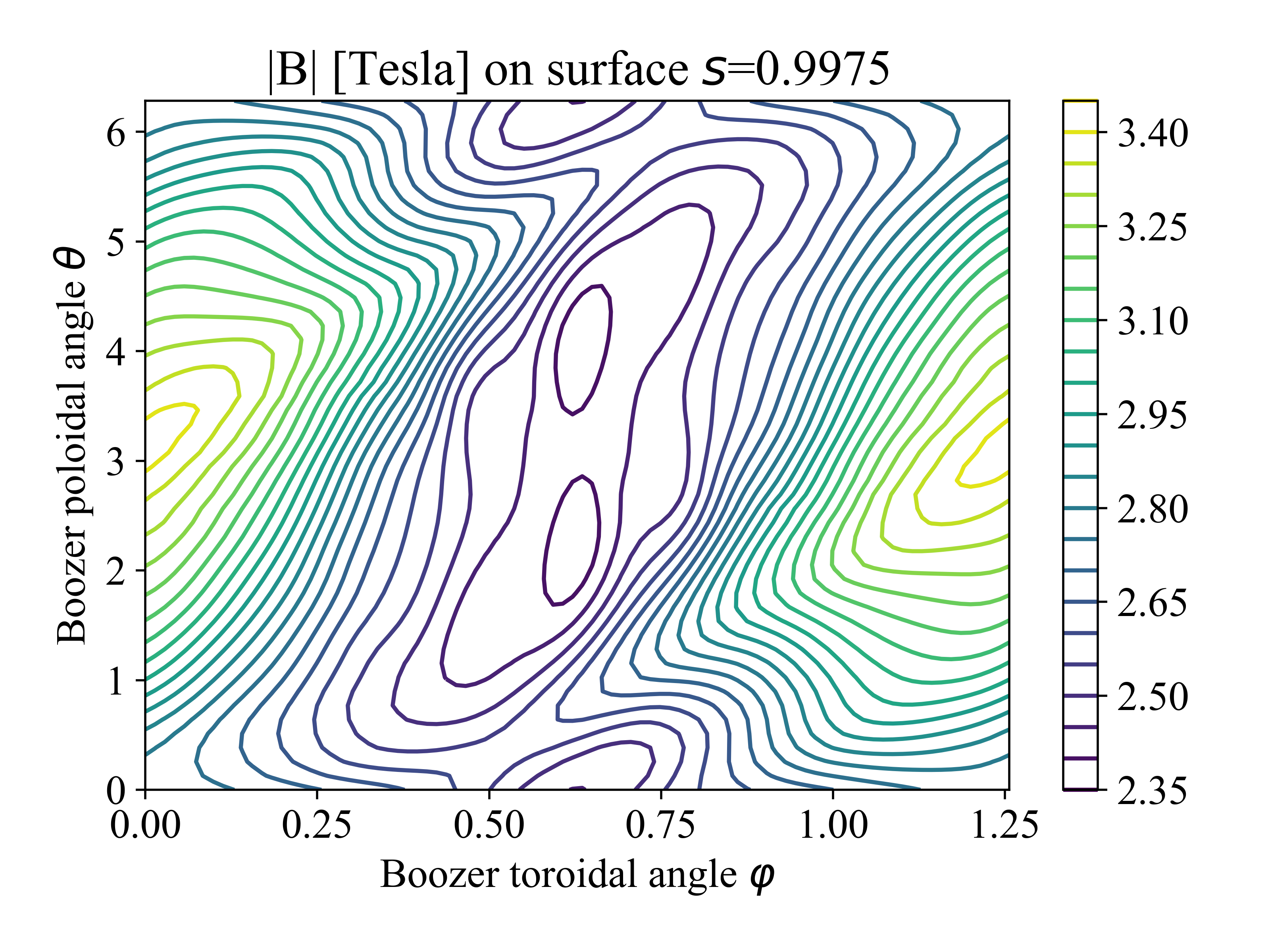}
    
    \includegraphics[width=0.47\linewidth]{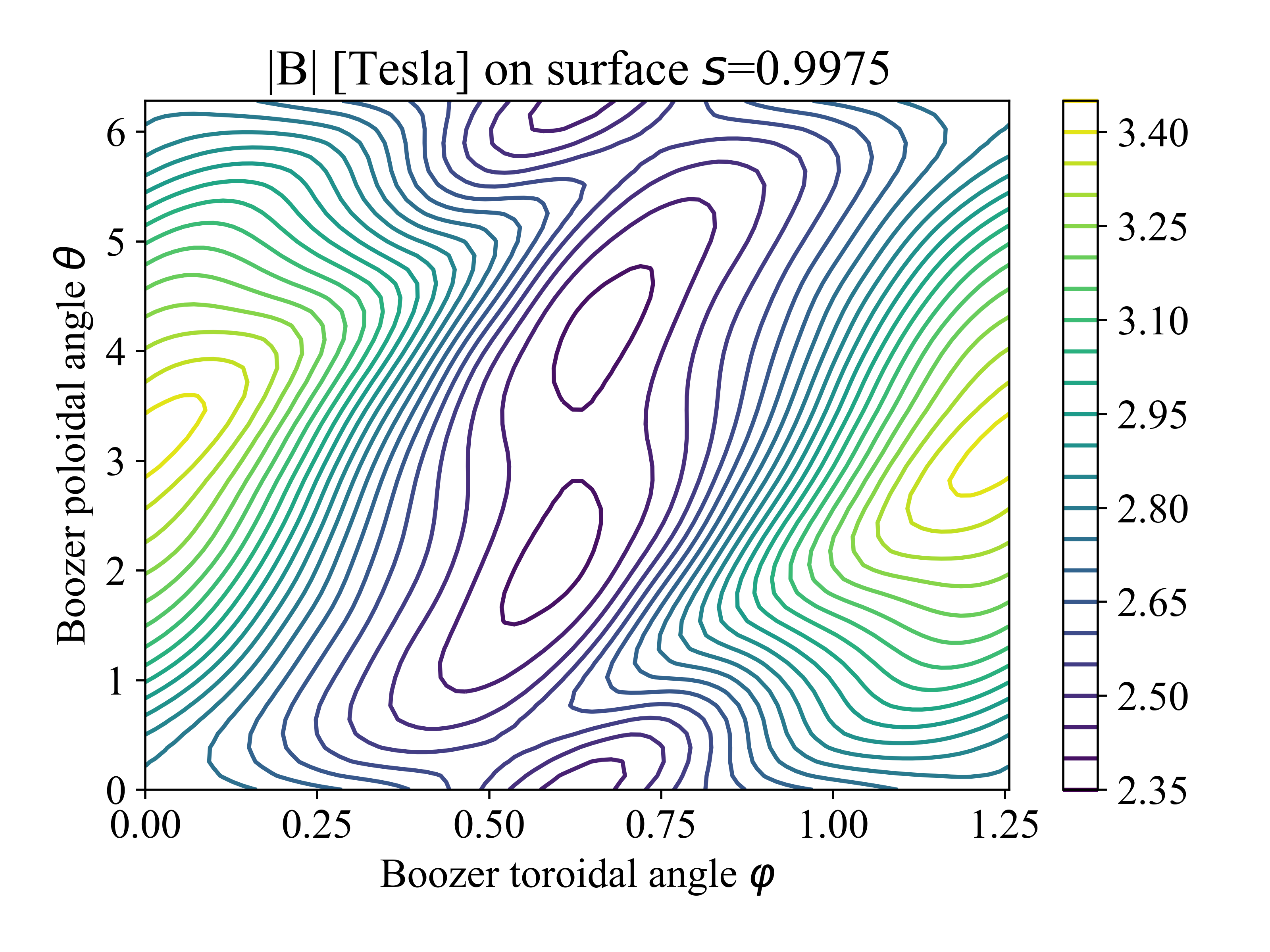}
    \includegraphics[width=0.47\linewidth]{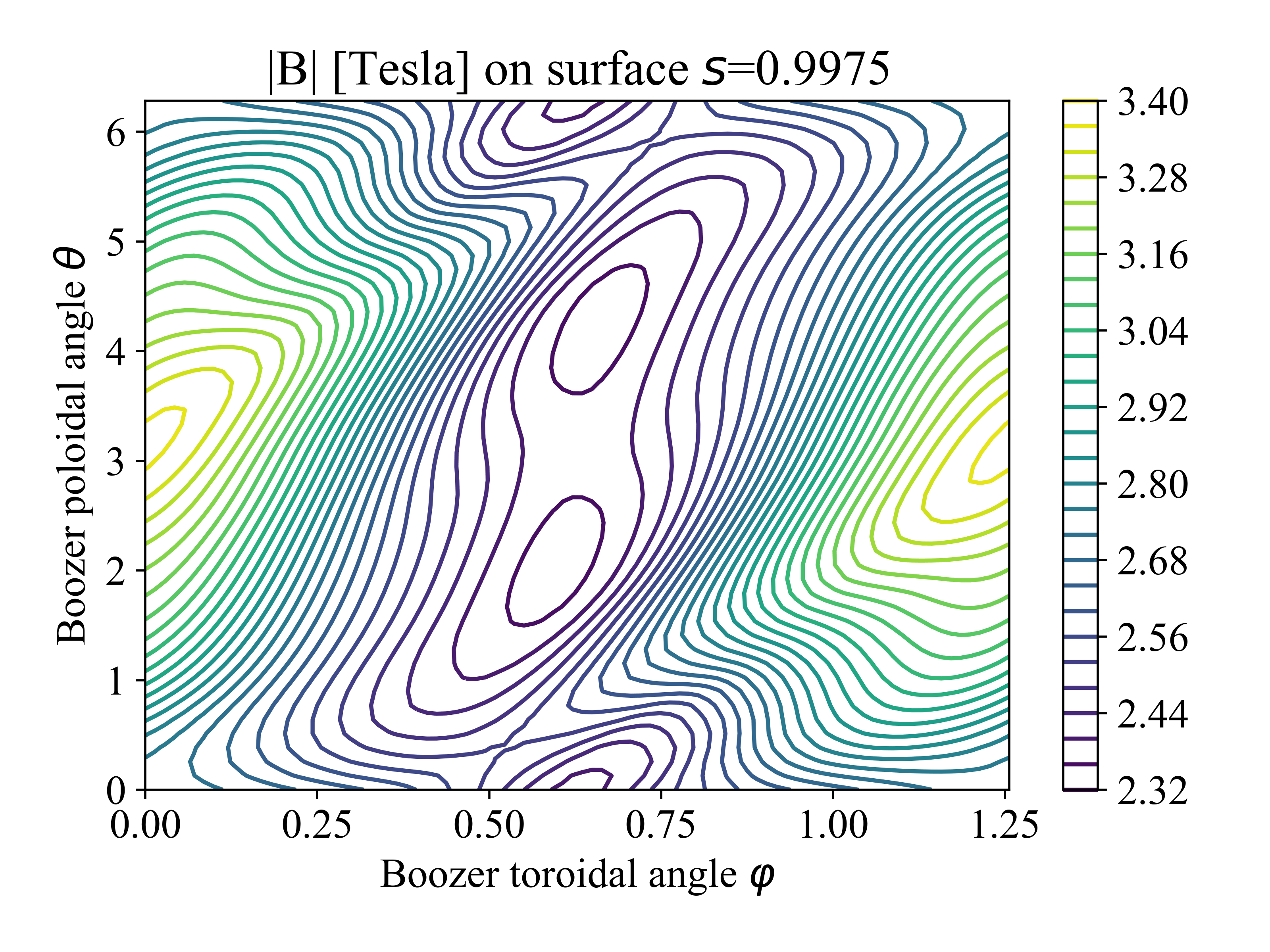}
    \caption{Boozer plots of the four W7-X configurations presented in this work showing the quality of the omnigenous field generated by the coils: four coils per half field period (\#1 Top Left), five coils per hfp with 45 m coil length and reduced forces (\#2 Top Right), five coils per hfp with 43 m coil length and best field accuracy (\#3 Bottom Left) and the target W7-X field in fixed-boundary (Bottom Right).}
    \label{fig:boozer_w7x}
\end{figure*}

\begin{figure*}[t]
    \centering
    \includegraphics[width=0.47\linewidth]{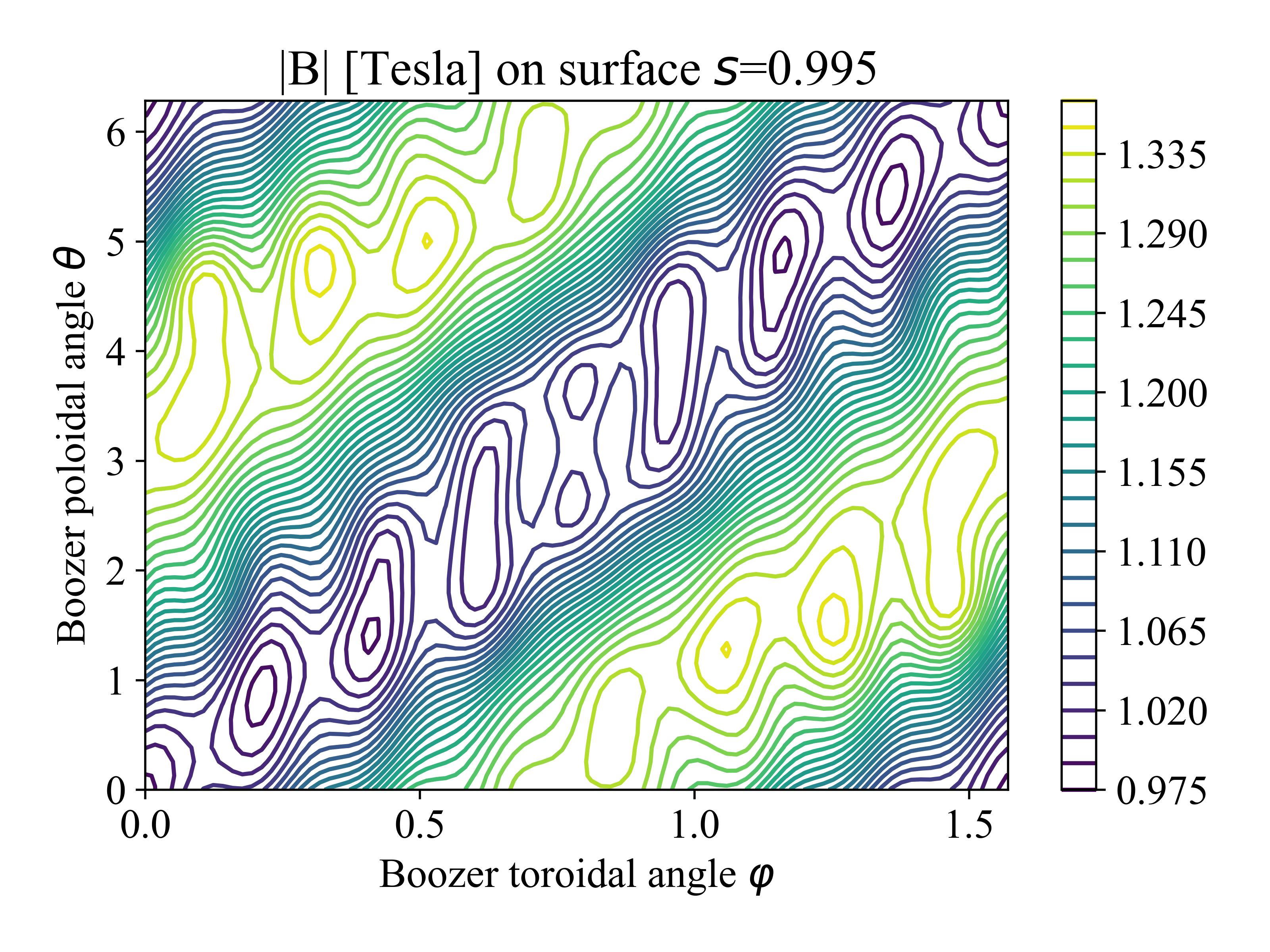}
    \includegraphics[width=0.47\linewidth]{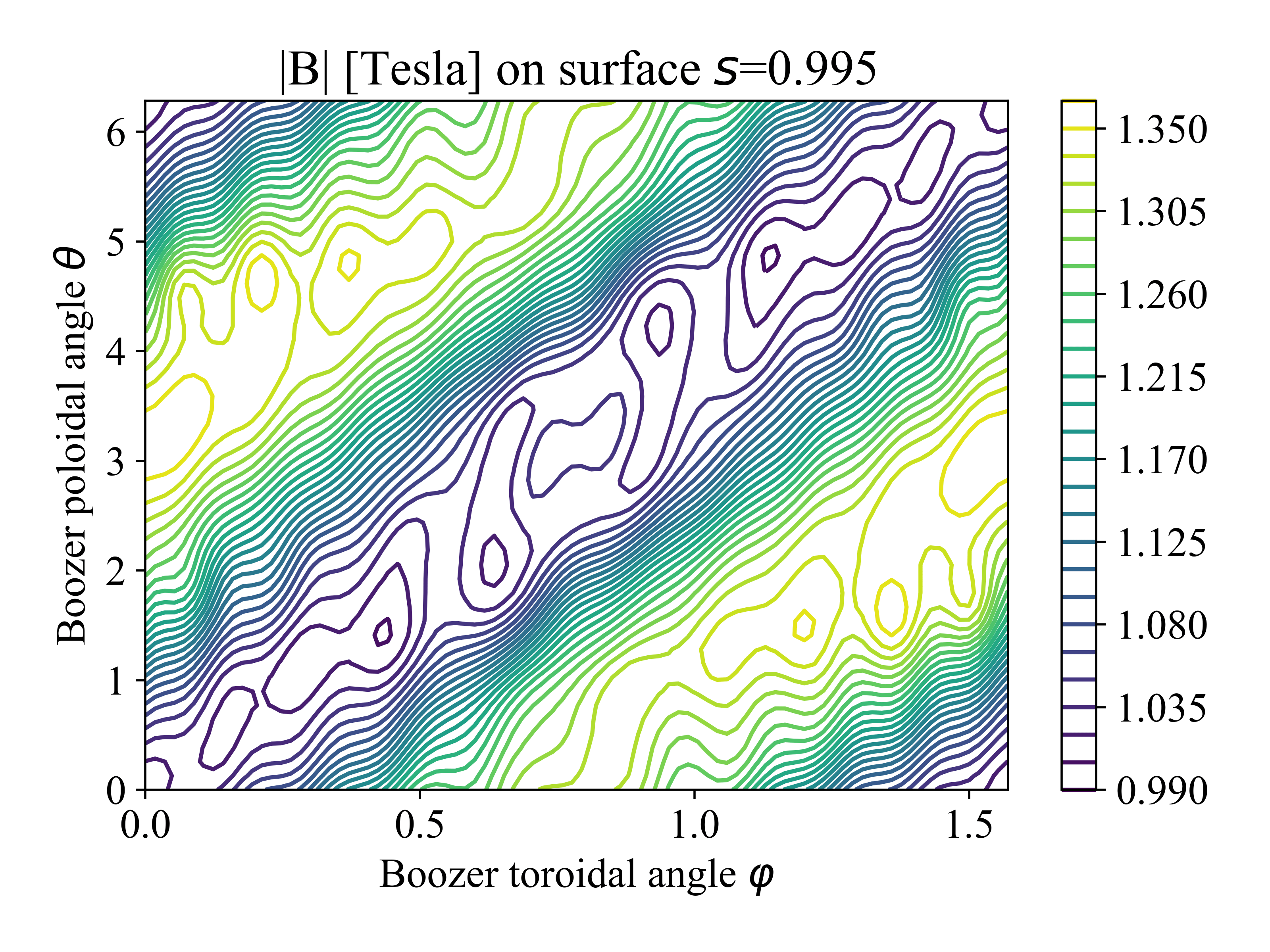}
    
    \includegraphics[width=0.47\linewidth]{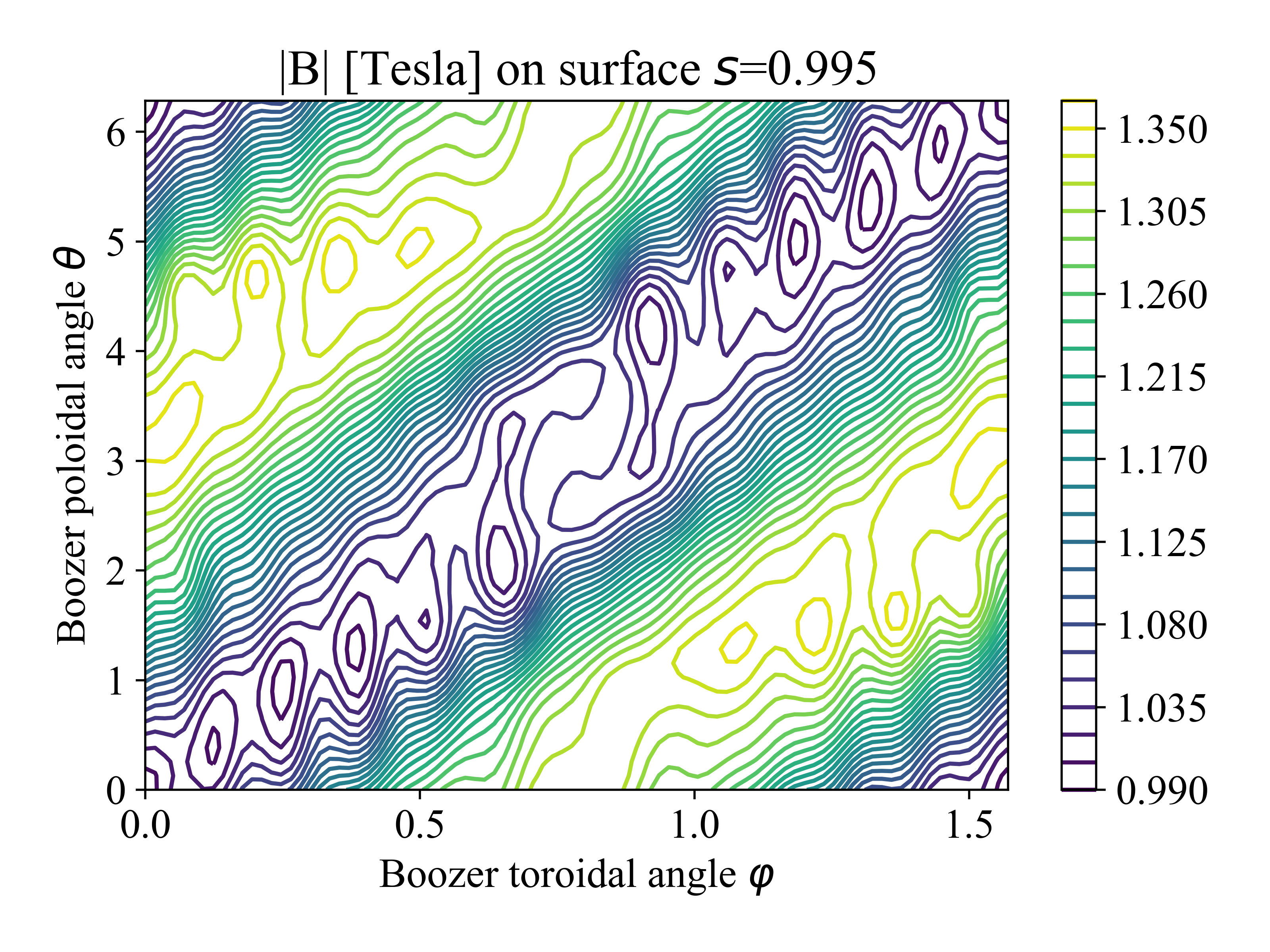}
    \includegraphics[width=0.47\linewidth]{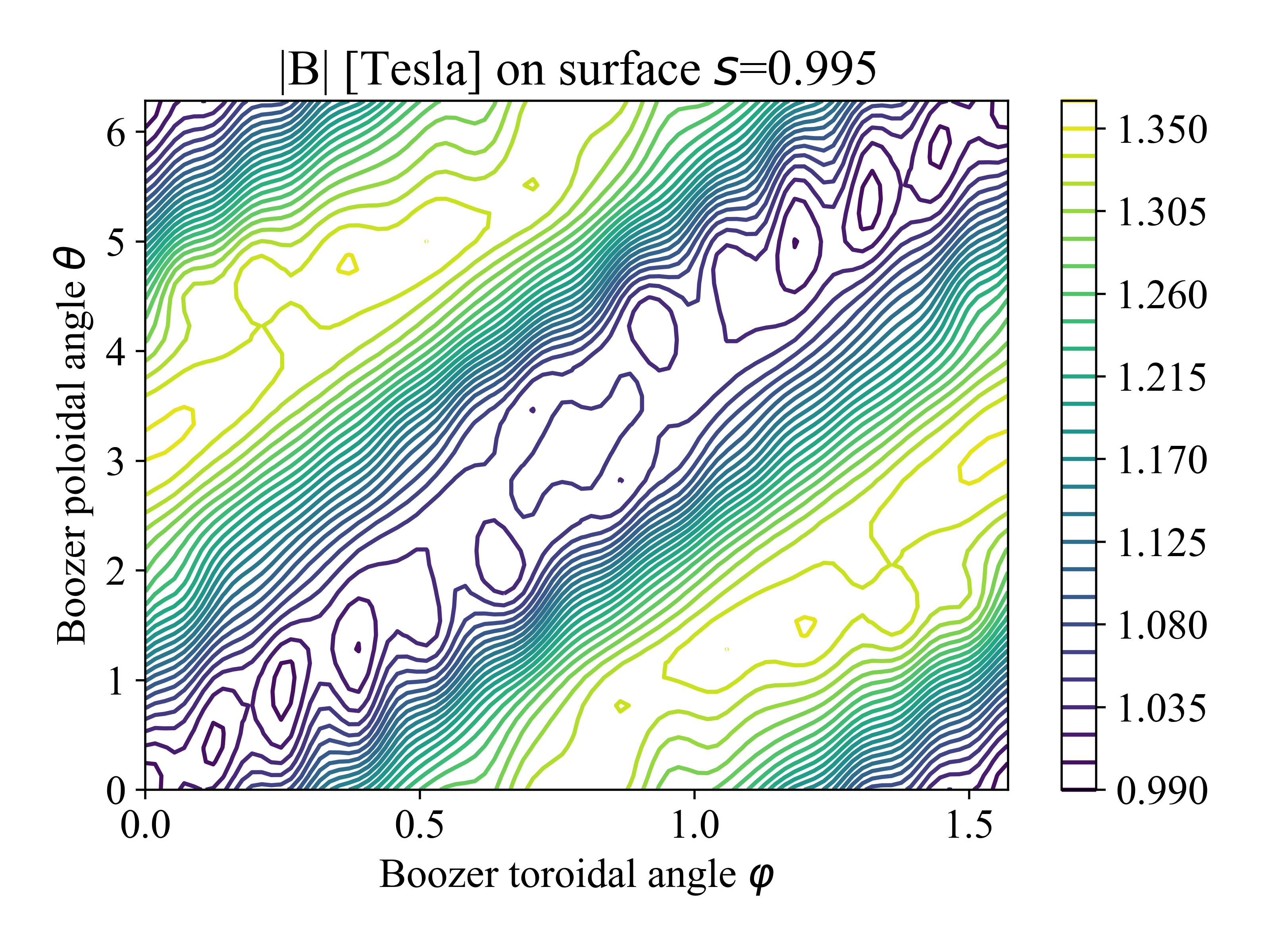}
    \caption{Boozer plots of the four HSX configurations presented in this work showing the quality of the quasi-helically symmetric field generated by the coils: four coils per half field period (\#1 Top Left), five coils per hfp (\#2 Top Right), six coils per hfp (\#3 Bottom Left) and the original HSX coil set (Bottom Right).}
    \label{fig:boozer_hsx}
\end{figure*}

This section provides post-processing illustrations demonstrating that the physics properties of the plasma are well-conserved for the coil solutions identified in this work. Fig.~\ref{fig:qa_boozer_poincare} illustrates Boozer and Poincaré plots for a Landreman\&Paul QA coil solution, and similarly for Landreman\&Paul QH in Fig.~\ref{fig:qh_boozer_poincare}, Stellaris in Fig.~\ref{fig:boozer_stellaris}, W7-X in Fig.~\ref{fig:boozer_w7x}, and HSX in Fig.~\ref{fig:boozer_hsx}.

\section{Fast Particle Simulations}
\label{sec:fp}
\begin{figure}[h!]
    \centering
    \includegraphics[width=\linewidth]{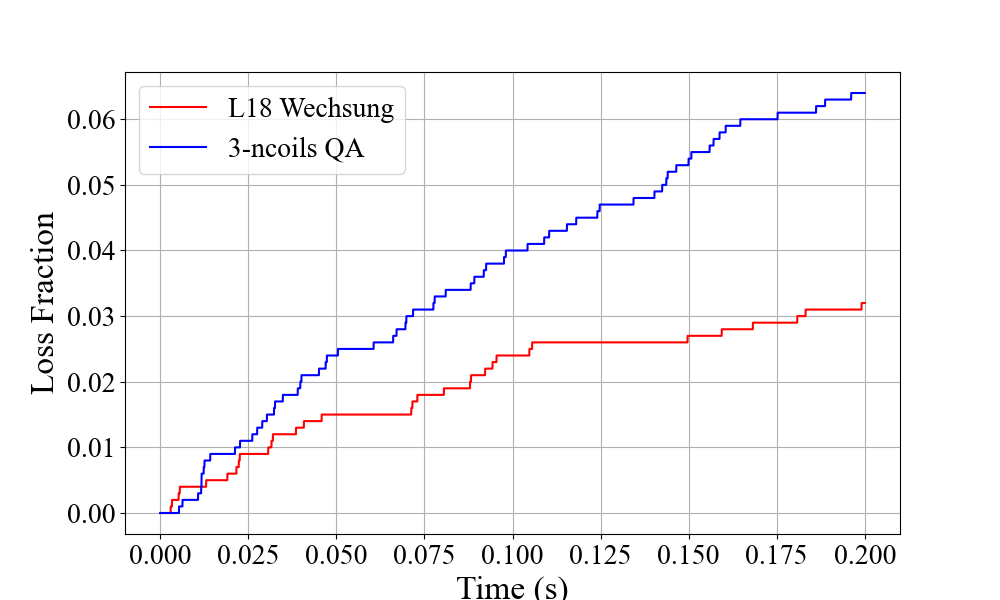}
    \caption{Loss Fractions for the Landreman\&Paul QA 3 coils per hfp solution versus the Wechsung L18 with 4 coils per hfp after 0.2s.}
    \label{fig:qa_simple}
\end{figure}

\begin{figure}[h!]
    \centering
    \includegraphics[width=\linewidth]{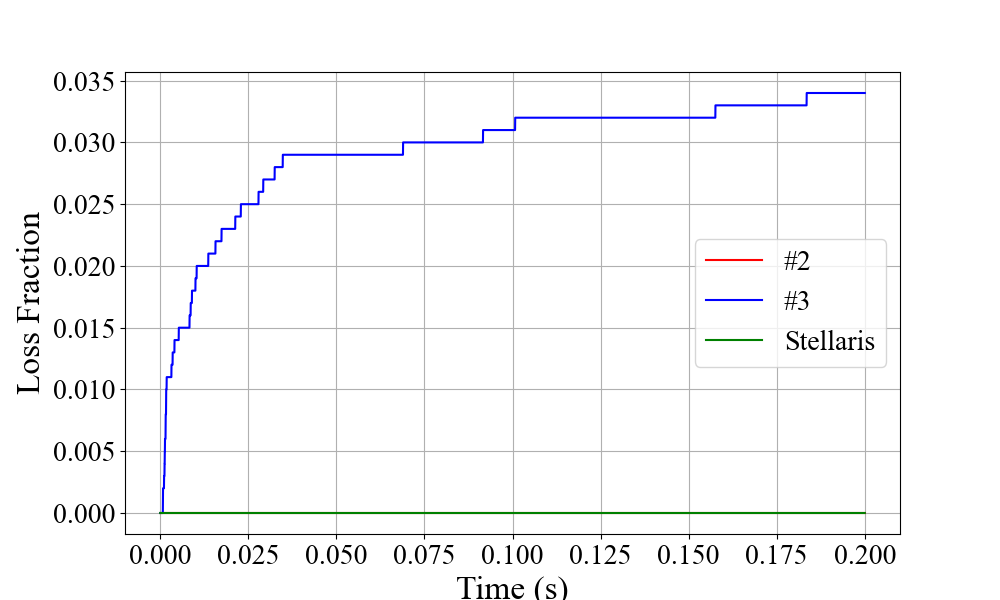}
    \caption{Loss Fractions for configurations \#2, \#3 and the original Stellaris coilset after 0.2s. Note that both the red and the green line have losses of 0\%.}
    \label{fig:qi_simple}
\end{figure}

Additional checks are performed regarding the confinement of $\alpha$-particles. This is performed for the Landreman\&Paul QA 3-coils per hfp solution versus the Wechsung L18 with 4 coils per hfp configuration in Fig.~\ref{fig:qa_simple}. Moreover, our Stellaris configurations \#2 and \#3 are also compared with the original Stellaris coilset in Fig.~\ref{fig:qi_simple}. The results here displayed correspond to the results of the companion paper \cite{auglag_prl}. The simulations are performed with SIMPLE \cite{ALBERT2020109065, Albert_Kasilov_Kernbichler_2020}, a simplectic integrator to calculate fast ion orbits and losses. The simulations are performed by initializing 1000 particles with random pitch angles at 3.5MeV at the radial coordinate of 0.2 and are followed for 0.2s.

While configurations \#2 and Stellaris display no losses after 0.2 seconds, configuration \#3 performs worse with 3.5\% losses. This is expected as this configuration is intentionally optimized to present simpler coils and therefore a worse field accuracy. Nonetheless, the losses remain very low and comparable to other top candidates for reactor designs self-consistently modeled with a coil set. 

\section{Dipole Optimization}
\label{sec:dipole_opt}
\begin{figure}[h!]
    \centering
    \includegraphics[width=\linewidth]{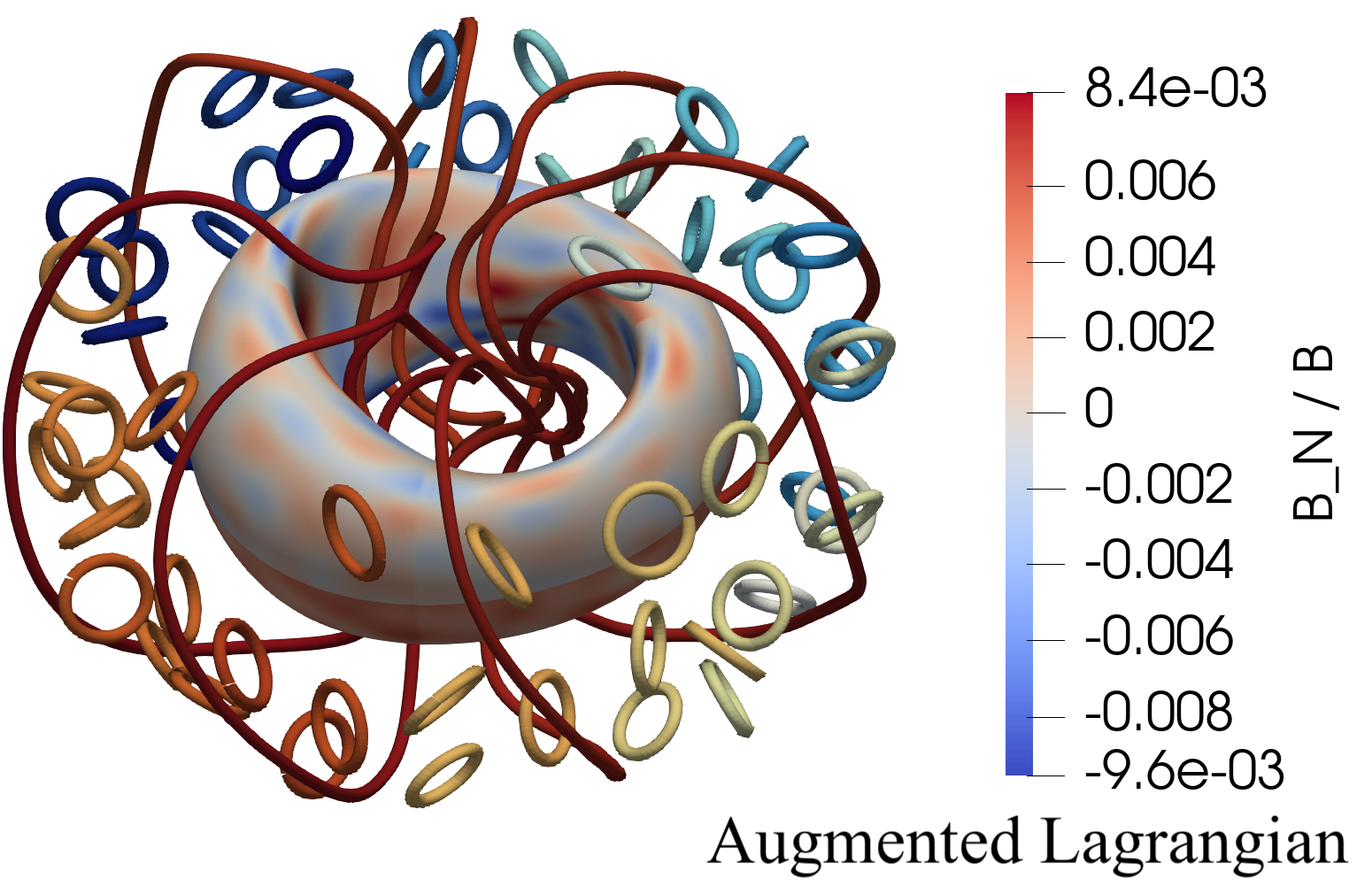}
    
    \includegraphics[width=\linewidth]{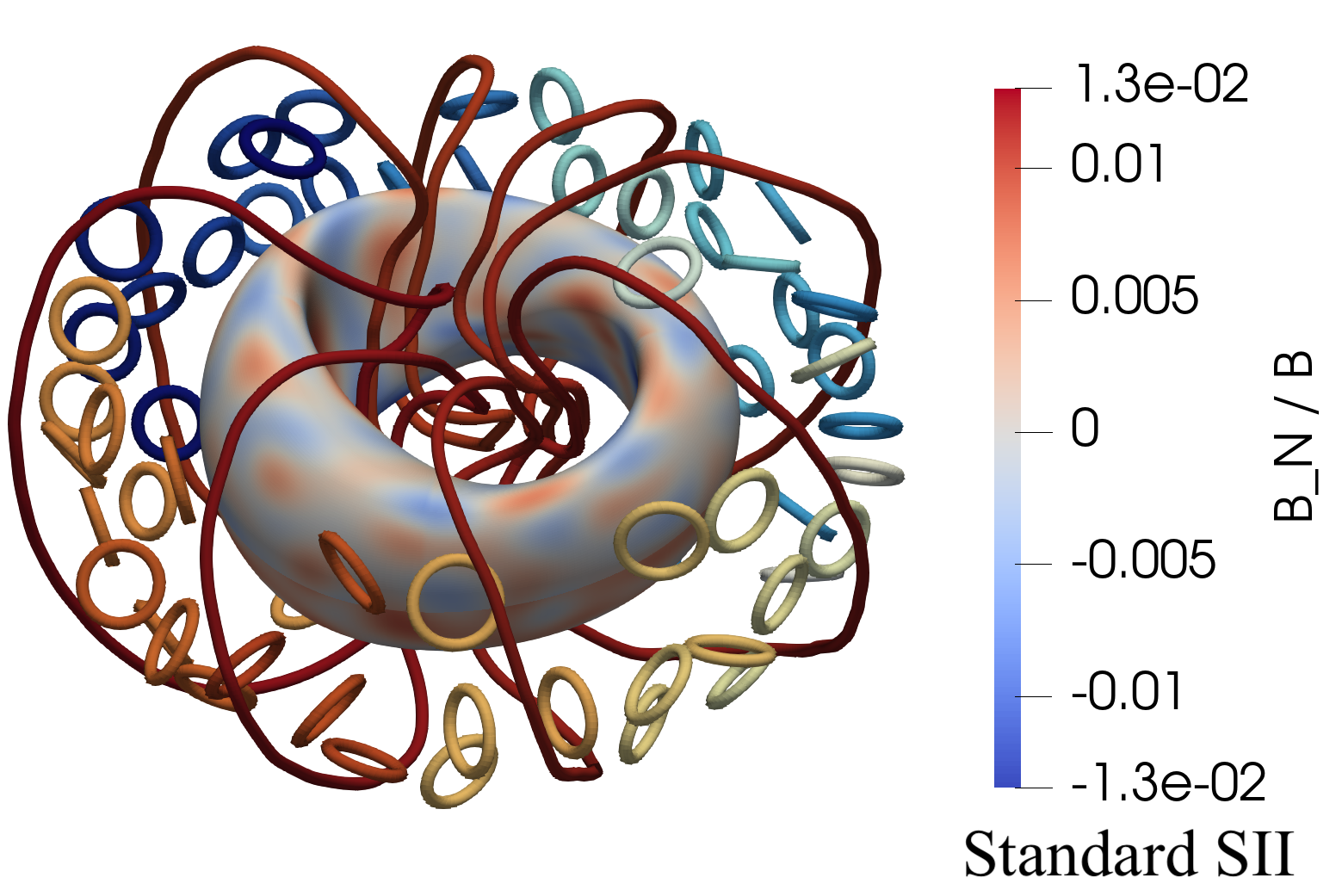}
    \caption{Coils and target surfaces for the Schuett-Henneberg tokamak-stellarator hybrid.}
    \label{fig:schuetthenneberg_auglag}
\end{figure}

In order to illustrate that our method can easily integrate with any existing functionality in the coil optimization literature, we now present dipole optimizations using the Schuett-Henneberg (tokamak-stellarator hybrid) QA equilibrium \cite{schuett_henneberg} as target field following a similar set-up as \cite{Kaptanoglu_2025}. Two optimizations are performed, one with the augmented Lagrangian Method and one with a standard stage two script. Both use 16 planar coils per half-field period as field shaping coils. Additionally, two non-planar toroidal field coils per half-field period are optimized. 
The toroidal field and dipole coils are jointly optimized and the dipole coils can move in space and change current, but are identical fixed-shape circular coils.
The optimization targets and results used are presented in Table \ref{tab:schuett_henneberg}. The resulting configurations are shown in Figure \ref{fig:schuetthenneberg_auglag}.

\begin{table}[ht!]
\centering
\begin{tabular}{|l|c|c|c|}
\hline
Property                              
& Aug. Lag. 
& Std. & Thresh.\\ 
\hline
$\#$ of coils per hfp $\downarrow$                                   
& 18 & 18 & - \\ 
$\langle \bm B\cdot \bm n\rangle / \langle B \rangle \times 10^{-3}$  $\downarrow$ 
& $\bm{1.9}$ & 2.6 & 0\\ 
Max $(\bm B\cdot \bm n / B)\times 10^{-3}$  $\downarrow$                          
& $\bm{9.6}$ & 13 & 0\\ 
TF Coil Length per hfp [m]   $\downarrow$                       
& $\bm{80.04}$ & 80.47 & 80  \\ 
Min CC Dist. (All coils) [m]   $\uparrow$                       
& 0.46 & $\bm{0.5}$ &  0.5   \\
Min CC Dist. (TF coils) [m]   $\uparrow$                       
& 0.93 & $\bm{1}$ &  1   \\ 
Min CS Dist. [m] $\uparrow$                 
&1.43& $\bm{1.48}$ & 1.5 \\ 
Max $\kappa$ [$\text{m}^{-1}$] $\downarrow$       
&  $\bm{0.59}$  & 1.68  & 0.5 
\\ 
Max MSC [$\text{m}^{-1}$]     $\downarrow$                   
& $\bm{0.13}$ & 0.14 & 0.05\\
\hline
\end{tabular}
\caption{Comparison of magnetic field and coil properties for the two field period Schuett-Hennerg [1] QA equilibrium. Apart from the coil-to-coil and coil-to-surface distances which remain within 10\% tolerance of the target field, the augmented Lagrangian outperforms the standard stage II optimization in most of the metrics.
}
\label{tab:schuett_henneberg}
\end{table}

Note that the standard stage II optimization is not performed with random weights, but with already chosen weights to give better performance following the script used in \cite{Kaptanoglu_2025}, however without force and torque optimizations for the coils. The sole modification was to replace the minimizer of the original script the augmented Lagrangian method and it performed as expected: apart from the coil-to-coil and coil-to-surface distances which remain within 10\% tolerance of the target field, the augmented Lagrangian outperforms the standard stage II optimization in most of the metrics. Notably it reaches lower field errors with curves that exhibit much smaller maximum curvature and even 50 cm less conductor length.
\section{Comparison with Quadratic Penalty Methods}\label{sec:QPM_comparison}

In order to benchmark our work against standard penalty methods, we decided to perform a comparison between the augmented Lagrangian algorithm in question and the Quadratic Penalty Method (QPM) by Nocedal \& Wright Framework 17.1 \cite{NocedalWright2006}:

For each outer iteration the inner loop minimises the augmented objective

\begin{equation}
    Q(x,\mu) = f(x) +\frac{1}{2}\sum_i\mu_i c_i(x)^2
\end{equation}
with L-BFGS-B to a tolerance $\tau_{\kappa}$ and between outer iterations the penalty grows: if the previous inner solve converged in at most 1000 iterations, $\mu_{\kappa+1} = 10\mu_{\kappa}$; otherwise $\mu_{\kappa+1} = 2\mu_{\kappa}$, $\tau_{\kappa}$ is halved each outer step toward $10^{-8}$, this follows strictly the implementation presented in Framework 17.1. All the solves had 100 outer loop iterations ensuring that both algorithms converged.

The constraints are the same ones as used previously in this work with addition of the arclength penalty that penalizes the quadrature point arclength variation.

We therefore present the results of the two mentioned coil optimization algorithms applied to the same stellarator coil-design problem, with exactly the same constraints, run on two target surfaces (for a QA and a QI stellarator) and a sweep of total coil-length thresholds. Both algorithms start from the same physical coilset (the same build problem instance, with coils placed along the magnetic axis for the QI surface) and use identical initial penalty parameters ($\mu_0 = 10$ broadcast scalar), so any difference observed is a consequence of the update rule, not of the starting point.

As a filtering policy, we only include here configurations that did not exhibit an individual constraint violation above 20\% in order to be able to perform the comparison. As a figure of merit, we compare the maximum force on coil versus the normalized squared flux just like in the 
submitted manuscript. 

Each data-point corresponds to one coilset from a full ALM or QPM run, the thresholds are fixed during each optimization. Increasing the coil length upper threshold yields the different data-points for both the ALM and QPM. The arrows indicate how the data-points move by increasing upper length threshold for the ALM (red) and the QPM (blue) respectively. The runs were done for a maximum [15, 17.5, 20, 22.5, 25]m total coil length per hfp for the QA equilibrium and [12, 13, 14, 15, 16]m for the QI equilibrium, both with 4 coils per half-field period.  So, the first data-points of the QA ALM and QPM series were optimized for a maximum coil length of 15m, the second for 17.5m and so on. 

The colorbar indicates the actual realized total coil length per half-field period in the final optimized coilsets.

\begin{figure}[h!]
    \centering
    \includegraphics[width=\linewidth]{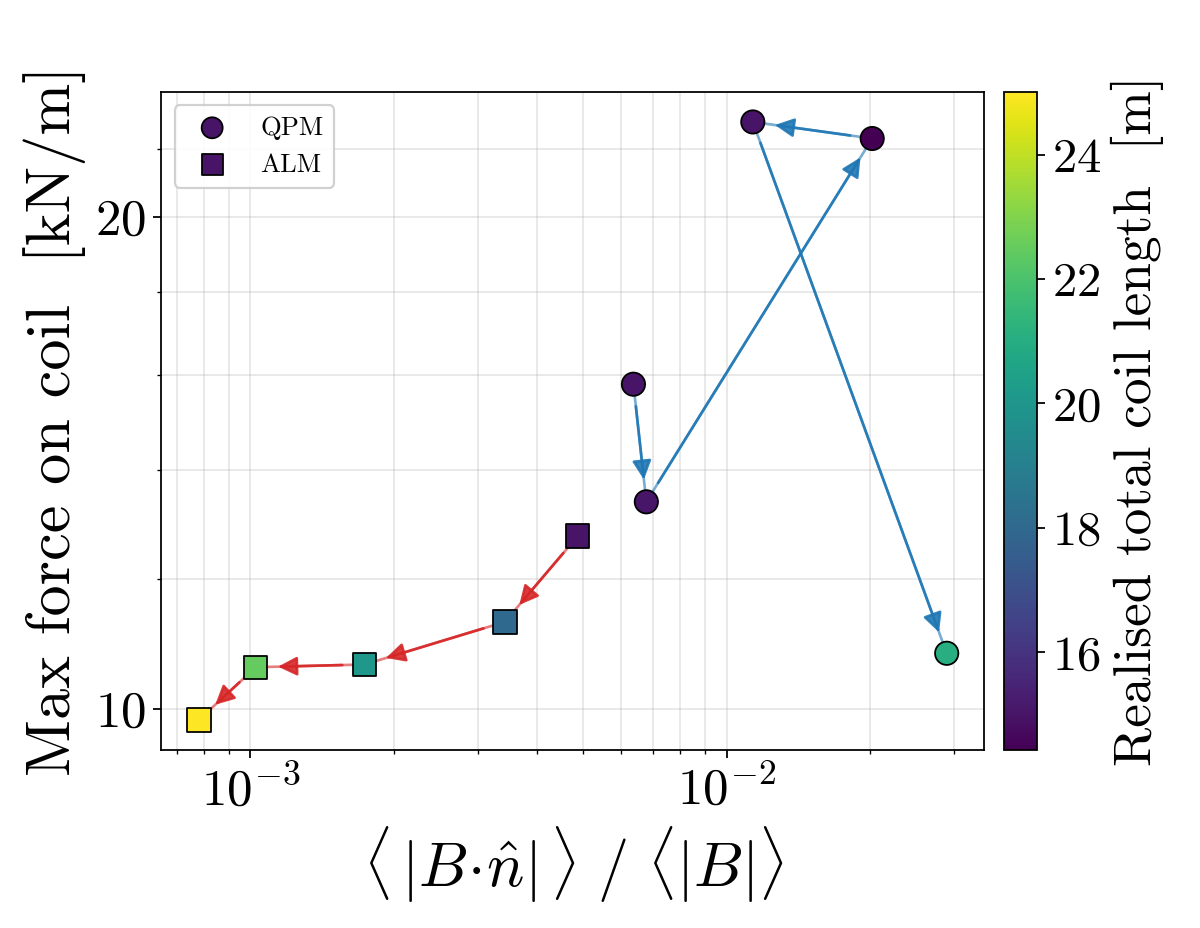}
    \includegraphics[width=\linewidth]{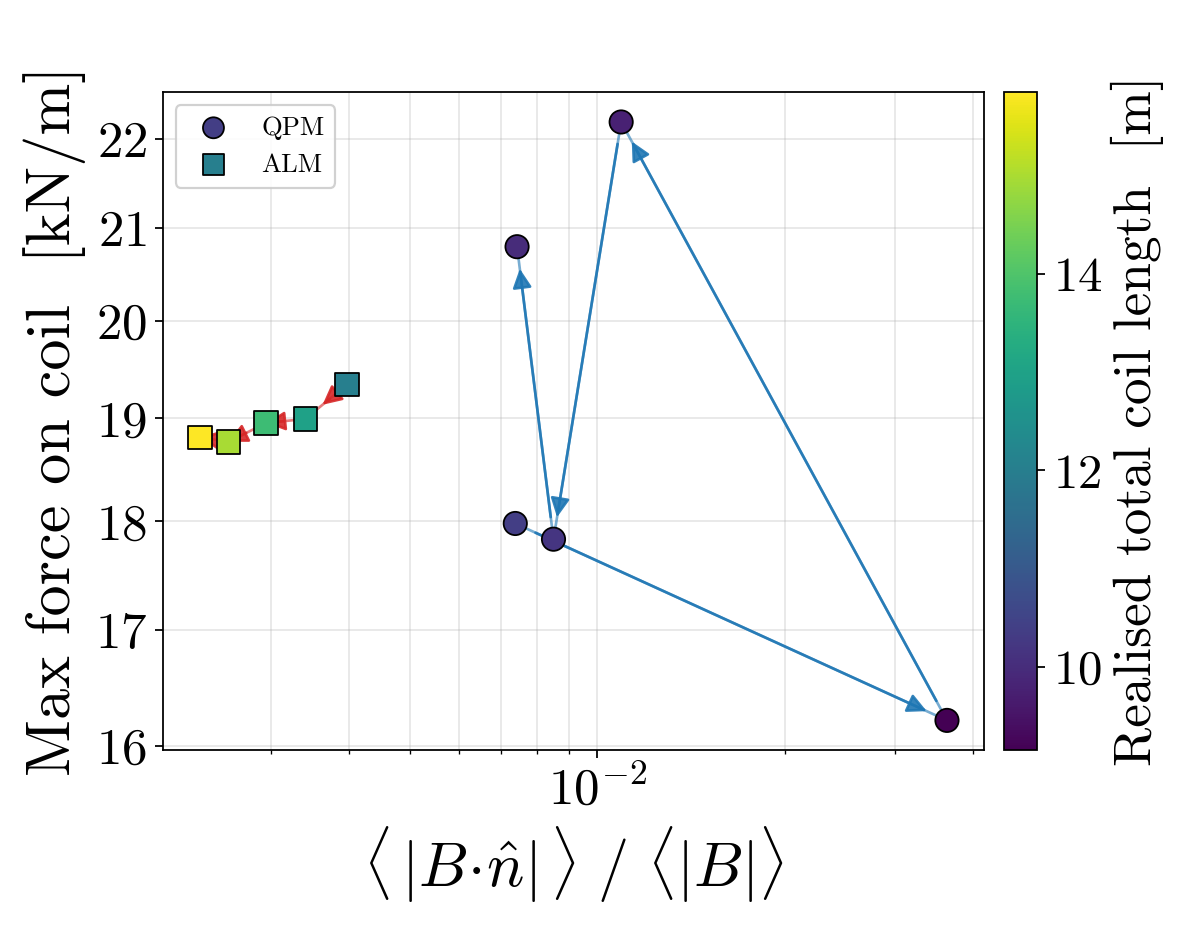}
    \caption{\textbf{Top}: Comparison between the Quadratic Penalty Method (QPM) and the Augmented Lagrangian Method (ALM), shown is the maximum force on coil versus normalized squared flux for varying coil lengths thresholds for the LandremanPaul QA configuration \cite{LandremanPaul}. \textbf{Bottom}: Same plot, but for completeness, the optimizations are performed on a Quasi-Isodynamic configuration \cite{Goodman_Camacho}. Note that the colorbar corresponds to the realized total coil lengths of the coilsets.}
    \label{fig:comparison_plots}
\end{figure}

The augmented Lagrangian (red arrows) appears in both cases to improve consistently both in squared flux and in maximum force as the length constraint is relaxed. The QPM trajectory however appears to be erratic as the length threshold is relaxed. As a reminder, the arrows in Figure \ref{fig:comparison_plots} show how the algorithm behaves by progressively relaxing the coil length constraint, i.e. by allowing the coils to get bigger. The irregular behavior of the QPM trajectory can be explained by observing that the realized coil length for the QPM appears to stay stuck even as the coil length threshold is relaxed. Moreover, the ALM pushes the coils to match the exact coil length threshold each time, as they are tight constraints. This is a revealing difference in behavior, and these finds strongly suggest the ALM algorithm regularly outperforms QPM on this class of optimization problems.

One possible explanation behind the QPM getting stuck is that it does not possess the additional decision node regarding the constraint violation that the ALM has, that comes after one minimizer solve. This results in always increasing magnitudes of the penalty parameters $\mu_p$ of the violated constraints, leaving the unviolated constraints, like the coil length, untouched. However in the ALM, depending on a true or false constraint violation up to a certain threshold $\eta_p$, it has the additional freedom of updating both the Lagrange multipliers and the tightening the constraint violation threshold $\eta_p$ while keeping the penalty
parameters $\mu_k$ untouched. The QPM algorithm does not possess \textit{adaptive constraint violation thresholds}.

Lastly, we make no claim that ALM \textit{always} outperforms QPM. This depends on the nonconvexity and structure of the problem, and we have only shown evidence that on typical apples-to-apples problems in stellarator coil optimization, ALM appears to outperform QPM.

\bibliography{thebibliography}

\end{document}